\documentstyle[twoside,epsfig,amstex,amssymb,rotating,hhline]{article}

\catcode`\@=11
\long\def\@makefntext#1{
\protect\noindent \hbox to 3.2pt {\hskip-.9pt  
$^{{\eightrm\@thefnmark}}$\hfil}#1\hfill}               

\def\thefootnote{\fnsymbol{footnote}}
\def\@makefnmark{\hbox to 0pt{$^{\@thefnmark}$\hss}}    
        
\def\ps@myheadings{\let\@mkboth\@gobbletwo
\def\@oddhead{\hbox{}
\rightmark\hfil\eightrm\thepage}   
\def\@oddfoot{}\def\@evenhead{\eightrm\thepage\hfil
\leftmark\hbox{}}\def\@evenfoot{} 
\def\sectionmark##1{}\def\subsectionmark##1{}}

\oddsidemargin=\evensidemargin
\renewcommand{\thefootnote}{\fnsymbol{footnote}}

\newcounter{sectionc}\newcounter{subsectionc}\newcounter{subsubsectionc}
\renewcommand{\section}[1] {\vspace{12pt}\addtocounter{sectionc}{1} 
\setcounter{subsectionc}{0}\setcounter{subsubsectionc}{0}\noindent 
        {\tenbf\thesectionc. #1}\par\vspace{5pt}}
\renewcommand{\subsection}[1] {\vspace{12pt}\addtocounter{subsectionc}{1} 
        \setcounter{subsubsectionc}{0}\noindent 
        {\bf\thesectionc.\thesubsectionc. {\kern1pt \bfit #1}}\par\vspace{5pt}}
\renewcommand{\subsubsection}[1] {\vspace{12pt}\addtocounter{subsubsectionc}{1}
        \noindent{\tenrm\thesectionc.\thesubsectionc.\thesubsubsectionc.
        {\kern1pt \tenit #1}}\par\vspace{5pt}}
\newcommand{\nonumsection}[1] {\vspace{12pt}\noindent{\tenbf #1}
        \par\vspace{5pt}}

\newcounter{appendixc}
\newcounter{subappendixc}[appendixc]
\newcounter{subsubappendixc}[subappendixc]
\renewcommand{\thesubappendixc}{\Alph{appendixc}.\arabic{subappendixc}}
\renewcommand{\thesubsubappendixc}
        {\Alph{appendixc}.\arabic{subappendixc}.\arabic{subsubappendixc}}

\renewcommand{\appendix}[1] {\vspace{12pt}
        \refstepcounter{appendixc}
        \setcounter{figure}{0}
        \setcounter{table}{0}
        \setcounter{lemma}{0}
        \setcounter{theorem}{0}
        \setcounter{corollary}{0}
        \setcounter{definition}{0}
        \setcounter{equation}{0}
        \renewcommand{\thefigure}{\Alph{appendixc}.\arabic{figure}}
        \renewcommand{\thetable}{\Alph{appendixc}.\arabic{table}}
        \renewcommand{\theappendixc}{\Alph{appendixc}}
        \renewcommand{\thelemma}{\Alph{appendixc}.\arabic{lemma}}
        \renewcommand{\thetheorem}{\Alph{appendixc}.\arabic{theorem}}
        \renewcommand{\thedefinition}{\Alph{appendixc}.\arabic{definition}}
        \renewcommand{\thecorollary}{\Alph{appendixc}.\arabic{corollary}}
        \renewcommand{\theequation}{\Alph{appendixc}.\arabic{equation}}
        \noindent{\tenbf Appendix \theappendixc #1}\par\vspace{5pt}}
\newcommand{\subappendix}[1] {\vspace{12pt}
        \refstepcounter{subappendixc}
        \noindent{\bf Appendix \thesubappendixc. {\kern1pt \bfit #1}}
        \par\vspace{5pt}}
\newcommand{\subsubappendix}[1] {\vspace{12pt}
        \refstepcounter{subsubappendixc}
        \noindent{\rm Appendix \thesubsubappendixc. {\kern1pt \tenit #1}}
        \par\vspace{5pt}}

\newcommand{\textlineskip}{\baselineskip=13pt}
\newcommand{\smalllineskip}{\baselineskip=10pt}

\def\eightcirc{
\begin{picture}(0,0)
\put(4.4,1.8){\circle{6.5}}
\end{picture}}
\def\eightcopyright{\eightcirc\kern2.7pt\hbox{\eightrm c}} 

\newcommand{\copyrightheading}[1]
        {\vspace*{-2.5cm}\smalllineskip{\flushleft
        {\footnotesize International Journal of Modern Physics A, #1}\\
        {\footnotesize $\eightcopyright$\, World Scientific Publishing
         Company}\hfill OUNP-2001-03\hfill\\
         }}

\def\abstracts#1#2#3{{
        \centering{\begin{minipage}{4.5in}\baselineskip=10pt\footnotesize
        \parindent=0pt #1\par 
        \parindent=15pt #2\par
        \parindent=15pt #3
        \end{minipage}}\par}} 

\newcommand{\bibit}{\nineit}

\renewenvironment{thebibliography}[1]
        {\frenchspacing
         \ninerm\baselineskip=11pt
         \begin{list}{\arabic{enumi}.}
        {\usecounter{enumi}\setlength{\parsep}{0pt}
         \setlength{\leftmargin 12.7pt}{\rightmargin 0pt} 
         \setlength{\itemsep}{0pt} \settowidth
        {\labelwidth}{#1.}\sloppy}}{\end{list}}

\newcounter{itemlistc}
\newcounter{romanlistc}
\newcounter{alphlistc}
\newcounter{arabiclistc}

\newcommand{\fcaption}[1]{
        \refstepcounter{figure}
        \setbox\@tempboxa = \hbox{\footnotesize Fig.~\thefigure. #1}
        \ifdim \wd\@tempboxa > 5in
           {\begin{center}
        \parbox{5in}{\footnotesize\smalllineskip Fig.~\thefigure. #1}
            \end{center}}
        \else 
             {\begin{center}
             {\footnotesize Fig.~\thefigure. #1}
              \end{center}}
        \fi}

\newcommand{\tcaption}[1]{
        \refstepcounter{table}
        \setbox\@tempboxa = \hbox{\footnotesize Table~\thetable. #1}
        \ifdim \wd\@tempboxa > 5in
           {\begin{center}
        \parbox{5in}{\footnotesize\smalllineskip Table~\thetable. #1}
            \end{center}}
        \else
             {\begin{center}
             {\footnotesize Table~\thetable. #1}
              \end{center}}
        \fi}

\def\@citex[#1]#2{\if@filesw\immediate\write\@auxout
        {\string\citation{#2}}\fi
\def\@citea{}\@cite{\@for\@citeb:=#2\do
        {\@citea\def\@citea{,}\@ifundefined
        {b@\@citeb}{{\bf ?}\@warning
        {Citation `\@citeb' on page \thepage \space undefined}}
        {\csname b@\@citeb\endcsname}}}{#1}}

\newif\if@cghi
\def\cite{\@cghitrue\@ifnextchar [{\@tempswatrue
        \@citex}{\@tempswafalse\@citex[]}}
\def\citelow{\@cghifalse\@ifnextchar [{\@tempswatrue
        \@citex}{\@tempswafalse\@citex[]}}
\def\@cite#1#2{{$\null^{#1}$\if@tempswa\typeout
        {IJCGA warning: optional citation argument 
        ignored: `#2'} \fi}}

\def\pmb#1{\setbox0=\hbox{#1}
        \kern-.025em\copy0\kern-\wd0
        \kern.05em\copy0\kern-\wd0
        \kern-.025em\raise.0433em\box0}

\def\fnt#1#2{\footnotetext{\kern-.3em
        {$^{\mbox{\scriptsize #1}}$}{#2}}}

\def\fpage#1{\begingroup
\voffset=.3in
\thispagestyle{empty}\begin{table}[b]\centerline{\footnotesize #1}
        \end{table}\endgroup}

\def\runninghead#1#2{\pagestyle{myheadings}
\markboth{{\protect\footnotesize\it{\quad #1}}\hfill}
{\hfill{\protect\footnotesize\it{#2\quad}}}}
\font\tenrm=cmr10
\font\tenit=cmti10 
\font\tenbf=cmbx10
\font\bfit=cmbxti10 at 10pt
\font\ninerm=cmr9
\font\nineit=cmti9

\font\eightrm=cmr8

\def\qed{\hbox{${\vcenter{\vbox{                        
   \hrule height 0.4pt\hbox{\vrule width 0.4pt height 6pt
   \kern5pt\vrule width 0.4pt}\hrule height 0.4pt}}}$}}

\renewcommand{\thefootnote}{\fnsymbol{footnote}}        

\def\nm{\nu_\mu}

\def\ne{\nu_e}

\def\nt{\nu_\tau}

\def\Dm2{\Delta m^2}

\def\evolt{e\text{V}}
\def\kev{\text{k}e\text{V}}
\def\ev2c4{e\text{V}^2/c^4}
\def\mev{\text{M}e\text{V}}
\def\gev{\text{G}e\text{V}}

\begin{document}

\runninghead{An Experimentalist's View of $\ldots$} {An Experimentalist's View of $\ldots$}

\normalsize\textlineskip
\thispagestyle{empty}
\setcounter{page}{1}

\copyrightheading{}                     

\vspace*{0.88truein}

\fpage{1}
\centerline{\bf AN EXPERIMENTALIST'S VIEW OF NEUTRINO OSCILLATIONS}
\vspace*{0.37truein}
\centerline{\footnotesize ANTONELLA DE SANTO}
\vspace*{0.015truein}
\centerline{\footnotesize\it  Sub-Department of Nuclear and Particle Physics,
  University of Oxford, 1 Keble Road}
\baselineskip=10pt
\centerline{\footnotesize\it Oxford, OX1 3RH, United Kingdom}
\vspace{0.5cm}
\centerline{Invited paper to appear in {\bf Intern. J.
    Mod. Phys. A} (2001)}

\vspace*{0.21truein}
\abstracts{
Neutrinos, and primarily neutrino oscillations, have undoubtedly been one
of the most exciting topics in the field of high-energy physics over the past
few years. 
The existence of neutrino oscillations would require an
extension of the currently accepted description 
of sub-nuclear phenomena beyond the Standard Model. 
Compelling evidence of new physics,
which seems to be pointing towards neutrino oscillations, is coming
from the solar neutrino deficit and from the atmospheric neutrino
anomaly. More controversial
effects have been observed with artificially produced neutrinos. The
present experimental status of neutrino oscillations is reviewed, 
as well as the planned future experimental
programme, which, it is hoped, will solve most of the outstanding puzzles.
}{}{}

\vspace*{1pt}\textlineskip      
\section{Introduction}    
\vspace*{-0.5pt}
\noindent
Since the 
discovery of the electron neutrino in 1955\cite{reines_cowan}, 
the study of neutrino physics
has often led to great progress in the unravelling
of sub-nuclear phenomena. However, although they have been used as 
powerful probes of elementary particles interactions, 
very little is known about the general properties of neutrinos: we
know neither their mass, 
nor their magnetic moment; we don't know whether they are
Dirac or Majorana particles; we don't know how many neutrino
species exist in nature. In many respects, neutrinos perhaps remain
the most intriguing mystery in particle physics.

In more recent years most of the research activity in 
neutrino physics
has focused on the phenomenon of neutrino oscillations, 
originally postulated by Bruno Pontecorvo.\cite{pontecorvo}
Interest in this field has been revived by several very interesting
experimental results, which in some cases very strongly support the
hypothesis of neutrino oscillations. This is only possible if, 
contrary to what assumed in the Standard Model
of electro-weak interactions, neutrinos have non-zero and
non-degenerate masses. Thus, evidence 
for neutrino oscillations would represent a major discovery in
elementary particle physics and would then require a complete revolution in our
understanding of the natural world. 

This is a very exciting time for neutrino physics, with new
data becoming available in a relatively short 
period of time. Whilst this paper was being completed, the
long-awaited first results from the SNO experiment were
presented.\cite{sno_first_results} We hope that many others will follow,
from all the running or planned experiments, and that more data
will soon lead to the solution of this fascinating puzzle of particle physics.

\textheight=7.8truein
\setcounter{footnote}{0}
\renewcommand{\thefootnote}{\alph{footnote}}

In this paper, after a brief introduction to massive neutrinos and to
the formalism of neutrino oscillations, 
the latest experimental results will be reviewed.
The focus will be on those experiments which have added pieces of
information to the current picture of neutrino oscillation
phenomenology, either by observing a signal or by setting a limit in
an interesting portion of the relevant parameter space. Planned
experiments, designed to resolve some of the as yet unsolved ambiguities,
will also be described.

\section{Neutrino Masses}

The present experimental limits on neutrino masses,
as obtained from direct kinematic searches, under the assumption that
flavour and mass eigenstates are essentially coincident, 
are the following :\cite{troitsk,mainz,pdg,lep_nutau}
\begin{alignat}{3}\label{eq:direct_limits}
m_{\ne} &< 2.2~\evolt &
\qquad&\text{ (95\%~C.L., from $^3\text{H}\rightarrow ^3\hspace{-3pt}{\text{He}} + e^-+\bar{\nu}_e$);} \notag\\
m_{\nm} &< 170~\kev &
\qquad&\text{ (90\%~C.L., from $\pi^+\rightarrow\mu^++\nu_\mu$)}\\
m_{\nt} &< 15.5~\mev &
\qquad&\text{ (95\%~C.L., from $\tau\rightarrow 5\pi(\pi^0)+\nu_\tau$)\,.}\notag
\end{alignat}

Although the smallness of these limits might lead to the naive conclusion that
neutrino masses are exactly vanishing, there is no theoretical reason
to assume that this is indeed the case. On the contrary, it is generally
believed that particle masses can be identically equal to zero only if 
they are associated with an exact gauge symmetry. This is not true
for the lepton quantum number, which is then expected not to be
conserved.

Perhaps the most popular expression for the neutrino mass term is
the one associated with the so-called {\it see-saw}\cite{see-saw_refs} 
form of the mass matrix. In the simple case of one generation this
mass term is given by: 
\begin{equation}\label{eq:see-saw_matrix}
  \begin{pmatrix}\nu_L, & \nu_R\end{pmatrix} 
  \begin{pmatrix}0 & m_D \\ m_D & M_M\end{pmatrix} 
  \begin{pmatrix}\nu_L\\ \nu_R\end{pmatrix}\,,
\end{equation}

\noindent
where a right-handed neutrino singlet, $\nu_R$, has been
introduced. Assuming that the Dirac mass $m_D$ is much smaller than the
Majorana mass $M_M$\footnote{in the generic $n-$generation case $m_D$
  and $M_M$ are $n\times n$ matrices: then $m_D\ll M_D$ means
  that all eigenvalues of $m_D$ are small compared to all eigenvalues
  of $M_M$.}, 
the diagonalisation of (\ref{eq:see-saw_matrix})
gives a heavy eigenstate, of mass $\sim M_M$, and a light 
eigenstate, which can be identified with the standard left-handed
neutrino and whose mass is given by:
\begin{equation}\label{eq:see-saw_numass}
  m_\nu \simeq \frac{m_D^2}{M_M}\,.
\end{equation}

If the Dirac mass is of the same order as that of a fundamental
fermion, be it a quark or a charged lepton ($m_D\sim O(m_{q,\ell})$),
and if $M_M\sim \text{O}(M_{GUT})$, then the
neutrino mass turns out to be naturally very small, and in a range of
values compatible with those indicated by experimental results.

Several versions of the see-saw mechanism are considered in
literature.\cite{see-saw_quadr_lin} Among them, the so-called 
{\it quadratic see-saw}, 
in which all the Majorana masses are similar to each other, and therefore
the light neutrino masses scale as the Dirac masses squared:
\begin{equation}
m_{\nu_1} : m_{\nu_2} : m_{\nu_3} ~\simeq ~ m_{q_1,\,\ell_1}^2  :
m_{q_2,\,\ell_2}^2 :  m_{q_3,\,\ell_3}^2 
\end{equation}

\noindent
Another possibility is the {\it linear see-saw}, where $M_M \propto m_D$ and then:
\begin{equation}
m_{\nu_1} : m_{\nu_2} : m_{\nu_3} ~\simeq ~ m_{q_1,\,\ell_1}  :
m_{q_2,\,\ell_2} :  m_{q_3,\,\ell_3}\,,
\end{equation}

\noindent
Both versions imply a hierarchy between the
light neutrino mass eigenstates:
\begin{equation}\label{eq:mass_hierarchy}
m_{\nu_1} \ll m_{\nu_2} \ll m_{\nu_3}\,.
\end{equation}

\section{Neutrino Oscillations in Vacuum}

Ordinary neutrinos are produced in weak processes together with 
their charged partner leptons ($W^+ \rightarrow \ell^+\nu_\ell$), as
for example in pion decays or in nuclear beta decays, 
and they are therefore eigenstates of the weak interactions. On the
other hand, weak interaction eigenstates and mass eigenstates are not
necessarily coincident.
With a change of basis, neutrinos produced as flavour eigenstates can be
represented as a coherent linear superposition of the mass eigenstates:
\begin{equation}
|\,\nu_\alpha\,\rangle = \sum_{i}U^{\, *}_{\alpha i}\ |\,\nu_i\,\rangle
\hspace{2cm} (\alpha = e, \mu, \tau)
\end{equation}

\noindent
where $i$ runs over all the existing mass eigenstates and 
$U_{\alpha i}$ are the elements of a 
unitary matrix.\footnote{
The convention of using Greeks letters to identify flavour eigenstates 
and Latin letters to identify mass eigenstatates has been adopted
throughout this paper.}

Although from measurements of the $Z^0$ width at LEP we know that the number of
light active neutrinos is ${\mathcal{N}}_\nu = 2.994\pm 0.012$
\cite{pdg}, nothing forbids the existence of more than 
three neutrinos of definite mass. If, for example, there were
four independent mass eigenstates, 
a change of basis from the mass to the flavour eigenstates would
necessarily lead to one linear combination, 
\begin{equation}\label{eq:sterile}
|\,\nu_s\,\rangle = \sum_{i}U^{\, *}_{si}\ |\,\nu_i\,\rangle\,,
\end{equation}
\noindent
which has no normal weak coupling. The linear combination
(\ref{eq:sterile}) is therefore referred to as a ``sterile neutrino'',
as opposed to the standard ``active'' neutrino. 

As we shall see in the next sections, the existence of at
least one sterile neutrino might be the only way to accommodate all
the experimental results on neutrino oscillations. Nonetheless, for
the sake of simplicity, in the
following discussion we will
only consider the three neutrino flavours 
whose existence has been firmly established.\footnote{the existence of
  the $\tau$ neutrino, of
  which only indirect evidence was available until very recent times,
  has been proven by the emulsion-based DONUT experiment,\cite{donut}
  at Fermilab.}\hspace{6pt}However, it must be noted that the same formalism
would be valid for any number of neutrinos.

After a neutrino has been prepared in the state
$|\,\nu_\alpha\,\rangle$  at $t=0$, each of its
components will evolve according to the Schr$\ddot{\text o}$dinger
equation. At a time $t$ after production it will be:
\begin{equation}
  |\,\nu_\alpha\,(t>0)\rangle = 
\sum_{i}e^{-iE_it}\ U^{\, *}_{\alpha i}\ |\,\nu_i\,\rangle = 
\sum_{\beta}\sum_{i}U_{\beta i}\ e^{-iE_it}\ U^{\, *}_{\alpha i}\ |\,\nu_i\,\rangle\,,
\end{equation}

Thus, after the
neutrino produced in the flavour eigenstate $\alpha$ 
has travelled a distance $L\simeq t$, the probability of finding
it in a different flavour state $\beta$ is non-zero and is given by:
\begin{equation}\label{eq:osc_prob_1}
{\mathcal{P}}(\nu_\alpha\rightarrow\nu_\beta) = |\langle\,\nu_\beta(t)|
\,\nu_\alpha\,\rangle|^2 = 
\begin{vmatrix} \sum_i U_{\beta i}\ e^{-iE_it}\ U^{\, *}_{\alpha
    i}\end{vmatrix}^2\,.
\end{equation}

Using the unitarity condition for $U$,
after some algebra 
the probability (\ref{eq:osc_prob_1}) takes the following oscillatory form:
\begin{equation}\label{eq:osc_prob_2}
{\mathcal{P}}(\nu_\alpha\rightarrow\nu_\beta) = 
\sum_i |U_{\beta i}|^2 |U_{\alpha i}|^2 +
2Re\sum_{j>i}U_{\beta i}U_{\beta j}^{*}U_{\alpha i}^{*}U_{\alpha j}
\exp\Big(-i\frac{\Delta m_{jk}^2 L}{2E}\Big)\,,
\end{equation}

\noindent
where $\Delta m_{ij}^2 = |m_i^2-m_j^2|$ and 
the ultra-relativistic expansion for the neutrino energy has been used.\footnote{ 
$E_i=\sqrt{p^2+m_i^2}\simeq p+m_i^2/2p\simeq p+m_i^2/2E$. 
}
Thus the transition probability 
${\mathcal{P}}(\nu_\alpha\rightarrow\nu_\beta)$  depends on 
a combination of the mixing matrix elements, on the squared difference
between the mass eigenvalues of the mass eigenstates, and on two parameters, 
the neutrino energy $E$ and the distance $L$ between the production and the
observation points, which are determined by the experimental conditions.

Although from the simplicity of the preceding discussion 
it might seem 
that neutrino oscillations can be obtained as
a straightforward consequence of elementary quantum
mechanics, a rigorous treatment of the
problem would require that a wave packet formalism be considered.
We shall not discuss this approach here and 
the interested reader is remanded, for example, to 
ref.\cite{giunti_1} and refs. therein.

The lepton mixing  matrix $U$, 
also known as the Maki-Nakagawa-Sakata (MNS) matrix,\cite{mns_matrix}
is the leptonic analogue of the CKM matrix for the quark sector.
In the case of three neutrino flavours the explicit relation between the
flavour and the mass eigenstates is given by:
\begin{equation}
  \begin{pmatrix}\nu_e\\ \nu_\mu \\ \nu_\tau\end{pmatrix} = 
  \begin{pmatrix}
    U_{e1} & U_{e2} & U_{e3}\\
    U_{\mu 1} & U_{\mu 2} & U_{\mu 3}\\
    U_{\tau 1} & U_{\tau 2} & U_{\tau 3}\\
  \end{pmatrix}
\begin{pmatrix}\nu_1 \\ \nu_2 \\ \nu_3\end{pmatrix}
\,.\
\end{equation}

It is sometimes convenient to parametrize $U$ in a form analogous to
that of the CKM
matrix,\cite{pdg} which for Dirac neutrinos is:
\begin{equation}\label{eq:lepton_mixing}
  U = 
  \begin{pmatrix}
    c_{12}c_{13} & 
    s_{12}c_{13} & 
    s_{13}e^{-i\delta}  \\
    -s_{12}c_{23}-c_{12}s_{23}s_{13}e^{i\delta} &
    c_{12}c_{23}-s_{12}s_{23}s_{13}e^{i\delta}  &
    s_{23}c_{13}\\
    s_{12}s_{23}-c_{12}c_{23}s_{13}e^{i\delta}  &
    -c_{12}s_{23}-s_{12}c_{23}s_{13}e^{i\delta} &
    c_{23}c_{13}
  \end{pmatrix}\,,
\end{equation}
\noindent
with the standard convention $c_{ij} = \cos\theta_{ij}$ and $s_{ij} =
\sin\theta_{ij}$ ($i,j=1,2,3$ are the generation labels). 
Thus in this formalism the mixing matrix depends only on three mixing angles,
$\theta_{12}$, $\theta_{13}$ and $\theta_{23}$, and a CP-violating
phase, $\delta$.\footnote{in the case of Majorana
neutrinos there would be two additional phases.}

Although the most general description of the mixing among all
neutrinos implies the three-flavour treatment, most experimental results
are expressed in the two generation mixing representation. 
More recent analyses, however, generally attempt a global fit
to all data assuming three-flavour mixing.

In the case of transitions between two flavours the mixing matrix has
the form of a rotation in a two-dimensional space:
\begin{equation}
  U = 
  \begin{pmatrix}
    \cos\theta & \sin\theta\\
    -\sin\theta & \cos\theta 
  \end{pmatrix}\,,
\end{equation}

\noindent
where $\theta$ is the mixing angle, analogous to the Cabibbo angle for 
the quarks. The relation between the flavour and the mass eigenstate
is then given by
\begin{equation}\label{eq:mixing_two_flav_vacuum}
  \begin{matrix}
    |\,\nu_\alpha\,\rangle & = & \,\,\,\,\,\cos\theta\,|\,\nu_1\,\rangle & + &
    \sin\theta\,|\,\nu_2\,\rangle \\
    |\,\nu_\beta\,\rangle & = & -\sin\theta\,|\,\nu_1\,\rangle & + &
    \cos\theta\,|\,\nu_2\,\rangle \\
  \end{matrix}\,.
\end{equation}

\noindent
The oscillation probability in this case can be written as follows:
\begin{equation}\label{eq:osc_prob_2flav_1}
  {\mathcal{P}}(\nu_\alpha\rightarrow\nu_\beta) = 
  \sin^22\theta~\sin^2\Big(\pi\frac{L}{\lambda_{osc}}\Big)
\end{equation}

\noindent
where $\lambda_{osc}$ is the oscillation wavelength,
defined as:
\begin{equation}
\lambda_{osc} = \frac{4\pi E}{\Dm2}\,,
\end{equation}
\noindent
where $E$ is again the neutrino energy and $\Dm2$ is the squared
neutrino mass difference.

Substituting the expression of $\lambda_{osc}$ in
(\ref{eq:osc_prob_2flav_1}), the oscillation probability takes its
usual form:
\begin{equation}\label{eq:osc_prob_2flav_2}
{\mathcal{P}}(\nu_\alpha\rightarrow\nu_\beta) = 
  \sin^22\theta~\sin^2\Bigg(\frac{1.27~\Dm2 ~L}{E}\Bigg)\,,
\end{equation}

\noindent
where $\Dm2$ is expressed in $\evolt^2$, 
the detector-to-source distance $L$ is in 
meters ($k$m) and the neutrino energy $E$ in $\mev$ ($\gev$).

The oscillation probability (\ref{eq:osc_prob_2flav_2}) is the product 
of two factors. The first one, $\sin^22\theta$, does not depend on the 
experimental conditions, but is an intrinsic parameter which describes the strength of the coupling
between the two neutrino flavour; it gives the maximum amplitude of
the neutrino oscillations. The mixing is maximal when
$\sin^22\theta=1$, namely when $\theta=\pi/4$. 

The second factor is another 
oscillatory term, the period of which is determined by  
$\Dm2\,L/E$. In order to have sensitivity to small values of
$\Dm2$, the ratio $E/L$ must be also small. This can be achieved by
having low energy neutrino beams and large
distances between the neutrino source and the detector. On the other
hand, when the
oscillation phase becomes too large, that is to say when $E/L$ is too
small compared to $\Dm2$, the oscillations occur very rapidly and, due 
to the finite energy resolution of the experiments,
the term $\sin^2(1.27\,\Dm2\, L/E)$ averages to $1/2$.

Before concluding this overview of the phenomenology of 
neutrino oscillations in
vacuum, it is interesting to consider the effect of the CPT, CP and T
symmetries on the oscillation probabilities. 
Without entering in any details, we shall only say that 
for both Dirac and Majorana neutrinos, CPT invariance implies:
\begin{equation}
{\mathcal{P}}(\nu_\alpha\rightarrow\nu_\beta) =
{\mathcal{P}}(\bar{\nu}_\beta\rightarrow\bar{\nu}_\alpha)
\qquad\text{(CPT conserved).}
\end{equation}

In particular, this means that the neutrino and the antineutrino
survival probabilities are the same:
\begin{equation}
{\mathcal{P}}(\nu_\alpha\rightarrow\nu_\alpha) =
{\mathcal{P}}(\bar{\nu}_\alpha\rightarrow\bar{\nu}_\alpha)
\qquad\text{(CPT conserved).}
\end{equation}

On the other hand, the oscillation probabilities for neutrinos and
antineutrinos are generally different. They coincide only if CP,
the symmetry which converts a left-handed neutrino into a right-handed 
antineutrino, is conserved:
\begin{equation}
{\mathcal{P}}(\nu_\alpha\rightarrow\nu_\beta) =
{\mathcal{P}}(\bar{\nu}_\alpha\rightarrow\bar{\nu}_\beta) 
\qquad\text{(CP conserved).}
\end{equation}

Finally, if the time-reversal symmetry T is conserved, the oscillation 
probability is invariant under interchange of the initial and the
final states:
\begin{equation}
{\mathcal{P}}(\nu_\alpha\rightarrow\nu_\beta) =
{\mathcal{P}}(\nu_\beta\rightarrow\nu_\alpha)  
\qquad\text{(T conserved).}
\end{equation}

\section{Neutrino Oscillations in Matter}

In the previous section we have seen that the probability for
neutrino oscillations in vacuum cannot exceed $\sin^2 2\theta$,
$\theta$ being the mixing angle in vacuum between the two flavours under consideration.
On the contrary, the situation can be very different for 
neutrino oscillations in matter. Matter effects can 
greatly enhance neutrino mixing, resulting in a large oscillation
probability even for very small vacuum mixing angles.

When they travel through matter, neutrinos of all flavours 
can have neutral-current interactions with the protons,
neutrons and electrons of the medium.
However only electron neutrinos can interact
with the electrons, undergoing a coherent forward
scattering via a $W$ boson exchange.
The consequence of this asymmetry between neutrino flavours is known as the
{\it Mikheyev-Smirnov-Wolfenstein (MSW) effect}.\cite{wolfenstein,mikheyev_smirnov} 

At low neutrino energies, 
for electron, muon and tau neutrinos
traversing an electrically neutral and unpolarised medium, 
the matter-induced potentials are given by:
\begin{equation}
  \begin{matrix}
    V_e ~=~\sqrt{2}G_F\Big({\mathcal{N}}_e-\displaystyle{\frac{{\mathcal{N}}_n}{2}}\Big)\, \\
    \\
    V_\mu ~=~ V_\tau ~=~ -\,\sqrt{2}G_F\displaystyle{\frac{{\mathcal{N}}_n}{2}}\,,
  \end{matrix}
\end{equation}

\noindent
where $G_F$ is the Fermi constant and ${\mathcal{N}}_e $ and
${\mathcal{N}}_n$ are the electron and the neutron number densities
respectively. For antineutrinos the potentials have opposite signs.

For simplicity, the discussion will be restricted to the two-flavour
scenario, more specifically to the mixing between $\nu_e$ and
$\nu_\mu$. In this case, using the ultra-relativistic expansion for
the neutrino energy and neglecting all common phases (which would not
affect the result for mixing between active neutrinos), the time 
evolution for the flavour eigenstates is given by:
\begin{equation}\label{eq:matter_evolution}
i\frac{d}{dt}\,\begin{pmatrix}\nu_e \\\nu_\mu\end{pmatrix} = 
\begin{pmatrix}
-\displaystyle{\frac{\Dm2}{4E}}\cos 2\theta+\sqrt{2}G_F{\mathcal{N}}_e 
& &
\displaystyle{\frac{\Dm2}{4E}}\sin 2\theta \\
\displaystyle{\frac{\Dm2}{4E}}\sin 2\theta & & 
\displaystyle{\frac{\Dm2}{4E}}\cos 2\theta 
\end{pmatrix}
\begin{pmatrix}\nu_e\\ \nu_\mu\end{pmatrix}\,, 
\end{equation}
\noindent
where the common term due to neutral-current interactions,
proportional to the identity matrix, has been ignored.

The simplest case of matter distribution is that of constant
matter density, ${\mathcal{N}}_e=const$. Although not very realistic,
this is however quite an instructive example and we shall discuss it
here in some detail.

As for the mass and flavour eigenstates for oscillations in
vacuum, the mass and flavour
eigenstates in matter are also connected by a two-dimensional rotation. The
relation between the two bases is given by:
\begin{equation}\label{eq:mixing_two_flav_matter}
  \begin{matrix}
    |\,\nu_1^M\,\rangle & = & \,\,\,\,\,\cos\theta_M\,|\,\nu_e\,\rangle & + &
    \sin\theta_M\,|\,\nu_\mu\,\rangle \\
    |\,\nu_2^M\,\rangle & = & -\sin\theta_M\,|\,\nu_e\,\rangle & + &
    \cos\theta_M\,|\,\nu_\mu\,\rangle \,.\\
  \end{matrix} 
\end{equation}
\noindent
The vacuum mixing angle, $\theta$, has been replaced by the
mixing angle in matter, $\theta_M$, given by:
\begin{equation}\label{eq:mixing_angle_matter}
\displaystyle{\sin^22\theta_M = }
\dfrac
{\displaystyle{\Big(\frac{\Dm2}{2E}\Big)^2}\sin^22\theta}
{
\Big(
\displaystyle{\frac{\Dm2}{2E}}\cos 2\theta - \sqrt{2}G_F{\mathcal{N}}_e 
\Big)^2 +
\displaystyle{\Big(\frac{\Dm2}{2E}\Big)^2}\sin^2 2\theta
}\,.
\end{equation}
\noindent
It has to be noticed that, 
since the mixing angle is not the same as in vacuum, the matter
eigenstates $|\,\nu_1^M\,\rangle$ and $|\,\nu_2^M\,\rangle$ are not
coincident with the mass eigenstates in vacuum, $|\,\nu_1\,\rangle$
and $|\,\nu_2\,\rangle$.

The oscillation probability is of the same form as in 
Eq.(\ref{eq:osc_prob_2flav_1}), but the vacuum mixing angle and the
vacuum oscillation wavelength are now replaced by those in matter:
\begin{equation}\label{eq:matter_prob}
  {\mathcal{P}}(\nu_e\rightarrow\nu_\mu) = 
  \sin^22\theta_M~\sin^2\Big(\pi\frac{L}{\lambda_{M}}\Big)\,,
\end{equation}
\noindent
with $\lambda_M$ given by:
\begin{equation}\label{eq:matter_lambda}
\lambda_M = 
\dfrac{2\pi}
{\sqrt{\Big(
\displaystyle{\frac{\Dm2}{2E}}\cos 2\theta - \sqrt{2}G_F{\mathcal{N}}_e 
\Big)^2 +
\displaystyle{\Big(\frac{\Dm2}{2E}\Big)^2}\sin^2 2\theta
}}\,.
\end{equation}

From Eq.(\ref{eq:mixing_angle_matter}) it follows that, regardless of
the smallness of 
the mixing angle in vacuum, the mixing angle in matter can be very
large. In particular, maximal mixing can be achieved if the medium
density is such that the following condition is satisfied:
\begin{equation}\label{eq:msw_resonance}
\sqrt{2}\,G_F\,{\mathcal N}_e = \dfrac{\Dm2}{2E}\cos2\theta\,.
\end{equation}
\noindent
This is the the so-called {\it MSW resonance condition}. Using
the characteristic solar electron density, 
${\mathcal N}_e\sim 10^{26}~\text{cm}^{-3}$, and the characteristic
value for solar neutrino energies, $E\sim 1~\mev$, assuming small
mixing in vacuum (hence $\cos2\theta\sim 1$), one obtains 
$\Dm2\sim 10^{-5}~\evolt^2$.

Since ${\mathcal N}_e>0$, Eq.(\ref{eq:msw_resonance}) is 
fulfilled only if $\Dm2\cos\theta>0$. 
Once a convention on the phase has been chosen, 
Eq.(\ref{eq:msw_resonance}) implies that resonant oscillation
enhancement is possible only for one particular sign of $\Dm2$. For
example, if $\cos\theta>0$, neutrino oscillations are enhanced if
$\Dm2=m_2^2-m_1^2>0$. On the other hand, from the change of sign in
the matter-induced potentials, it follows that for antineutrinos the resonance
condition would require $\Dm2<0$. This means that, for a given sign of 
$\Dm2$, matter effects cannot enhance neutrino and antineutrino oscillations at
the same time: if neutrino oscillations are enhanced, antineutrino
oscillations will be suppressed, and vice-versa.

The evolution equation (\ref{eq:matter_evolution}) cannot be solved 
analytically for any non-uniform 
matter distribution. In general
a numerical solution of the system of differential 
equations (\ref{eq:matter_evolution}) has
to be computed, accounting for the density profile and the
neutrino energy distribution. 
An interesting case to consider is the so-called 
{\it adiabatic approximation} for matter density
monotonically decreasing along the neutrino path.
Without entering into the details of the calculation, 
we shall only qualitatively discuss the results for 
the case of two-flavour neutrino mixing 
(for example between $\nu_e$ and $\nu_\mu$),
which is useful to describe neutrino oscillations in the Sun.

Electron neutrinos produced in the core of the Sun, where the 
density is well above that corresponding to the MSW resonance, 
will initially
see a mixing angle $\theta_{core}\approx \pi/2$.
From Eq.(\ref{eq:mixing_two_flav_matter})
it follows that, 
at production point, a neutrino born as a 
$\nu_e$ will essentially
coincide with one of the matter eigenstates $\nu_2^M$. 
The adiabaticity condition, that requires slowly changing matter distribution,
guarantees that the neutrino system can
gradually adjust to the changing density of the environment and
therefore the system 
will not make any transition to the other matter eigenstate.
As $\nu_2^M$ propagates through the mantle, it
encounters regions of smaller densities: the effective mixing
angle decreases and the strength of the mixing increases, 
until the resonance condition is fulfilled and maximal mixing is reached.
As the neutrino travels further, the mixing angle becomes smaller
and smaller, approaching the value of the mixing angle in vacuum, $\theta$.
If $\theta$ is very small, the
$\nu_e$ component of $\nu_2^M$ is very small
at the final point and $\nu_2^M$ is mainly composed of
$\nu_\mu$. The survival and the oscillation probabilities
for electron neutrinos are given by:
\begin{equation}
\begin{matrix}
{\mathcal{P}}(\nu_e\rightarrow\nu_e) &= &\sin^2\theta\\
{\mathcal{P}}(\nu_e\rightarrow\nu_\mu) &=& \cos^2\theta\,.
\end{matrix}
\end{equation}
\noindent
Thus, in the range of validity of the adiabatic approximation and 
for small values of the mixing angle in vacuum, if the depth
of the traversed matter is large enough,
the probability of finding the neutrino in the flavour
state $\nu_e$ when it gets outside the Sun is tiny and 
a complete conversion of
$\nu_e$ to $\nu_\mu$ is possible.

At this
point it becomes natural to discuss the solar neutrino deficit and to present the 
status of the search for neutrino oscillations in the Sun.

\section{The Solar Neutrino Problem}

\subsection{Solar Models}

Historically the first hint for neutrino oscillations came from the
observation of neutrinos from the Sun. In 1968 Ray Davis and his
collaborators published the first results of the
Homestake chlorine experiment,\cite{davis68,homestake}
showing that the measured 
flux of solar electron neutrinos was significantly lower than the
expected value. This was the beginning of the {\it solar neutrino deficit} 
saga.

The Homestake results were later confirmed by four other
experiments:  SAGE,\cite{sage} GALLEX,\cite{gallex}
Kamiokande,\cite{kamiokande_solar1} and
Super-Kamiokande.\cite{superk_solar1} Both SAGE and GALLEX
are radiochemical experiments, like Homestake,
while Kamiokande and
Super-Kamio-kande are water-Cherenkov experiments. Before we discuss
them, it is important to introduce the so-called Standard Solar
Model (SSM), used to predict the solar neutrino flux under the
assumption that no exotic phenomena, such as neutrino oscillations, can 
affect the nature of the produced neutrinos before they reach the Earth.

Our Sun belongs to the category of the main
sequence stars, 
which produce energy in their 
interiors via thermonuclear reactions.
The two reaction chains responsible for the energy production are known as
the $pp$-cycle and the CNO-cycle.
For both cycles the overall result
is the fusion of hydrogen nuclei into helium, with the emission of electron 
neutrinos:
\begin{equation}\label{eq:fusion_reaction}
4p + 2e^- \rightarrow ^4\text{He} + 2\nu_e + 26.73~\mev\,.
\end{equation}
\noindent
Only $\sim2\%$ of the solar energy is emitted in the form of neutrinos,
while the rest is radiated through photons.

In the three decades since the first experimental results on solar
neutrinos, 
there has been a flourishing of
solar models, all of which use the present values of some
fundamental solar parameters  (such as radius, mass, luminosity,
He/H ratio) as constraints on the stellar evolution. They all 
make essentially the same assumptions on the
nature of the solar energy and of the solar energy transfer mechanism.  
The most famous and accredited model is certainly
that of 
Bahcall-Pinsonneault (BP SSM).\cite{bahcall_1964}
In the following we shall be referring to its two most recent
versions, the 
BP98\cite{bp98} and BP2000,\cite{bp2000} in terms
of which most published experimental results are interpreted.
All reactions belonging to the $pp$ and 
the CNO cycles, together with the corresponding average and maximum neutrino 
energies are listed in
Tab.\ref{tab:bp98_fluxes_energies} (from ref.\cite{bilenky_giunti_grimus}).
The CNO cycle
contributes less than $2\%$ to the total neutrino flux.

\begin{table}[h]
\tcaption{Average and maximum neutrino energy for the different
  reactions contributing to the solar neutrino flux (tab. from ref.\cite{bilenky_giunti_grimus}).}
\centerline{\footnotesize\smalllineskip
\begin{tabular}{c c c c}\label{tab:bp98_fluxes_energies}\\
\hline
{} & {}& {}& {}\\
Source & Reaction & $<E_\nu>$ ($\mev$) & Max. $E_\nu$ ($\mev$) \\
{} & {}& {}& {}\\
\hline
\hline\\
$pp$ & $p+p\rightarrow d+e^++\nu_e$ & $0.2668$ & $0.423\pm0.03$  \\
$pep$ & $p+e^-+p\rightarrow d+\nu_e$ & $1.445$ & $1.445$ \\
$hep$ & $^3{\text{He}}+p\rightarrow 4^{\text{He}}+ e^+ +\nu_e$ & $9.628$ &
$18.778$ \\
$^7\text{Be}$ & $e^-+^7\text{Be}\rightarrow^7\text{Li}+\nu_e$ & 0.3855 
& 0.3855 \\
{} & {} & $0.8631$ & $0.8631$ \\
$^8{\text{B}}$ & $^8\text{B}\rightarrow ^8\hspace{-3pt}\text{Be}^*+e^++\nu_e$ &
$6.735\pm 0.036$ & $\sim 15$ \\
$^{13}{\text{N}}$ & $^{13}{\text{N}}\rightarrow
^{13}\hspace{-2pt}\text{C}+e^++\nu_e$ & $0.7063$ & $1.1982\pm 0.0003$ \\ 
$^{15}{\text{O}}$ & $^{15}{\text{O}}\rightarrow
^{15}\hspace{-2pt}\text{N}+e^++\nu_e$ &$0.9964$ & $1.7317\pm 0.0005$ \\
$^{17}{\text{F}}$ & $^{17}{\text{F}}\rightarrow
^{17}\hspace{-2pt}\text{O}+e^++\nu_e$ & $0.9977$ & $1.7364\pm 0.0005$ \\
{} & {}& {}& {}\\
\hline\\
\end{tabular}}
\end{table}

The neutrino fluxes predicted by BP2000, as well as the predicted neutrino capture rates in the chlorine and gallium experiments,
are listed in Tab.\ref{tab:bp2000_predictions}.
The uncertainties on the neutrino absorption cross sections represent the
dominant contribution to the error on the gallium
rate predictions, while for the chlorine rate those
uncertainties are relatively small compared to other systematic components
intrinsic to the experimental technique. 
The rates are measured in Solar Neutrino Units (SNU), a convenient
unit to describe the rates of solar neutrino experiments 
($1~\text{SNU} = 10^{-36}\text{events atoms}^{-1}\text{s}^{-1}$).

\begin{figure}[p]
\vspace*{13pt}
\begin{center}
\epsfig{figure=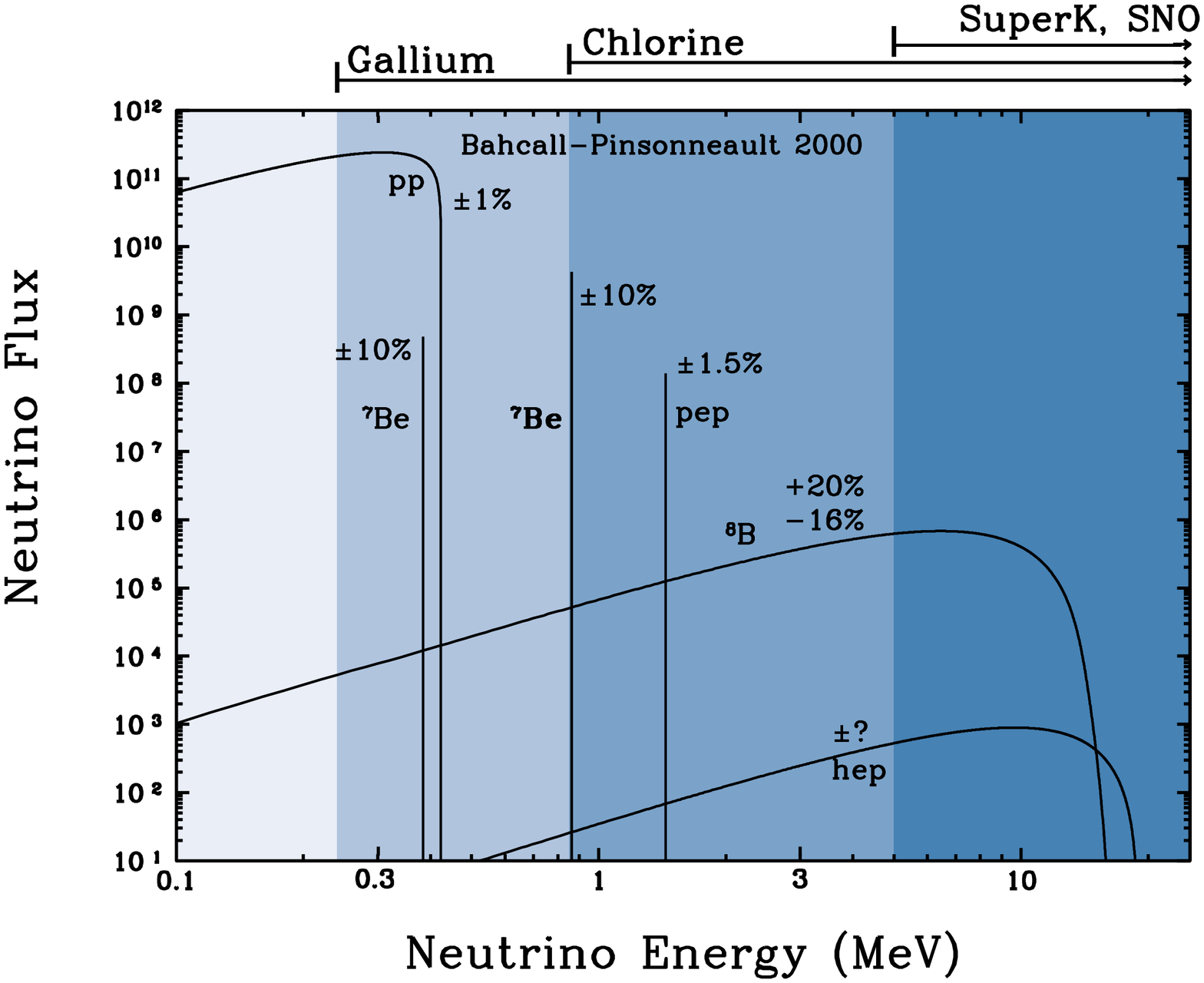,width=6cm,angle=270}
\vspace*{13pt}
\fcaption{BP2000\cite{bp98} predictions for the energy spectra of solar neutrinos
  produced in the $pp$-cycle reactions, which are responsible for more 
  than $98\%$ of the energy balance in the Sun
  (fig. from\cite{bahcall_web_page}). The continuum
spectra are expressed in events
$\text{cm}^{-2}\text{s}^{-1}\mev^{-1}$ at one astronomical unit,
while the monochromatic lines ($pep$ and $^7\text{Be}$) are given in
events $\text{cm}^{-2}\text{s}^{-1}$.}\label{fig:bahc_nuspectrum}
\vspace{1.5cm}
\psfig{file=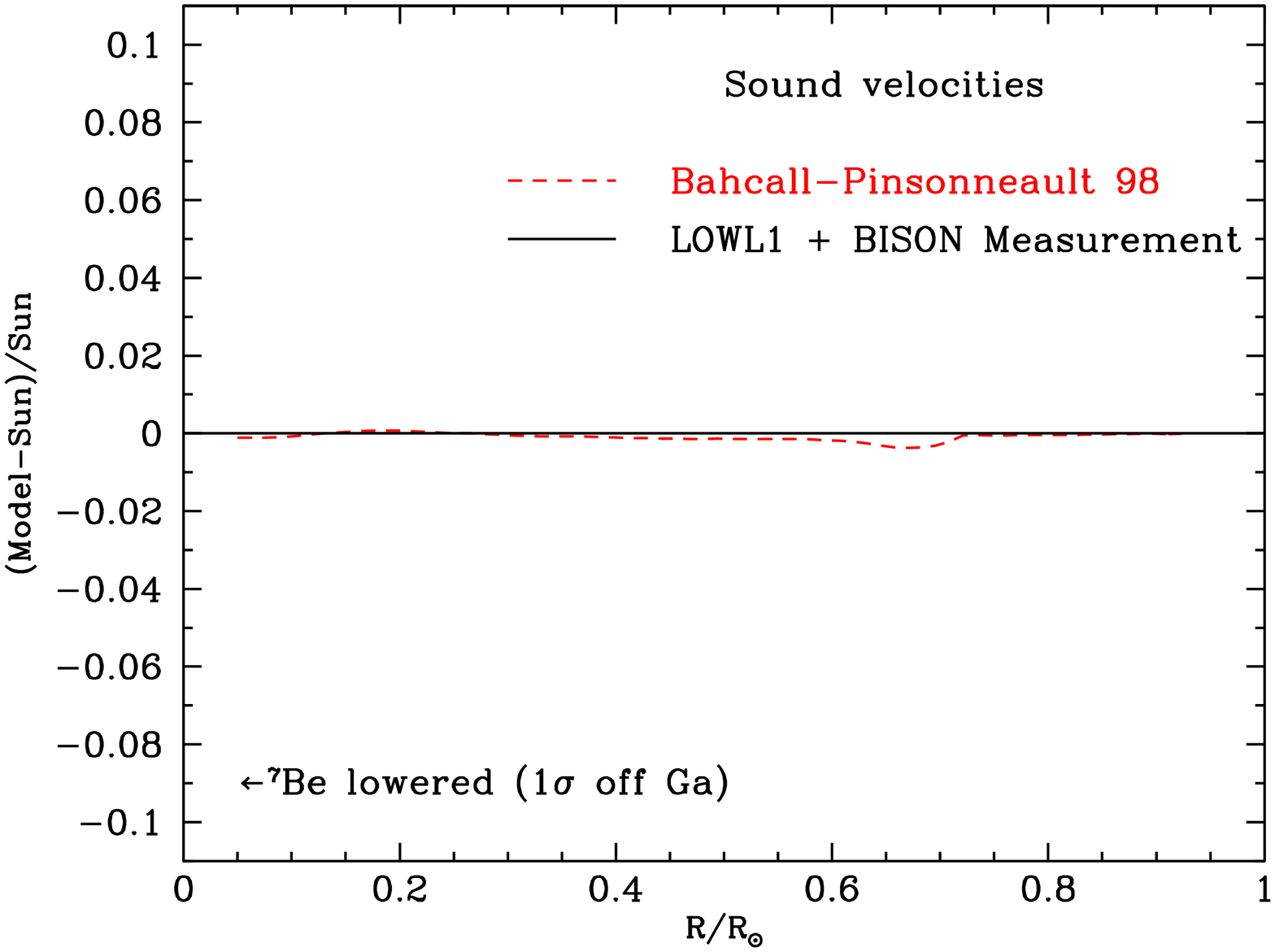,width=6cm,angle=270}
\vspace*{13pt}
\fcaption{Comparison between the predicted (Model) and measured (Sun)
  sound speeds in the Sun. Predictions come from the BP98 model,\cite{bp98} while 
  measurements come from the helioseismological data on the sound speeds.\cite{helioseismology}}\label{fig:bahc_helioseism}
\end{center}
\end{figure}

The energy spectra of neutrinos from the $pp$-cycle are shown in
Fig.\ref{fig:bahc_nuspectrum} (BP2000 predictions, from ref.\cite{bahcall_web_page}). The uncertainties on the
different flux components are also shown and the energy thresholds
of the different experiments are indicated at the top of the
plot.

\begin{table}[h]
\tcaption{BP2000\cite{bp2000} predictions for the solar neutrino
  fluxes and the neutrino capture rates in chlorine and gallium experiments.}
\centerline{\footnotesize\smalllineskip
\begin{tabular}{c c c c}\label{tab:bp2000_predictions}\\
\hline
{} & {}& {}& {}\\
Source & Flux ($10^{10}~\text{cm}^{-2}\text{s}^{-1}$) & Cl (SNU) & Ga (SNU)\\
{} & {}& {}& {}\\
\hline\hline
{} & {}& {}& {}\\
$pp$ & $5.95\big(1.00^{+0.01}_{-0.01}\big)$ & $-$ & $69.7$\\
$pep$ & $1.40\times 10^{-2}\big(1.00^{+0.015}_{-0.015}\big)$ & $0.22$ & $2.8$\\
$hep$ & $9.3\times 10^{-7}$ & $0.04$ & $0.1$\\
$^7\text{Be}$ & $4.77\times 10^{-1}\big(1.00^{+0.10}_{-0.10}\big)$ & $1.15$ & $34.2$\\
$^8\text{B}$ & $5.05\times 10^{-4}\big(1.00^{+0.20}_{-0.16}\big)$ & $5.76$ & $12.1$\\
$^{13}\text{N}$ & $5.48\times 10^{-2}\big(1.00^{+0.21}_{-0.17}\big)$ & $0.09$ & $3.4$\\
$^{15}\text{O}$ & $4.80\times 10^{-2}\big(1.00^{+0.25}_{-0.19}\big)$ & $0.33$ & $5.5$\\
$^{17}\text{F}$ & $5.63\times 10^{-4}\big(1.00^{+0.25}_{-0.25}\big)$ & $-$ & $0.1$\\
{} & {}& {}& {}\\
\hline
Total& {} & $7.6^{+1.3}_{-1.1}$ & $128^{+9}_{-7}$\\
\hline\\
\end{tabular}}
\end{table}

There are several reasons, other than the high precision of the
input data and
calculations, why the BP SSM is believed to be
robust. First of all, there is what is known as the ``luminosity
constraint'' argument. 
It assumes that the Sun is in a stationary state and then
infers a strong correlation between the solar neutrino flux and the
solar luminosity (${\mathcal{L}}_\odot=2.4\times
10^{39}~\mev~\text{s}^{-1}$, known to an accuracy of $\sim 0.4\%$). 
From Eq.(\ref{eq:fusion_reaction})
we see that, in the nuclear reactions occurring in the Sun, 
the production of two neutrinos is accompanied by
an energy release of $26.73~\mev$. Neglecting the small amount of energy
carried away by the neutrinos, this means that the electron neutrino
flux in the absence of oscillations can be estimated as
\begin{equation}
\Phi_{\nu_e} \simeq 2\times\dfrac{{\mathcal{L}}_\odot/(4\pi
  R_\odot^2)}{26.73~\mev} = 6.4\times10^{10}~\text{cm}^{-2}\text{s}^{-1}\,,
\end{equation}
\noindent
where $R_\odot$ is one astronomical unit ($= 1.496\times 10^{13}~\text{cm}$).

Another model-independent constraint, although looser than the
luminosity one, is given by the
fact that the helium nuclei, necessary for the boron and
beryllium production, are created in the $pp$ and $pep$ reactions. Thus:
\begin{equation}
\Phi_{\nu_e}(^7\text{Be}) + \Phi_{\nu_e}(^8\text{B}) < 
\Phi_{\nu_e}(pp) + \Phi_{\nu_e}(pep)\,.
\end{equation}

One the most impressive checks of the BP SSM come perhaps from the
comparison between the BP98 model predictions,
obtained without any
adjustment of the parameters, and 
the most accurate helioseismology measurements of the sound speed.\cite{helioseismology}
Fig.\ref{fig:bahc_helioseism} shows the excellent agreement between
the calculated and the measured values:
the size of the fractional difference between measurements and predictions
($0.001$ standard deviation for radii between
$5\%$ and $95\%$ of the Sun radius)
is much smaller than any
generic change in the
model with an impact on the predicted neutrino flux.

\newpage
\subsection{Experimental Results}
We shall now describe the experimental techniques which have been used to
detect solar neutrinos. The experiments can be
subdivided into two categories: radiochemical experiments (Homestake, 
SAGE, GALLEX and its successor, the GNO\cite{gno} project) and
water-Cherenkov experiments (Kamiokande, Super-Kamiokande and SNO).
One of the main differences between these two categories is the fact
that, while for radiochemical experiments only an integrated
measurement of the fluxes is possible, water-Cherenkov detectors can perform
real-time measurements and therefore can study differential 
distributions (i.e. energy spectrum, angular distribution, correlation 
with the Sun position in the sky).

The Homestake\cite{davis68,homestake} chlorine experiment is located
$1480$~m underground ($4200$ m.w.e., meters water equivalent), 
at the Homestake gold mine, Lead, South Dakota, USA. A tank of
$6\times 10^5$ litres volume is filled with tetrachlorethylene
($\text{C}_2\text{Cl}_4$). The detection principle is
based on the reaction:
\begin{equation}\label{eq:nue_cl}
\nu_e+^{37}\hspace{-2pt}\text{Cl}\rightarrow e^- +{^{37}\text{Ar}}(\text{T}_{1/2}=35~\text{d})\,,
\end{equation}
\noindent
which has a threshold of $814~\kev$, above the end-point of the
$pp$-cycle energy spectrum. After an exposure time of $1-3$
months, the $^{37}\text{Ar}$ atoms produced in (\ref{eq:nue_cl}) are extracted
by purging the detector with $^4\text{He}$ and detecting the Auger
electron produced in the electron capture of the radioactive
$^{37}\text{Ar}$ nuclei. The measured rate, which mainly arises from
of $^8\text{B}$ and $^7\text{Be}$ neutrinos, plus a small contribution
from $pep$ neutrinos, is $2.56\pm 0.23$~SNU,\cite{homestake}
significantly below the predicted value of $7.6^{+1.3}_{-1.1}$~SNU.
Although 
the Homestake detector is not calibrated, for lack of a 
suitable artificial neutrino source, the
efficiency of the extraction technique 
has been checked by doping the detector with a known amount of radioactive
argon atoms.

All gallium experiments, SAGE,\cite{sage} GALLEX\cite{gallex} and
GNO,\cite{gno} detect neutrinos from the Sun by means of the  
process:
\begin{equation}\label{eq:nue_ga}
\nu_e+^{71}\text{Ga}\rightarrow
e^-+^{71}\text{Ge}(\text{T}_{1/2}=11.43\text{d})\,,
\end{equation}
for which the energy threshold ($233.2~\kev$) is well below the
maximum energy of the $pp$ neutrinos. This means that gallium
experiments, by measuring low-energy solar neutrinos, have the ability
to prove the thermonuclear nature of the energy production mechanism
in the Sun.

The GALLEX experiment, now discontinued as such, 
was located at the Gran Sasso Underground Laboratory (LNGS),
Assergi, Italy. It consisted of $30.3$~tons of $^{71}\text{Ga}$, in
the form of a concentrated solution of gallium chloride
($\text{GaCl}_3-\text{HCl}$) in water. The $^{71}\text{Ge}$ atoms form 
the volatile compound $\text{GeCl}_4$ which, at the end of each run
($3-4$ weeks) is swept out of the solution by means of a nitrogen
stream. The $\text{GeCl}_4$ is absorbed in water in a gas scrubber,
where the nitrogen is filtered out, and then converted to
$\text{GeH}_4$. This compound 
is finally introduced, together with xenon,
into a proportional counter where the number of 
$^{71}\text{Ga}$ atoms is determined by counting the radioactive decays.
The final results from GALLEX(I-IV)\cite{gallex} give a solar
neutrino flux of $77.5^{+7.6}_{-7.8}$~SNU, again well below
theoretical predictions of $128^{+9}_{-7}$~SNU.
The GALLEX detector was calibrated using two independent
methods. The first one made use of
neutrinos from two
intense ($>60$~PBq) $^{51}\text{Cr}$ sources: the combined value of
the ratio $R$ between the neutrino source strength as derived from the 
measured rate of $^{71}\text{Ga}$ production and the directly
determined source strength is $R=0.93\pm 0.08$.\cite{gallex_calibration1}
The other calibration was performed by injecting, under 
varying conditions, a known amount of
$^{71}\text{As}$ into the full-scale detector.\cite{gallex_calibration2} 
The arsenic isotope decays by electron capture and
positron emission to $^{71}\text{Ge}$ ($\text{T}_{1/2}=2.72$~d),
producing radioactive atoms which mimic the solar neutrino capture
kinematics. Although neutrinos from the $^{51}\text{Cr}$ source
provide a better match to solar neutrino energies, the second method has the great advantage of large
statistics. The measured recovery rate of $^{71}\text{Ge}$ from
gallium is $1\pm 0.01$. The two
results\cite{gallex_calibration1,gallex_calibration2} together rule
out the possibility that the solar neutrino deficit observed by GALLEX 
can be attributed to
systematic errors in the radiochemical extraction procedure.

The SAGE\cite{sage} experiment is carried out by a Russian-American
collaboration at the underground 
Baksan Neutrino Observatory ($4715$~m.w.e.), in the Northern
Caucasus mountains. The detector, which weighs $57$~tons,  
uses gallium in its metallic form. The germanium produced in
(\ref{eq:nue_ga}) is removed from the
metallic gallium by a liquid-liquid extraction into a
$\text{HCl}-\text{H}_2\text{O}_2$ phase. After this initial step,
the experimental technique is very similar to that followed by GALLEX.
The results reported by the SAGE collaboration are based on a
$10$~years exposure time (from January 1990 to October 1999),
yielding a measured solar neutrino flux of $74^{+7.8}_{-7.4}\text{(stat.+sys.)}$~SNU,\cite{sage} in very
good agreement with the GALLEX results. 
The SAGE detector has been calibrated, exposing $13$~tons of metallic
gallium to an intense $^{51}\text{Cr}$
source.\cite{sage_calibration}
As for GALLEX, the ratio of the measured production rate to that
predicted from the source activity, $R=0.95\pm 0.11
\text{(stat.)}^{+0.05}_{-0.08}$, has confirmed that the extraction
procedure is reliable.

In 1998, after an upgrade of the existing experimental setup,  
the GNO (Gallium Neutrino Observatory) experiment\cite{gno} 
started operation at
LNGS, as the successor project to the GALLEX experiment. The main aims
of GNO are to provide a long time record of low-energy solar
neutrinos, to determine the bulk production rate with an accuracy of
$5$~SNU and to monitor the time dependence of the $pp$-neutrino flux
during a whole solar cycle to a precision of $\sim 15\%$. The GNO
collaboration has recently reported the results from the GNO I phase
of the experiment, based on $19$ months of observation (from May 1998 to
January 2000). The measured solar neutrino flux is
$65.8^{+10.7}_{-10.2}\text{(stat.+sys.)}$~SNU,\cite{gno} which
confirms the GALLEX result. 
A combined analysis of GNO I and GALLEX gives $74.1^{+6.7}_{-6.8}$~SNU.\cite{gno}

Both Kamiokande\cite{kamiokande_solar1} and
Super-Kamiokande\cite{superk_solar1} are water-Cherenkov detectors,
located $1000$~m underground ($2700$~m.w.e.), in the Kamioka mine in
Japan.

Kamiokande started its operation in 1984 and was originally intended
to study nucleon decay.\cite{kamiokande_pdecay} It was later upgraded 
to detect also low energy events and succeeded in observing the first
solar neutrinos in 1987. Kamiokande was made of a cylindrical tank, of total 
volume $4.5$~kton, filled with pure water. 
An inner volume of $2.14$~kton was defined by $980$
inward-looking photo-multiplier tubes (PMTs), 
used to detect the Cherenkov light
produced by the relativistic particles traversing the water.
Kamiokande ended physics data-taking at the beginning of 1995 and
was completely stopped in summer 1997. Since then its physics
programme has been continued by its successor experiment,
Super-Kamiokande. 

Super-Kamiokande  
consists of a huge cylindrical
tank (volume $50$~kton) filled
with pure water (water transparency is $\sim 100$~m at $\lambda =
420$~nm). Also in this case an inner volume ($16.9$~m diameter,
$36.2$~m height, volume $32.5$~kton) is defined by an inner surface
equipped with 
a large number of PMTs ($11,146$ PMTs of $50$~cm diameter, $40\%$
surface coverage). The increased coverage of
PMTs makes Super-Kamiokande not only bigger than Kamiokande, but also better in 
terms of energy, position and angular resolution. An outer volume,
$2$~m thick, equipped with $1,185$ PMTs ($20$~cm diameter) 
surrounds the inner detector and serves as an active veto counter
against gamma-rays, neutrons and through-going cosmic muons.
For the solar
neutrino analysis, the fiducial mass is $22.5$~kton, the fiducial
volume boundaries being $2$~m inside the inner surface.

\begin{figure}[h]
\vspace*{13pt}
\begin{center}
\epsfig{figure=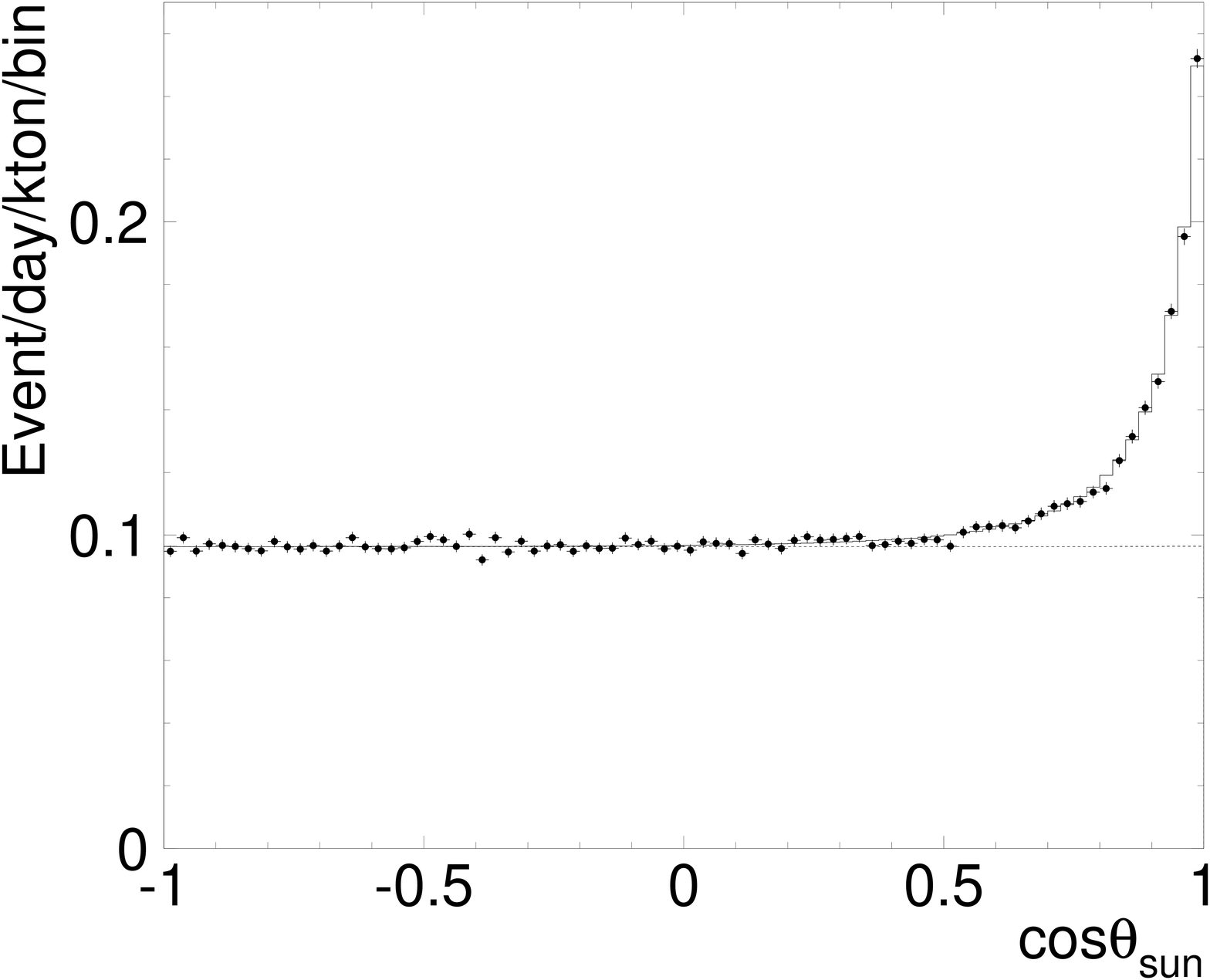,width=6cm}
\vspace*{13pt}
\fcaption{Super-Kamiokande ($1258$~days) distribution of the cosine of the angle 
  $\theta_{sun}$
  between the recoil electron momentum and the vector from the Sun to
  the Earth.\cite{superk_latest_b8} The dots are the experimental
  data, while the solid histogram is the best fit for signal plus
  background. The peak emerging at $\cos\theta_{sun}$ over a flat
  background is due to solar
  neutrinos.}\label{fig:superk_costhetasun}
\end{center}
\end{figure}

In both Kamiokande and Super-Kamiokande the 
solar neutrinos are detected through the observation of the
Cherenkov rings produced by the electrons emitted in the elastic
process:
\begin{equation}\label{eq:elastic_nue_scatt}
\nu_x + e^- \rightarrow \nu_x + e^-\hspace{2cm} (x=e,\mu,\tau)\,.
\end{equation} 
\noindent
Although the cross-sections for this process are very small,\footnote{
$0.920\times 10^{-43}\Big(\dfrac{E_\nu}{10\mev}\Big)~\text{cm}^2$ for electron
neutrinos and 
$0.157\times 10^{-43}\Big(\dfrac{E_\nu}{10\mev}\Big)~\text{cm}^2$ for
muon neutrinos.}\hspace{3pt} elastic interactions turn out to be very useful, thanks 
to the nice correlation between the recoil electron momentum direction 
and the direction
of the incoming neutrino ($E_e\theta_{rec}\le 2m_e$, $\theta\sim 18^0$ for
$10~\mev$ neutrinos). Therefore the direction of the recoil
electron can be used to correlate the direction of the impinging
neutrino with the Sun's position in the sky (see
Fig.\ref{fig:superk_costhetasun}). Moreover the energy of the recoiling
electron can be used to obtain a lower limit on 
the incoming neutrino
energy. 

The total released energy is correlated to the number of PMT hits,
which is in turn a function of the total emitted light, corrected for
light absorption through water, PMT geometry and 
overlapping hits. The energy scale, however, must be determined in an
independent manner. 

Kamiokande used gamma-rays from the
$\text{Ni(n,}\gamma\text{)Ni}$ reaction to obtain
it.\cite{kamiokande_solar1} The measurement precision was at the
$1-2\%$ level, the main limitation being the knowledge 
of the branching ratios and
of the neutron absorption cross-sections for different nickel
isotopes. Only limited information could be extracted about the 
energy resolution, while no information at all could be obtained on the angular
resolution of the detector. 

Super-Kamiokande determines the 
neutrino energy scale by injecting, at various positions of the inner
detectors, electrons from an electron LINAC
placed near the detector tank.\cite{superk_linac}
Electrons from the LINAC span the energy range $5-16~\mev$, which
matches exactly the region relevant for solar neutrinos.  
By means of this procedure, the absolute energy scale
is known to better than $1\%$. Moreover, Monte Carlo studies show 
that the energy resolution of the detector can be reproduced to an
accuracy of $2\%$, while the angular resolution is reproduced to
better than $1.5^0$ for $10~\mev$ electrons. 

In order to check these results
and, what is possibly even more important, in order to reduce the
intrinsic 
systematic error inherent in the described procedure,\footnote{
electrons from the LINAC only move in a downward direction; due to
limited access, the beam
pipe and the LINAC calibration equipment can only be operated at a
finite number of positions; the presence of the beam pipe in the tank
limits the calibration procedure precision and is in fact the main
source of systematic uncertainties at low energy.} another calibration
technique has been introduced.\cite{superk_n16} A pulsed deuterium-tritium
neutron generator produces isotropically distributed neutrons of
$14.2~\mev$, via the reaction
$^3\text{H}+^2\text{H}\rightarrow ^4\text{He}+n$. The neutrons
create $^{16}\text{N}$ by the ($n,p$) reaction on $^{16}\text{O}$ in
the water. $^{16}\text{N}$ decays
($\tau = 7.13$~s, $Q=10.4~\mev$) are characterised by the emission of
a $4.3~\mev$ electron in coincidence with a $6.1~\mev$ gamma-ray and
they are 
therefore well suited to verify the solar neutrino absolute energy scale.
Additional $^{16}\text{N}$ is also naturally
produced in cosmic muon capture by
$^{16}\text{O}$. These events, which are selected 
out of the full data sample and which are
generally used to monitor the solar neutrino signal extraction method, 
can be also analysed to
check the absolute energy scale. Excellent agreement (within $1\%$) has 
been found between the data and the MC, previously tuned
on the LINAC data. Moreover it has been possible to show that the position and
the angular dependence of the energy scale are within systematic
uncertainties for the energy scale of the detector.

The energy threshold for Kamiokande and Super-Kamiokande is determined 
by the threshold for detecting the recoil electron in the elastic
scattering (\ref{eq:elastic_nue_scatt}), which is about
$5-7.5~\mev$. Thus the solar neutrino flux measured in those experiments is
essentially the $^8\text{B}$ flux (plus a small component from the
$hep$ reaction in Super-Kamiokande). 
The value measured by Kamiokande, based on $2079$ days of
data-taking (from January 1987 to February 1995), is 
$2.80\pm 0.19 \text{(stat.)} \pm 0.33\text{(sys.)}\times
10^6~\text{cm}^{-2}\text{s}^{-1}$.\cite{kamiokande_solar1} 
The Super-Kamiokande
collaboration has recently reported a result
based on $1258$~days of data-taking, for recoil
electrons in the energy range $5-20~\mev$. The measured neutrino flux
is $2.32\pm 0.03 \text{(stat.)} ^{+0.08}_{-0.07}\text{(sys.)}\times
10^6~\text{cm}^{-2}\text{s}^{-1}$.\cite{superk_latest_b8} For the $hep$ neutrinos they set an upper limit of $4\times 10^4
~\text{cm}^{-2}\text{s}^{-1}$, at $90\%$~C.L.

Before we move to other issues relevant for the understanding solar neutrino
puzzle, we summarize the results discussed up to now and 
their implications on neutrino physics.
In Tab.\ref{tab:rates_vs_bp2000} the measured values of the integrated 
solar neutrino fluxes are compared with the BP2000 model
predictions\cite{bp2000}.

\begin{table}[ht]
\tcaption{Comparison between the measured integrated solar neutrino
  fluxes and the corresponding BP2000 predictions.\cite{bp2000} Units
  are SNU for chlorine and gallium experiments and
  $10^6\text{cm}^{-2}\text{s}^{-1}$ for $^8\text{B}$ and $hep$ neutrinos.}
\centerline{\footnotesize\smalllineskip
\begin{tabular}{c c c c}\label{tab:rates_vs_bp2000}\\
\hline
{} & {} & {} & {} \\
Experiment & Measured & BP2000 & Measured/BP2000\\
{} & {} & {} & {} \\
\hline\hline
{} & {} & {} & {} \\
Chlorine\cite{homestake}& $2.56\pm 0.23$ & $7.6^{+1.3}_{-1.1}$ &
$0.34\times\big(1.00\pm 0.06\big)$\\
{} & {} & {} & {} \\
GALLEX+GNO\cite{gallex,gno} & $74.1^{+6.7}_{-7.8}$ & $128^{+9}_{-7}$ &
$0.58\times\big(1.00^{+0.11}_{-0.12}\big)$\\
{} & {} & {} & {} \\
SAGE\cite{sage} & $75.4^{+7.8}_{-7.4}$ & $128^{+9}_{-7}$ & 
$0.59\times\big(1.00\pm 0.12\big)$\\
{} & {} & {} & {} \\
$^8\text{B}$-Kamiokande\cite{kamiokande_solar1} & $2.80\times\big( 1.00\pm 0.14\big)$ &
$5.05\times\big(1.00^{+0.20}_{-0.16}\big)$ & $0.55\times\big(1.00^{+0.24}_{-0.21}\big)$\\
{} & {} & {} & {} \\
$^8\text{B}$-Super-Kamiokande\cite{superk_latest_b8} & $2.32\times\big(1.00\pm 0.03\big)$ & 
$5.05\times\big(1.00^{+0.20}_{-0.16}\big)$ & $0.46\times\big(1.00^{+0.20}_{-0.16}\big)$\\
{} & {} & {} & {} \\
$hep$-Super-Kamiokande\cite{superk_latest_b8} & $<40\times 10^{-3}$ &
$9.3\times 10^{-3}$ & $<4.3$\\
{} & {} & {} & {} \\
\hline\\
\end{tabular}}
\end{table}
In brief, all experiments see a deficit of solar neutrinos compared to
predictions. We have seen that the solar model's predictions appear to 
be robust, and so are 
the experimental results (although the Homestake experiment is 
sometimes still being questioned because of the lack of a calibration). 
Therefore neither an astrophysical solution to the solar neutrino
puzzle, nor attributing the discrepancy between theory and experiments to 
not completely understood systematic
uncertainties seem acceptable. There must be a particle physics
solution to the solar neutrino deficit, which is however more than a
sole normalisation problem. The impossibility of accommodating
all the experimental results in a consistent picture forces us to go
beyond the standard physics framework and to look for explanations
which involve new phenomenology.

\begin{figure}[h]
\vspace*{13pt}
\begin{center}
\psfig{figure=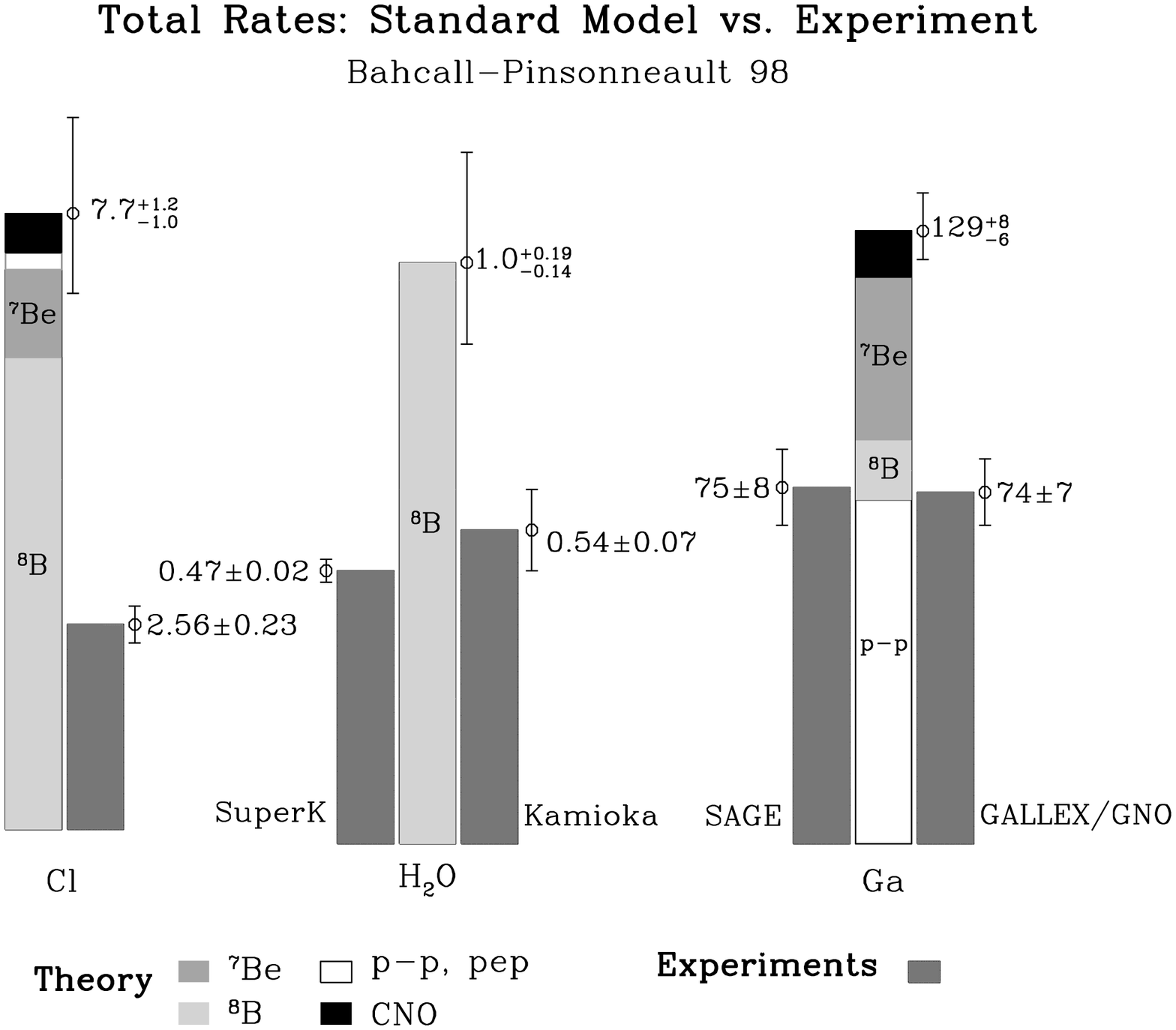,width=7cm,angle=270}
\vspace*{13pt}
\fcaption{Measured vs. predicted solar neutrino rates (fig. from
  ref.\cite{bahcall_web_page}) for all the solar neutrino
  experiments. The predicted rates for the various components of the
  neutrino spectrum are taken from the BP98 model.\cite{bp98}}\label{fig:bahc_theoryvsexp}
\end{center}
\end{figure}

The internal inconsistency of the solar neutrino results 
is very nicely summarised in Fig.\ref{fig:bahc_theoryvsexp} (from 
ref.\cite{bahcall_web_page}), where a breakdown of 
all the measured and calculated
event rates is shown as a function of the experimental technique and of the
different reactions contributing to the solar neutrino flux.
First of all there is the low rate observed in the chlorine
experiment, which basically measures $^8\text{B}$
neutrinos, with smaller contributions
from $^7\text{Be}$ and CNO neutrinos. Second, 
there is the incompatibility between the chlorine and the
water-Cherenkov experiments. In fact, 
since the shape of the 
$^8\text{B}$ neutrino energy spectrum is very stable under reasonable changes of
the solar parameters, in the absence of new physics it should be possible
to compute the $^8\text{B}$ rate in Cl from the spectrum observed by
Kamiokande and Super-Kamiokande, which essentially measure the
$^8\text{B}$ flux at high energies.
But the number obtained from the analysis of the water-Cherenkov data 
is by itself higher than the total flux measured by Homestake. This
result, taken as such, implies that the 
net contribution of the $pep$, $^7\text{Be}$ and CNO reactions to the
total chlorine experiment flux is negative. 
In other words,
the Homestake measurement is compatible with the $^7\text{Be}$
neutrinos being completely suppressed, and this is indeed a big puzzle, since 
beryllium is needed for the production of $^8\text{B}$ neutrinos.
Finally, examining the gallium experiments result in detail, one finds
that essentially all of the measured flux can be accounted for by only 
considering the
contribution from the $pp$-cycle, which is known within
$1\%$ and therefore is not questionable. Ignoring for simplicity the 
small contribution from $^8\text{B}$ neutrinos (which, by the way, can 
also be derived from water-Cherenkov results),
this means that there is no room to accommodate the significant 
contribution expected from the $^7\text{Be}$ lines.

\begin{figure}[h]
\vspace*{13pt}
\begin{center}
\epsfig{figure=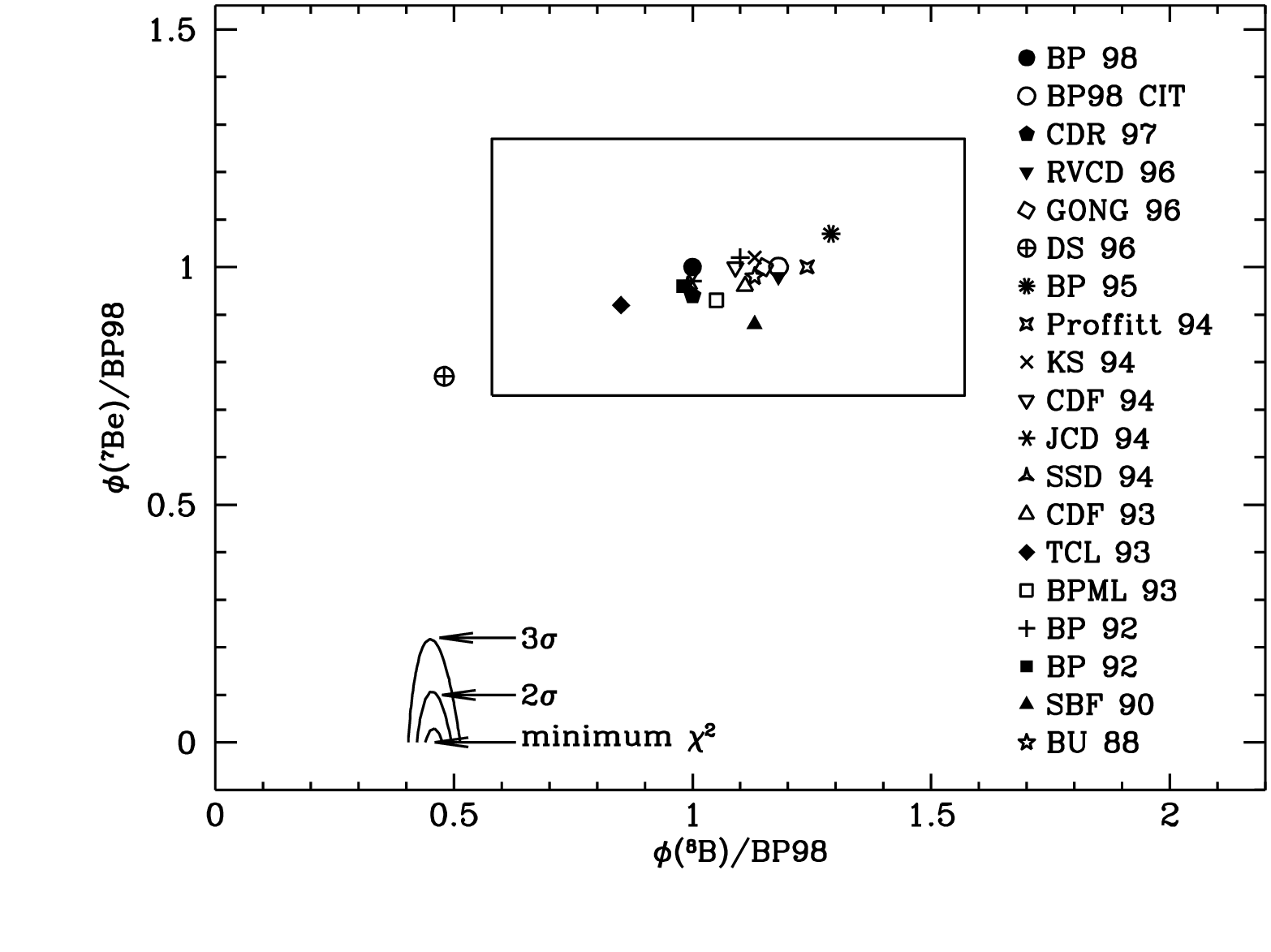,width=6cm}
\vspace*{13pt}
\fcaption{Correlation between the $^7\text{Be}$ and the $^8\text{B}$
  fluxes (normalised to BP98 predictions) as given by published solar
  models (fig. from ref.\cite{bahcall_krastev_smirnov}). The rectangle 
  corresponds to the $3\sigma$ contour for the BP98 model prediction
  and the ellipses are the $1,2,3\sigma$ contours for the best fit
  solution as obtained from all experimental data assuming no new physics.}\label{fig:bahc_allmodels}
\end{center}
\end{figure}

The strong correlation
between the predicted $^7\text{Be}$ and the $^8\text{B}$ fluxes is
shown in Fig.\ref{fig:bahc_allmodels}
for many different solar models, 
together with the best fit 
solution for the $^7\text{Be}$ flux as obtained from experimental
data. All the fluxes have been normalised to the predictions of the BP98
model and a $3\sigma$ rectangle has been drawn, showing that all
models but one (that of Dar-Shaviv\cite{dar-shaviv}) agree with each
other within $3\sigma$ of the BP98 model and are all far away from the 
value derived from experimental data.

In summary,  
if one believes the conclusions of the standard solar models, 
the solar neutrino puzzle can only be solved if one is 
either ready to discard at least three experimental results (the
chlorine experiment and either the gallium or the water-Cherenkov
measurements), which
seems unreasonable, or to accept that new physics, not accounted for in the
standard model of electro-weak interactions, 
is modifying the energy spectrum of solar neutrinos.

One very attractive possibility is that neutrino oscillations are changing the
neutrino flavour composition while they travel between the centre of
the Sun and the Earth. If solar electron 
neutrinos have transformed into muon or tau neutrinos before reaching
the detector, given their low average energy they would fail to produce 
muons or taus via charged-current interactions.
A ``disappearance'' of electron neutrinos in the
chlorine and gallium experiments would then become an expected
effect. The deficit in the water-Cherenkov
experiments, where in principle muon and tau neutrinos could be detected 
via their neutral-current scatters with electrons of the medium, could 
also be explained by the low cross-section for that process.

\begin{figure}[h] 
\setlength{\unitlength}{1cm}
\hbox{\hspace{-1.0cm}
\epsfxsize=7.0cm\epsfbox{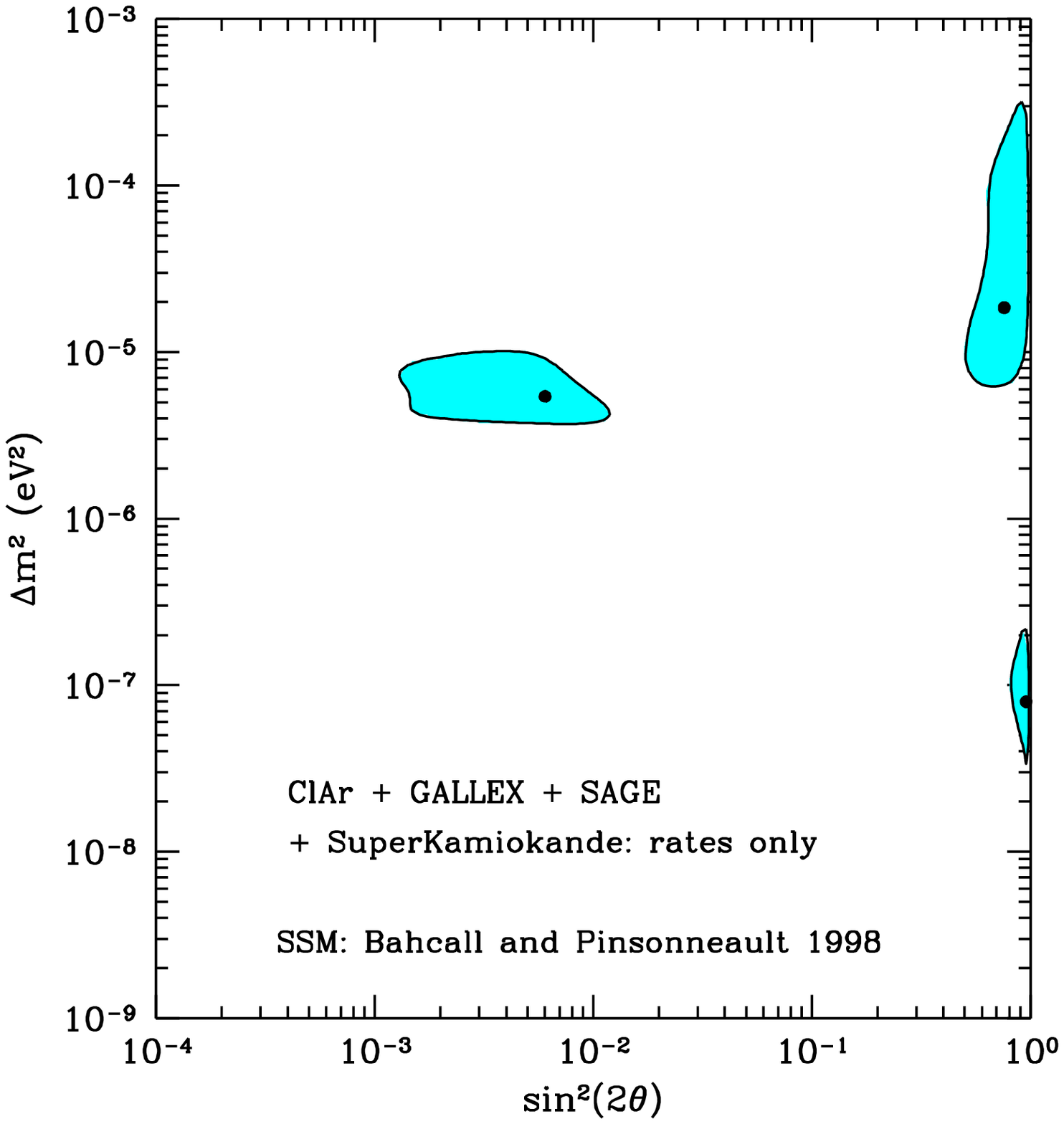}
\hspace{-0.5cm}
\epsfxsize=7.0cm\epsfbox{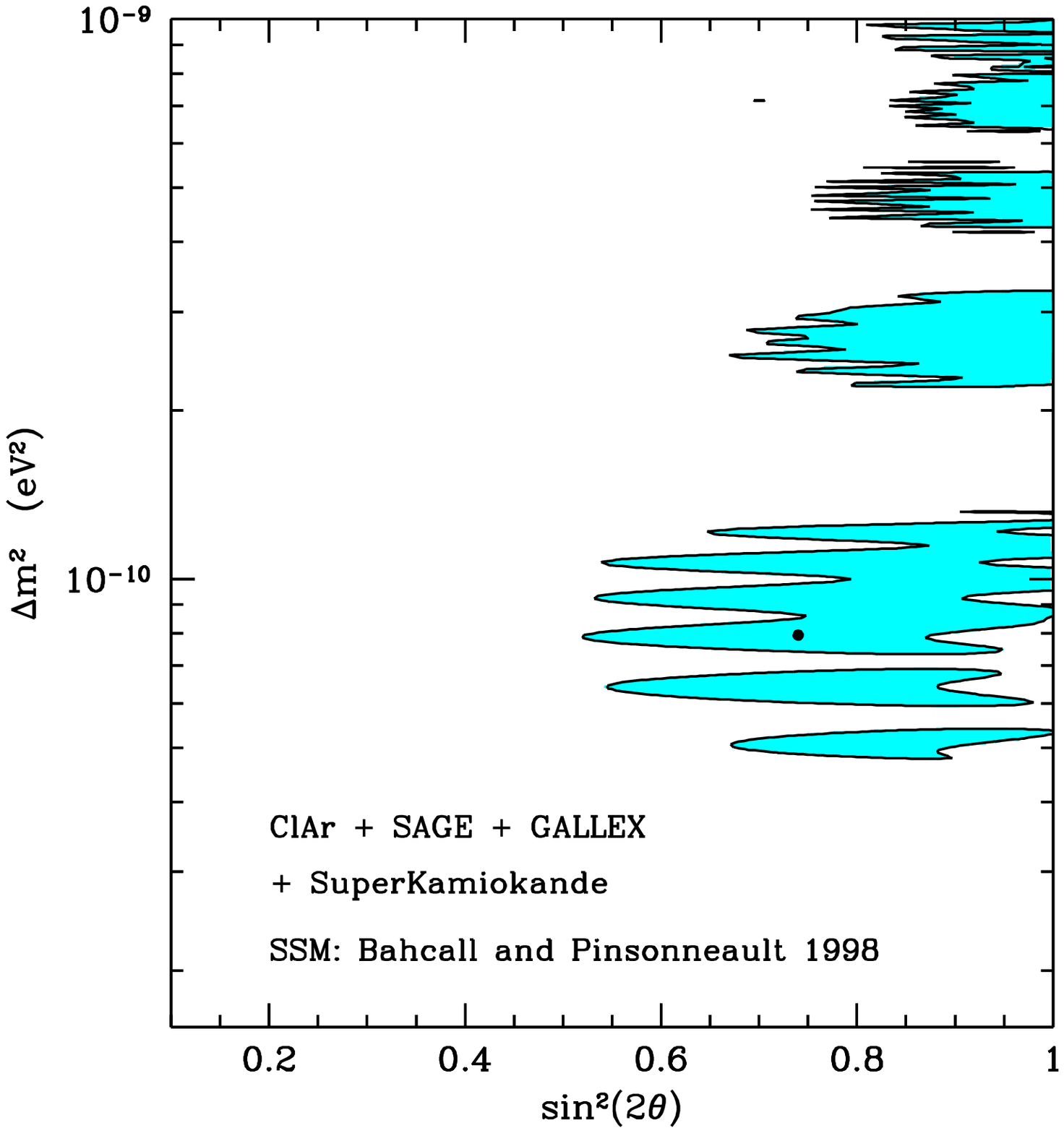}
}
\vspace*{13pt}
\fcaption{Allowed regions ($99~\%$~C.L.) on the plane
  $(\sin^22\theta,\Dm2)$, as obtained in
ref.\cite{bahcall_krastev_smirnov} from a combined two-flavour analysis of the 
total rate results in the chlorine, SAGE, GALLEX and Super-Kamiokande
experiments, for oscillations in matter (left) and in vacuum
(right).}
\label{fig:solar_solutions}
\end{figure}

Interpreting the solar neutrino deficit in the two-flavour mixing
framework, allowed regions can be drawn on the oscillation parameter space
$\{\sin^22\theta,\Dm2\}$ at a set
C.L..\footnote{As stressed in
  ref.\cite{sun_dark_side}, under certain conditions the MSW solutions 
  to the solar neutrino deficit can extend to the $\pi/4<\theta <
  \pi/2$ domain, which is
  not covered by the standard two-flavour analyses; to overcome this
  problem, results can be more conveniently presented in the
  $\{\tan^2\theta,\Dm2\}$ parameter space.} \hspace{3pt}For example,
Fig.\ref{fig:solar_solutions} shows
the $99~\%$~C.L. solutions as obtained in
ref.\cite{bahcall_krastev_smirnov} from a combined analysis of the 
total rate results in the chlorine, SAGE, GALLEX and Super-Kamiokande
experiments, for oscillations in matter (MSW solutions, left) and in vacuum
(``Just-so'' or VO solutions, right).
The matter solutions are conventionally named according to the allowed 
ranges of the oscillation parameters: small mixing angle (SMA,
$\Dm2\sim 10^{-5}~\evolt^2$, $\sin^22\theta\sim 10^{-3}-10^{-2}$), large
mixing angle (LMA, $\Dm2\sim 10^{-5}-10^{-4}~\evolt^2$, $\sin^22\theta 
>0.5$) and low $\Dm2$ (LOW, $\Dm2\sim 10^{-7}~\evolt^2$,
$\sin^22\theta >0.9$). The VO
solution ($\Dm2 <10^{-9}~\evolt^2$) 
corresponds to the case when one astronomical unit exactly matches the
neutrino oscillation length for typical solar energies.

One way to discriminate between small and large mixing angles 
is to consider the 
so-called ``day/night effect''. Neutrinos detected during night, as
opposed to neutrinos detected during day, are subject to matter
effects in the Earth, which are expected to be sizeable for large
values of the mixing angle. The solar neutrino flux can be studied as a
function of the solar zenith angle $\theta_z$ and an asymmetry
${\mathcal{A}}$ can be defined:
\begin{equation}
{\mathcal{A}} = \frac{\Phi_N-\Phi_D}{\Phi_N+\Phi_D}\,,
\end{equation}
\noindent 
where $\Phi_N$ and $\Phi_D$ are the night and day fluxes respectively. 
A non-zero
asymmetry would favour large mixing angles.
Another way to separate between the two regimes is to
consider the shape of the solar zenith angle
distribution, which is expected to be flat for 
large mixing angles, while for small mixing angles 
it should show an excess of events in the
upward-going direction (around 
$-1<\cos\theta_z < -0.8$).

\begin{figure}[h]
\vspace*{13pt}
\setlength{\unitlength}{1cm}
\hbox{\hspace{0.5cm}
\epsfxsize=5.5cm\epsfbox{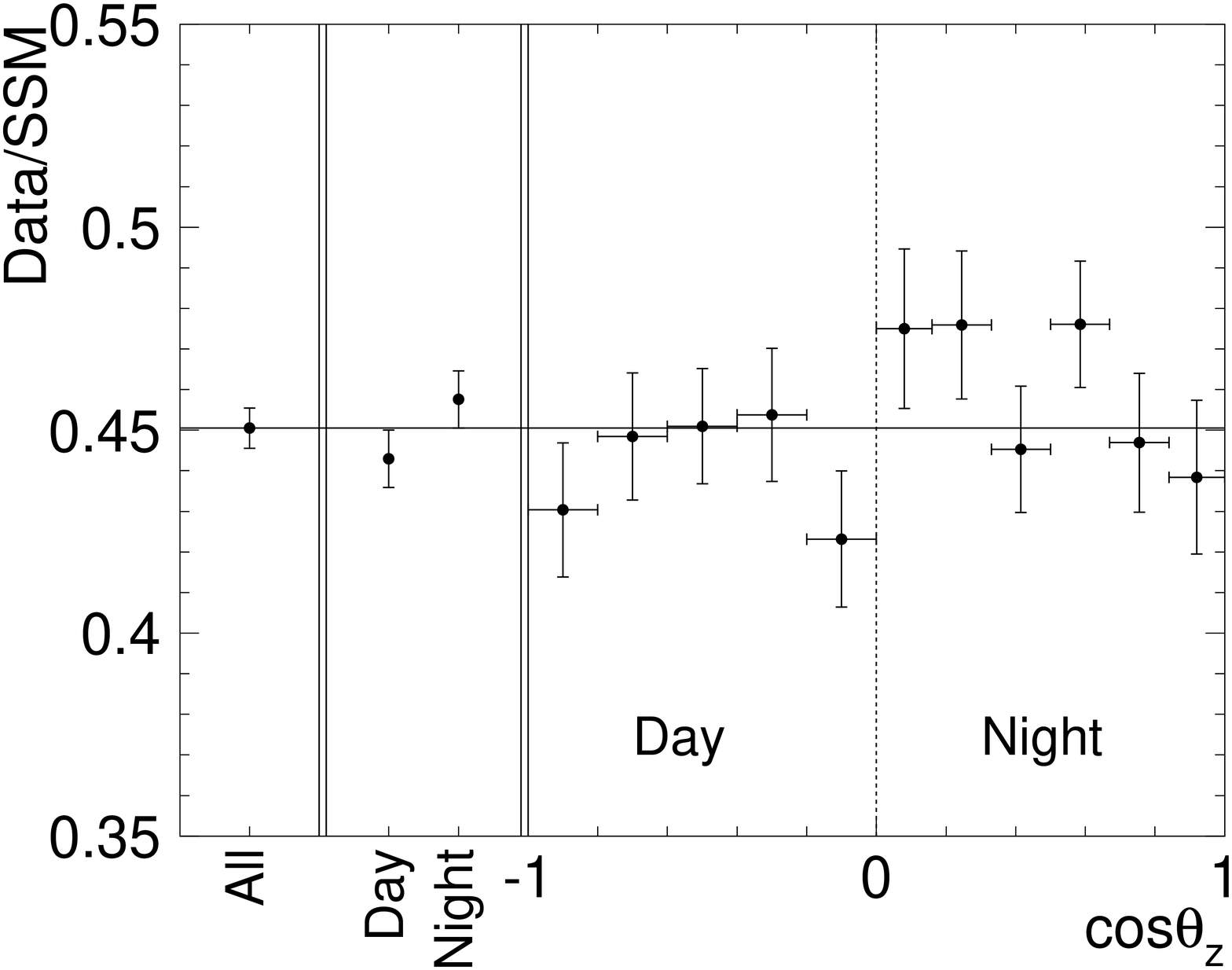}
\epsfxsize=5.5cm\epsfbox{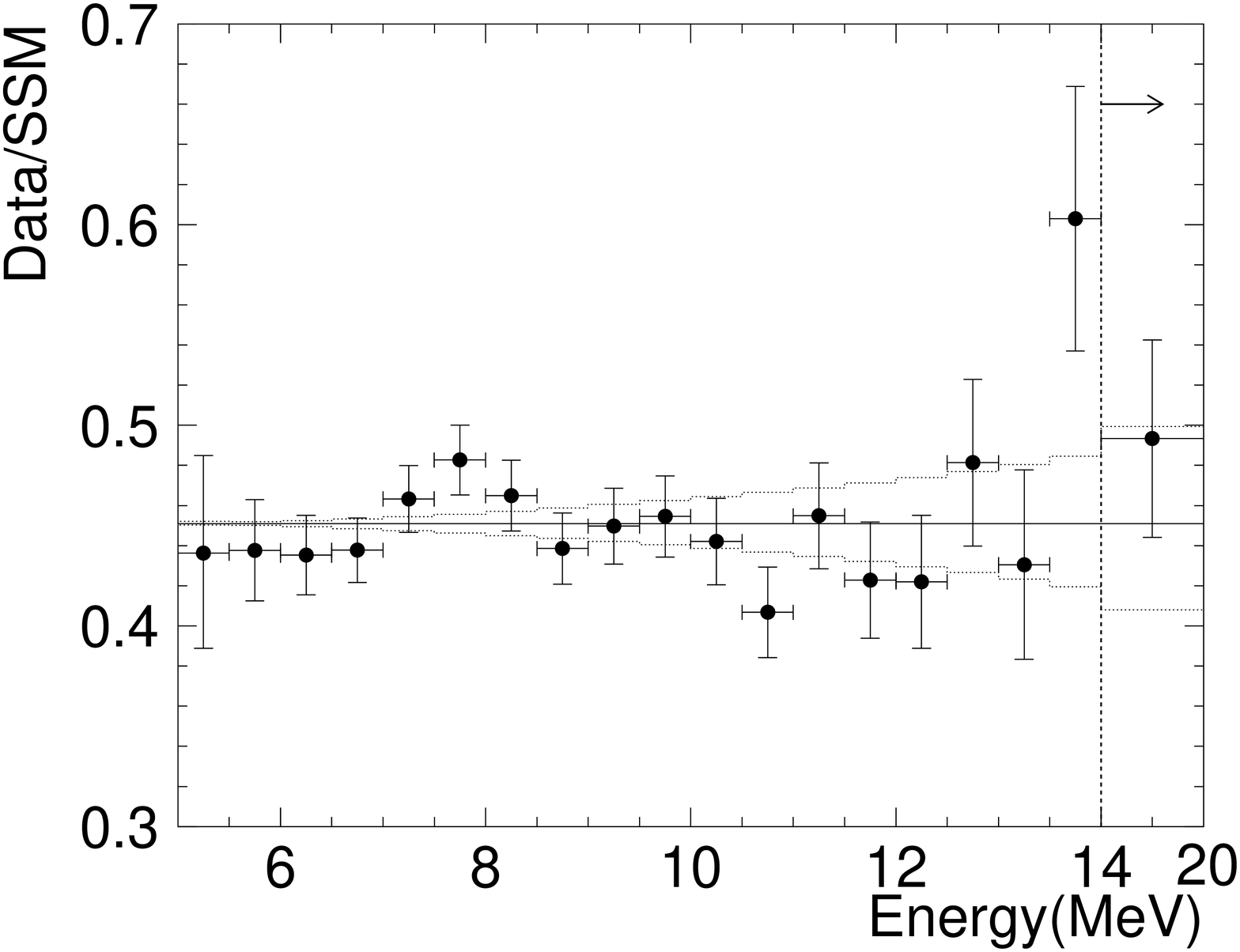}
}
\vspace*{13pt}
\fcaption{Super-Kamiokande ($1258$~days) measured distributions for 
  the solar zenith
  angle (left) and of the solar neutrino energy (right).\cite{superk_latest_b8}}\label{fig:costhetaz_energy_sun}
\vspace*{13pt}
\begin{center}
\epsfig{figure=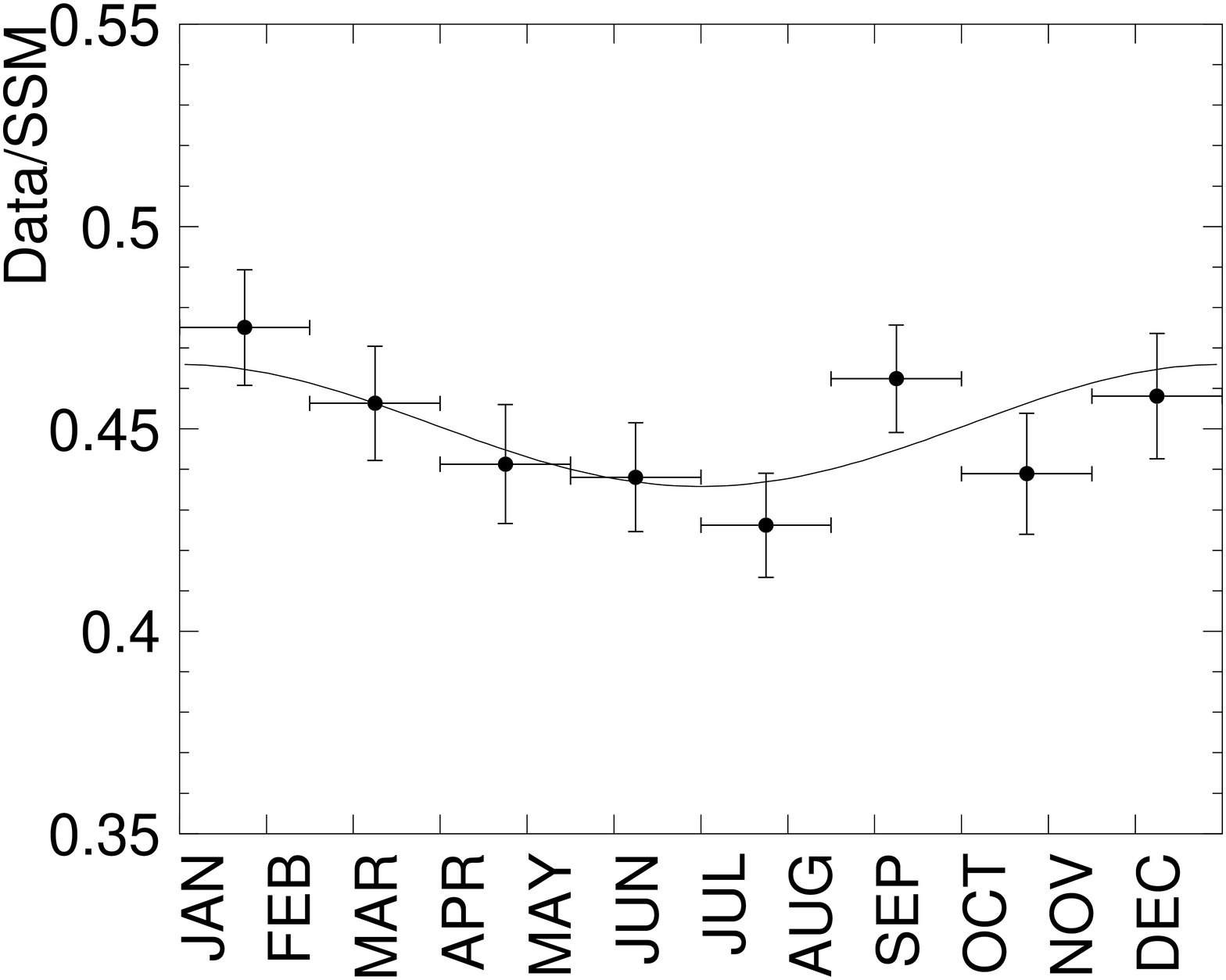,width=6.0cm}
\vspace*{13pt}
\fcaption{Super-Kamiokande ($1258$~days) measurement of the seasonal  variation of 
  the solar neutrino flux, not corrected for the effect due to the orbit
  eccentricity of the Earth.\cite{superk_latest_b8}}\label{fig:seasonal_variation}
\end{center}
\end{figure}

Oscillations would also manifest themselves as a distortion of
the neutrino energy spectrum and it is 
therefore important to study the
neutrino energy distribution, which for Super-Kamiokande is
approximated by the recoil electron energy. 
It turns out that, if the predicted $hep$
neutrino flux is correct, a distribution peaked at
high energies would favour the VO solution.

Finally, strong evidence for neutrino
oscillations in vacuum would come from the observation of a seasonal
variation of the solar neutrino flux, different from the standard
geometrical one due to the orbit eccentricity of the Earth($\epsilon =
0.0167$).\footnote{For
$\Dm2$ of the order of that suggested by the VO solution, and for solar 
neutrino energies of the order of $\sim \text{few}~\mev$, the seasonal
change in the distance between the Sun and the Earth ($\Delta R_\odot
= 2\epsilon R_\odot$) is a sizeable
fraction of the 
oscillation length in vacuum, $\lambda = 4\pi E/\Dm2$ (numerically,
$2\times 0.016\times
(1.496\times 10^{11}~\text{m})\times (2\times
10^{-10}~\evolt^2)/1~\mev\sim 1$).}

The day-night asymmetry obtained by Super-Kamiokande based on $1258$ days 
of data is ${\mathcal{A}}=0.033\pm
0.022\text{(stat.)}^{+0.013}_{-0.012}\text{(sys.)}$,\cite{superk_latest_b8}
which is only a
$1.3\sigma$ deviation from zero asymmetry. The zenith angle
distribution (Fig.\ref{fig:costhetaz_energy_sun}, left) is rather flat and
shows no feature which might lead to a preference for the SMA solution.
The
measured energy spectrum shows no 
statistically significant distortion relative to the predicted
$^8\text{B}$ spectrum (Fig.\ref{fig:costhetaz_energy_sun}, right). No deviation from the 
expected standard annual variation of the neutrino flux has been
observed (see Fig.\ref{fig:seasonal_variation}).

\begin{figure}[h]
\vspace*{13pt}
\begin{center}
\epsfig{figure=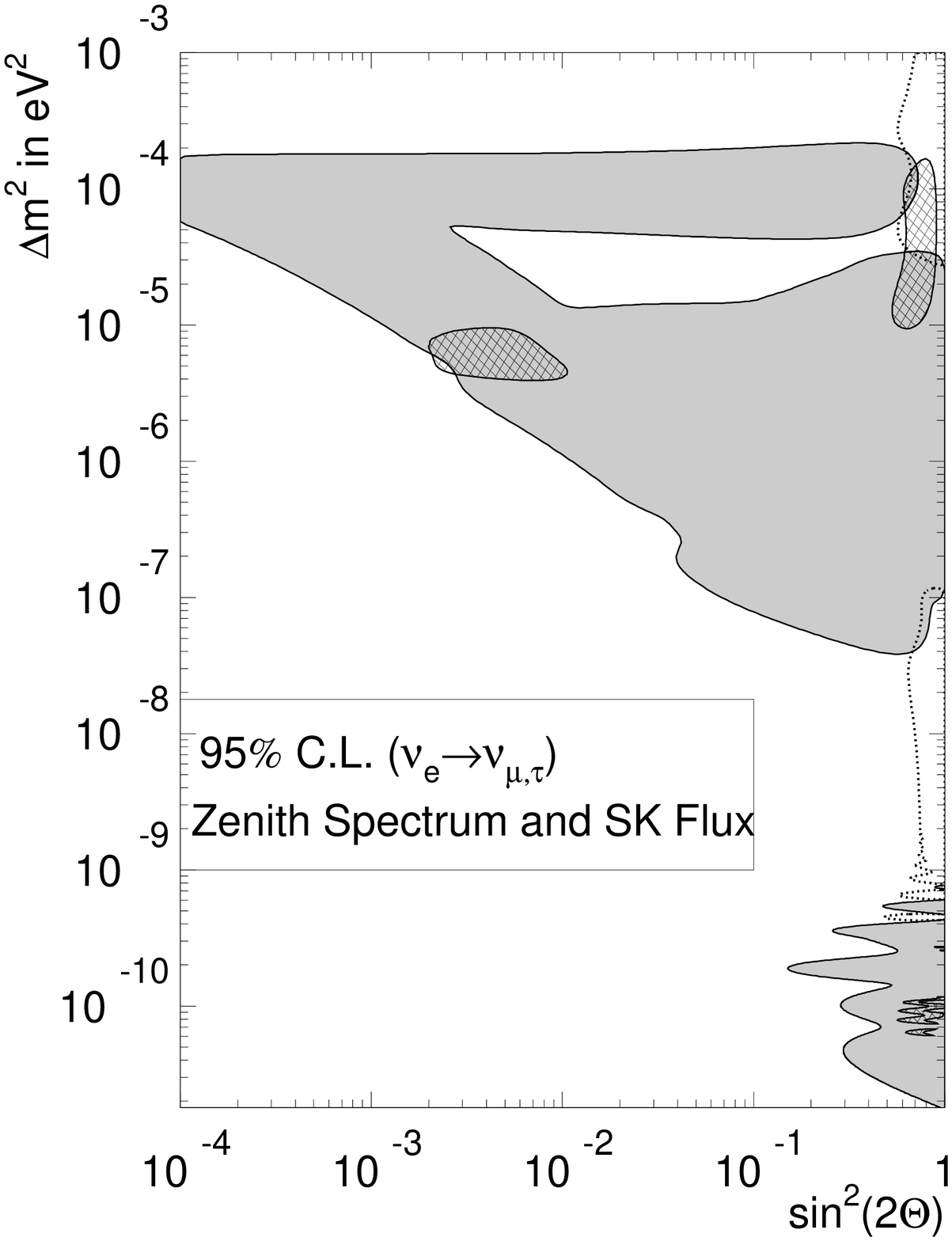,width=6cm}
\vspace*{13pt}
\fcaption{Super-Kamiokande exclusion areas for solar neutrino
  oscillations (shaded area, based on $1258$~days).\cite{superk_latest_b8} The allowed areas obtained
  using the zenith angle distribution and the SSM predictions (dotted
  line) and those obtained combining all the experimental results on
  integrated solar neutrino fluxes (hatched islands) are also shown.}\label{fig:superk_solar_exclusion}
\end{center}
\end{figure}

The Super-Kamiokande collaboration has recently performed a flux-independent
$\chi^2$ analysis of their data, assuming two-flavour mixing.
The resulting exclusion plot ($95\%$~C.L.) for the
$\nu_e\rightarrow\nu_{\mu,\tau}$ case is shown in
Fig.\ref{fig:superk_solar_exclusion} (shaded areas), together with the
allowed areas obtained using the zenith angle distribution and the SSM 
predictions (dotted) and the allowed areas obtained from a
combined analysis of the GALLEX, SAGE, Homestake and Super-Kamiokande
flux measurements (hatched). From this analysis the SMA and the
VO solutions seem to be disfavoured at $95\%$~C.L. The same analysis
repeated for pure $\nu_e\rightarrow\nu_s$ shows that all solutions 
are disfavoured at $95\%$~C.L. for the sterile neutrino hypothesis.

Although Super-Kamiokande has measured the energy dependence 
and the zenith angle dependence of the solar $^8\text{B}$ flux, and
extracted useful information about the MSW and the VO solutions out of
an impressively high-quality and high-statistics sample, the solar neutrino
problem can by no means be considered definitively solved.
Many questions still remain open and 
it is therefore fair to say that still lot of
information is missing and further experimental effort is
required.
Along this line there come the Sudbury Neutrino Observatory project
(SNO),\cite{sno} which has been running for about one year, as well as the
Borexino\cite{borexino} and the 
KamLAND\cite{kamland} experiments, 
which are both
scheduled to start data taking very soon.

SNO is a $1$~kton heavy-water Cherenkov detector located
$2$~km underground ($6010$~m.w.e), in the
Creighton mine, Sudbury, Ontario, Canada. 
The detector is made of a spherical acrylic vessel, $12$~m diameter,
containing ultra-pure $\text{D}_2\text{O}$ and surrounded by
an ultra-pure $\text{H}_2\text{O}$ shield, in turn contained in a cylindrical
cavity ($34$~m height, $22$~m maximum diameter). The light produced in 
the water is detected by $9,456$
PMTs ($20$~cm diameter) with light concentrator, installed on a stainless steel
structure surrounding the acrylic vessel.

As well as using the elastic
scattering (ES) reaction (\ref{eq:elastic_nue_scatt}), as Kamiokande and
Super-Kamiokande do, SNO can detect $^8\text{B}$
solar neutrinos via charged-current (CC) and neutral-current
(NC) interactions on deuterium:
\noindent
\begin{equation}\label{eq:sno_cc}
  \begin{matrix}
    \nu_e + d \rightarrow p + p + e^- & & \text{(threshold = $1.4~\mev$)}\\
  \end{matrix}
\end{equation}
\noindent
and:
\begin{equation}\label{eq:sno_nc}
  \begin{matrix}
    \nu_x + d \rightarrow n + p + \nu_x & & \text{(threshold = $2.2~\mev$)}\,.\\
  \end{matrix}
\end{equation}
\noindent
For completeness, we give again the ES reaction (\ref{eq:elastic_nue_scatt}):
\begin{equation}
\nu_x + e^- \rightarrow \nu_x + e^-\hspace{2cm} (x=e,\mu,\tau)\,.\tag{36}
\end{equation}

Neutral current interactions do not depend on the flavour 
of the incident neutrino and therefore
neutrino oscillations to active neutrinos should not affect the number of NC events. On
the other hand, given the energy range for solar neutrinos, $\nu_\mu$
and $\nu_\tau$ coming from oscillated $\nu_e$
would be below threshold for the production of 
their charged partners: the CC sample would then be depleted by
neutrino oscillations. This means that the CC/NC ratio can be used as
a powerful probe to test the oscillation hypothesis. Moreover,
following the same argument, any distortion in the measured CC
energy spectrum with respect to predictions 
would be another signature of new physics.

The electron produced in (\ref{eq:sno_cc}), 
as opposed to that produced in (\ref{eq:elastic_nue_scatt}), 
carries most of the neutrino kinetic energy in the final state. 
Therefore it provides a better
estimate of the initial neutrino energy (resolution of $\sim 20\%$ for
the range of interest) 
and can be used to 
observe possible distortions of the neutrino energy spectrum.

While electrons produced in CC and ES reactions are detected via the
associated Cherenkov light, 
neutrons from reaction (\ref{eq:sno_nc}) are detected
using different techniques during the three different phases of the
experiment. In the first phase, recently concluded, when only heavy
water was used,
they were measured through the $6.25~\mev$ photons produced
in the neutron capture ($25\%$ efficiency); in the current phase, when
$2.5$~tons of NaCl have been added
to the water, they are seen 
through the $8.6~\mev$ photons produced in the neutron
capture on Cl ($85\%$ efficiency); finally, and this will be SNO's third phase, the
salt will be removed and $^3\text{He}$ proportional counters will be
installed, allowing direct detection of the neutrons ($45\%$
efficiency)
and measurement of the spectra with completely different systematics. 

The sample collected during the first phase, when the sensitivity to
NC was lower, is essentially dominated by CC interactions, with a smaller
component from ES events.
Therefore the CC/NC comparison will not be possible until data from
the second phase, with the NC 
rate 
highly enhanced by the addition of salt, become available.
However some information on neutrino oscillations can already be
extracted from the first phase data, by 
comparing the CC and ES rates. In fact, while the CC reaction
(\ref{eq:sno_cc}) is only sensitive to $\nu_e$'s, because of
the low solar neutrino energy, the ES reaction
(\ref{eq:elastic_nue_scatt}) is sensitive to all active neutrinos,
although with a reduced sensitivity 
to $\nu_\mu$ and $\nu_\tau$. Thus, if neutrino oscillations change the 
flavour composition of the solar neutrino flux, transforming
a fraction of the original $\nu_e$'s into other active neutrino species, 
the CC sample will be depleted, while the ES sample remains
unchanged. In other words, if the CC and ES measurements of the 
$^8\text{B}$ electron neutrino fluxes give:
\begin{equation}
\Phi_{^8\text{B}}^{CC}(\nu_e) < \Phi_{^8\text{B}}^{ES}(\nu_e) \,,
\end{equation}
\noindent 
then, without having to refer to any particular solar model, the
inequality between the two measured values can by itself be interpreted as
an indication of neutrino oscillations to active neutrinos.

The SNO collaboration has very
recently published their first results on ES and CC
reactions, based on $240.95$~live days of
data.\cite{sno_first_results} In order to avoid biases in the event selection procedure, the
data has been sub-divided in two subsets, with $70\%$ of them used to
establish the analysis procedure and the remaining $30\%$ used as a
blind sample to validate the analysis itself. The final data set
consists of $1169$ events. 

The different contributions from CC, ES
and residual neutron events are extracted by means of a maximum
likelihood function, which combines the information from three variables:
the effective kinetic energy of the event, the angle
between the reconstructed direction of the event and the instantaneous 
Sun-to-Earth direction and a volume-weighted radial variable,
related to the position of the event inside the detector.  
The breakdown of the events as obtained from the likelihood 
maximisation is given in Tab.\ref{tab:events_sno}.
\begin{table}[h]
\tcaption{Contribution from CC, ES and neutron events to the 
  neutrino interaction sample extracted from 
  $240.95$~live days of SNO data, first phase.\cite{sno_first_results}
The quoted errors are statistical only.}
\centerline{\footnotesize\smalllineskip
\begin{tabular}{c c}\label{tab:events_sno}\\
\hline
Sample & Nr. of events\\
\hline
CC      & $975.4\pm 39.7$\\
NC      & $106.1\pm 15.2$\\
neutron & $87.5\pm 24.7$\\
\hline\\
\end{tabular}}
\end{table}

SNO data show that the shape of the CC energy spectrum is
consistent with the Bahcall SSM predictions within one standard
deviation. However, the low ratio of the measured $^8\text{B}$ flux 
to SSM predictions,\cite{bp2000} $0.347\pm 0.029$,\cite{sno_first_results}
confirms the solar neutrino deficit. Moreover it is smaller
than what obtained by Super-Kamiokande from the analysis of ES events.\cite{superk_latest_b8}

Using events with kinetic energy above $6.75~\mev$, the 
$^8\text{B}$ fluxes as measured from CC and ES interactions in SNO are:
\begin{align}
\Phi^{CC}_{SNO}(\nu_e) & = 
(1.75\pm 0.07 (stat.) ^{+0.12}_{-0.11} (sys.)\pm 0.05 (theor.)) 
\times 10^6~\text{cm}^{-2}\text{s}^{-1}\label{eq:sno_phi_cc}\\
\Phi^{ES}_{SNO}(\nu_x) & = 
(2.39\pm 0.34 (stat.) ^{+0.16}_{-0.14} (sys.)) 
\times 10^6~\text{cm}^{-2}\text{s}^{-1}
\end{align}
\noindent
where the theoretical uncertainty in (\ref{eq:sno_phi_cc}) is that on the total CC cross section.

The difference between $\Phi^{CC}_{SNO}(\nu_e)$ and $\Phi^{ES}_{SNO}(\nu_e)$ 
is $(0.64\pm 0.40)\times 10^6~\text{cm}^{-2}\text{s}^{-1}$, that is to say
a discrepancy of only $1.6~\sigma$. However the SNO measurement of the ES 
flux is consistent with that measured with higher precision by Super-Kamiokande:\cite{superk_latest_b8}
\begin{equation}
\Phi^{ES}_{SK}(\nu_x)  = 
(2.32\pm 0.03 (stat.) ^{+0.08}_{-0.07} (sys.)) 
\times 10^6~\text{cm}^{-2}\text{s}^{-1}\,,
\end{equation}
\noindent
which permits to make a direct comparison between $\Phi^{CC}_{SNO}(\nu_e)$ and $\Phi^{ES}_{SK}(\nu_x)$.

The difference between the ES flux measured by Super-Kamiokande and
the CC flux measured by SNO is $(0.57\pm 0.17)\times
10^6~\text{cm}^{-2}\text{s}^{-1}$, this time corresponding to a more
significant $3.3~\sigma$
discrepancy. This result, by showing that the electron neutrino 
flux from CC interactions is indeed smaller than that obtained
from ES events, 
provides quite a convincing hint for appearance
of non-electron active neutrino flavours in the solar neutrino flux.

If electron neutrinos oscillated to sterile neutrinos, 
the $^8\text{B}$ flux derived by SNO using CC
events with energy above $6.75~\mev$ should be consistent with 
the flux
measured by 
Super-Kamiokande using ES interactions above
$8.5~\mev$.\cite{fogli_sno_sk} Instead, correcting for 
the different energy threshold used in the latest Super-Kamiokande
analysis,\cite{superk_latest_b8} the two fluxes differ by $(0.53\pm
0.17)\times 10^6~\text{cm}^{-2}\text{s}^{-1}$, that is to say
$3.1~\sigma$. Therefore the SNO data exclude the possibility that the
solar neutrino deficit can be explained by pure
$\nu_e\rightarrow\nu_s$ oscillations.

\begin{figure}[h]
\begin{center}
\epsfig{file=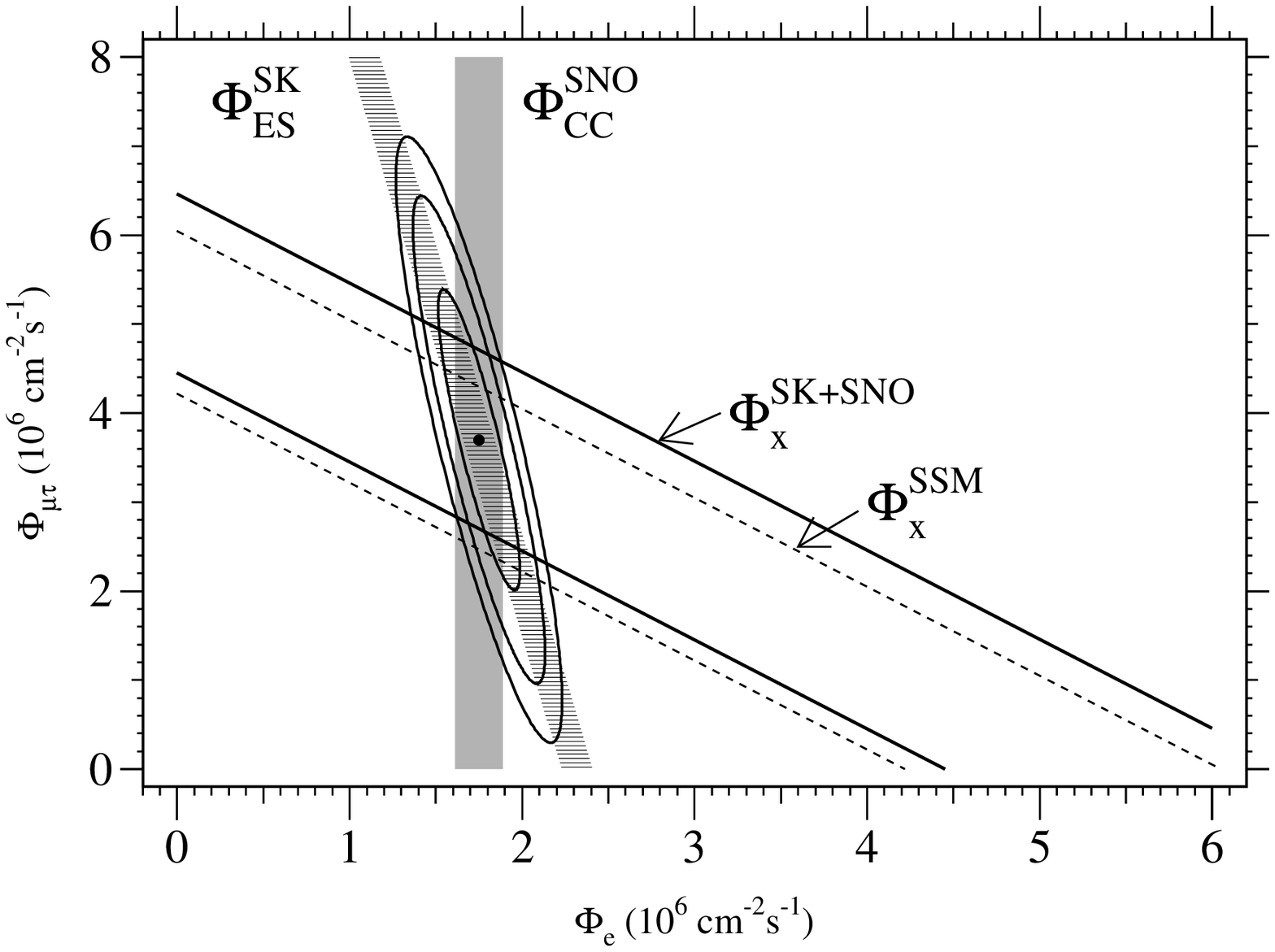,width=6cm}
\vspace*{13pt}
\fcaption{Correlation between the non-electron active component of the 
  solar $^8\text{B}$ flux, $\Phi_{\mu\tau}$, and the electron component 
  $\Phi_e$. 
The two shaded bands represent the 
$1\sigma$ bands of the neutrino fluxes as obtained from SNO CC\cite{sno_first_results} and
Super-Kamiokande ES measurements.\cite{superk_latest_b8} The ellipses are the $68\%$, $95\%$
and $99\%$ joint probability for the two fluxes.
The two diagonal bands are the measured (SK+SNO) and
predicted (SSM)\cite{bp2000} bands of constant total flux of active
$^8\text{B}$ neutrinos.}\label{fig:sno_fluxes}
\end{center}
\end{figure}

In the hypothesis that the observed solar neutrino  
deficit is due to oscillations of $\nu_e$'s to other
active neutrino species ($\nu_\mu$ and $\nu_\tau$), the flux of
non-electron active neutrinos, $\Phi_{\mu\tau}$,  
can be inferred. The obtained correlation between $\Phi_{\mu\tau}$
and $\Phi_e$ is shown 
in Fig.\ref{fig:sno_fluxes}. The best fit
corresponds to a non-electron active neutrino flux of:
\begin{equation}
\Phi_{\mu\tau} = (3.69\pm 1.13)\times 10^6~\times{cm}^{-2}\text{s}^{-1}\,.
\end{equation}
\noindent
The two diagonal bands correspond to constant total flux of active
$^8\text{B}$ neutrinos, as measured by SNO and Super-Kamiokande (solid
line) and as predicted by the SSM model (dashed line). The measured value,
$\Phi(\nu_x) = (5.44\pm 0.99)\times
10^6~\text{cm}^{-2}\text{s}^{-1}$, is in very good agreement with predictions.

In conclusion, SNO has presented a first direct indication for a
non-electron active component in the solar neutrino flux. Moreover
the experiment measures a 
$^8\text{B}$ neutrino flux consistent with the SSM expectations.
These are only the first of several very interesting measurements that
SNO should be able to perform over the next few years: we now look
forward to the results from the experiment's second phase.

Borexino,\cite{borexino} a $300$~ton liquid scintillator detector with PMT readout,
is being installed at the Gran Sasso Underground Laboratory. Thanks to 
the high radiopurity and to the high light yield of the scintillator, it will
be able to detect neutrino-electron scattering with an energy
threshold as low as $250~\kev$. This means that the $^7\text{Be}$
lines, especially that at $0.863~\mev$, will be within the reach of
the experiment. The monochromaticity of the beryllium line should
facilitate the observation of VO-induced variation of the
solar neutrino flux, which for Borexino would be a spectacular
effect. 
Moreover Borexino will
also be able to check the LOW solution, by studying the day/night
effect for beryllium neutrinos, which is expected to be strong if indeed
oscillations occur with the LOW
parameters.\cite{low_solution_borexino} 
Borexino is expected
to start data-taking in 2002.

A test of the LMA solution will be made by the KamLAND\cite{kamland} 
experiment, a
$1$~kton liquid scintillator detector, located in the Kamioka mine in
Japan. The
experiment will detect the low-energy ($1-8~\mev$) 
$\bar{\nu}_e$ flux emitted by the nuclear
reactors at several Japanese 
power stations, at distances between $150$ and $210$~km from the detector.
The estimated sensitivity of KamLAND is such that, at $90\%$~C.L. and
after three years of data taking, the experiment  should be able to cover
the entire domain defined by the LMA solution (down to $\Dm2\sim
4\times 10^{-6}~\evolt^2$). If oscillations are
observed, the oscillation parameters should be determined to a
precision of $20\%$, at $99\%$~C.L.. KamLAND is expected to start
data-taking at the end of this year.

More experiments are planned to study solar neutrinos in the
future. Among them one should certainly mention ICARUS,\cite{icarus_proposal}
a liquid-argon TPC detector, which will be mainly sensitive to
$^8\text{B}$ and of which the first $600$~tons half-module 
is currently under test,  HERON,\cite{heron}
which plans to study $pp$ and beryllium neutrinos using superfluid
helium as a target, and HELLAZ,\cite{hellaz} which intends to study
low-energy neutrinos, down to the $pp$ neutrino energies, using a
gaseous helium TPC.

In the previous sections we have outlined the solar neutrino problem, 
which historically represents the first experimental hint for neutrino 
oscillations. We have described the standard solar model results and 
the arguments which make the model convincing and robust. We have
discussed how experimental data are in striking contradiction with
expectations, if no new physics is taken into account, and we have
summarize the results obtained when the solar neutrino
deficit is interpreted as being due to neutrino oscillations. 
We now want to plunge into
another exiting domain, that of atmospheric neutrinos,
which undoubtedly provide the most convincing experimental 
evidence for neutrino oscillations.

\section{The Atmospheric Neutrino Anomaly}

\subsection{Atmospheric Neutrino Flux and Neutrino Oscillations}

Atmospheric neutrinos are produced in the decay of secondary
particles created in the interaction of primary
cosmic rays\footnote{mainly protons ($\sim 80\%$) and
  $\alpha$-particles ($\sim 15\%$),
plus a small contribution from heavier nuclei.\cite{gaisser_review_cosmic}}\hspace{3pt} with the
Earth's atmosphere.
If the energy of the secondary particles
is sufficiently low ($\lesssim 2~\gev$) that all of them decay, we have:
\begin{align}\label{eq:atmo_shower}
p + {\mathcal{N}} & \longrightarrow \pi^\pm + X \notag\\
\pi^\pm & \longrightarrow \mu^\pm + \nu_\mu (\bar{\nu}_\mu) \\
\mu^\pm & \longrightarrow e^\pm + \nu_e (\bar{\nu}_e) + \bar{\nu}_\mu (\nu_\mu)\notag
\end{align}
\noindent
Assuming that no 
effect can change the flavour composition of the shower before 
it is measured on Earth,
Eq.(\ref{eq:atmo_shower}) implies that:
\begin{equation}\label{eq:mu_e_ratio}
{\mathcal{R}} = \dfrac{{\mathcal{N}}_{\nu_\mu} + {\mathcal{N}}_{\bar{\nu}_\mu}}{{\mathcal{N}}_{\nu_e} + {\mathcal{N}}_{\bar{\nu}_e}}
\sim 2
\end{equation}
\noindent
where ${\mathcal{N}}_{\nu_\mu}$ (${\mathcal{N}}_{\nu_e}$) and 
${\mathcal{N}}_{\bar{\nu}_\mu}$ (${\mathcal{N}}_{\bar{\nu}_e}$) are the
number of muon (electron) neutrinos and anti-neutrinos respectively.
The exact value of ${\mathcal{R}}$ can in principle be affected by several 
effects, such as the primary spectrum composition, the geomagnetic
cut-off, solar activity and, of course, the details of
the model for the development of the hadronic shower. 
However, it has to be said that, although
the absolute neutrino fluxes are rather badly known (predictions from different
calculations disagree 
by $\sim 20-30\%$), the ratio (\ref{eq:mu_e_ratio}) is to
first order insensitive to such uncertainties and is 
known to $\sim 5\%$. Neutrino oscillations 
would manifest themselves as a discrepancy
between the measured and the expected value of the ratio
${\mathcal{R}}$. 

In order to quote a number which is independent of
experimental parameters, such as energy thresholds and detector
acceptances for signal and background, the result is generally
presented in terms of the double ratio:
\begin{equation}\label{eq:ratio_of_ratios}
{\mathcal{R}}'\equiv
\dfrac{{\mathcal{R}}_{DATA}}{{\mathcal{R}}_{MC}}=
\dfrac
{\ \ \Bigg[\dfrac{{\mathcal{N}}_{\nu_\mu} +
  {\mathcal{N}}_{\bar{\nu}_\mu}}{{\mathcal{N}}_{\nu_e} +
  {\mathcal{N}}_{\bar{\nu}_e}}\Bigg]_{DATA}}
{\Bigg[\dfrac{{\mathcal{N}}_{\nu_\mu} +
  {\mathcal{N}}_{\bar{\nu}_\mu}}{{\mathcal{N}}_{\nu_e} +
  {\mathcal{N}}_{\bar{\nu}_e}}\Bigg]_{MC}}\,,
\end{equation}
\noindent
where ${\mathcal{R}}_{DATA}$ is the measured ratio of muon-like
(i.e. $\nu_\mu$, $\bar{\nu}_\mu$) over
electron-like (i.e. $\nu_e$, $\bar{\nu}_e$) 
events and ${\mathcal{R}}_{MC}$ is the same ratio as
obtained from a Monte Carlo simulation. 
Assuming a correct modelling for ${\mathcal{R}}_{MC}$, ${\mathcal{R}}'<1$
could either mean that there is a deficit of
measured muon-like events, or that an excess of electron-like events
has been observed. The quantity (\ref{eq:ratio_of_ratios}) alone cannot
discriminate between these two possibilities. Moreover, a low value of 
${\mathcal{R}}'$ 
is not by itself a proof for neutrino
oscillations, since one could imagine other mechanisms (i.e. neutrino
decay) inducing a similar effect.

\begin{figure}[h]
\vspace*{13pt}
\begin{center}
\epsfig{file=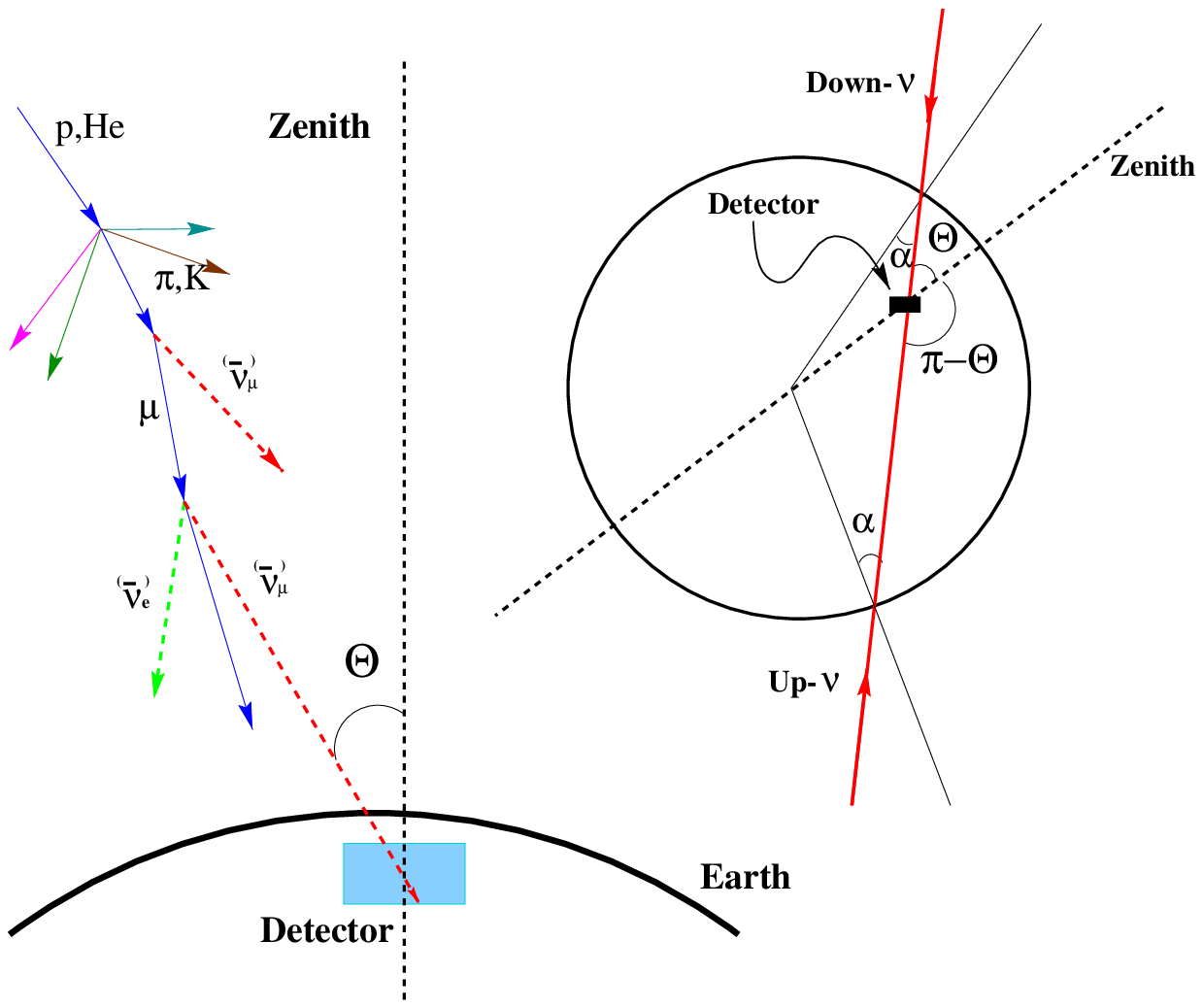,height=6cm}
\vspace*{13pt}
\fcaption{Sketch of the atmospheric neutrinos detection principles.}\label{fig:atmo_nu_sketch}
\end{center}
\end{figure}

Another, more sensitive way to detect neutrino oscillations is the
study of the zenith angle distribution of the incoming atmospheric neutrino.
Downward-going neutrinos, produced in the atmosphere above the
detector, will travel a path of the order of $10-20$~km, while
upward-going neutrinos, produced at the opposite side of the globe, 
will have travelled up to $12,000$~km before detection. 
This means that, by studying the neutrino flux as a function 
of the zenith angle, one has access to baselines spanning three orders 
of magnitude and can therefore hope to observe the modulation induced
on the flux by neutrino oscillations. 

If no oscillation occurs between the 
production point and the detector, assuming for simplicity that no
other effect alters the cosmic ray angular distribution,
the neutrino flux per unit area is the
same at any location on the Earth. If we consider the contribution to
the flux from a point at zenith angle $\Theta$ (downward-going
neutrinos) and we compare it to
the flux at zenith angle $\pi-\Theta$ (upward-going neutrinos), 
based on simple symmetry considerations (see
Fig.\ref{fig:atmo_nu_sketch}), we
can demonstrate that the two quantities must have the same value in
the absence of new physics. 
In summary, in the no-oscillation hypothesis, the zenith angle
distribution must be up-down symmetric, assuming no other phenomena
affecting the neutrino angular distribution relative to the local
vertical direction. Conversely, any deviation from up-down symmetry
could be interpreted as 
an indication for neutrino oscillations.

\begin{figure}[h]
\vspace*{13pt}
\begin{center}
\epsfig{file=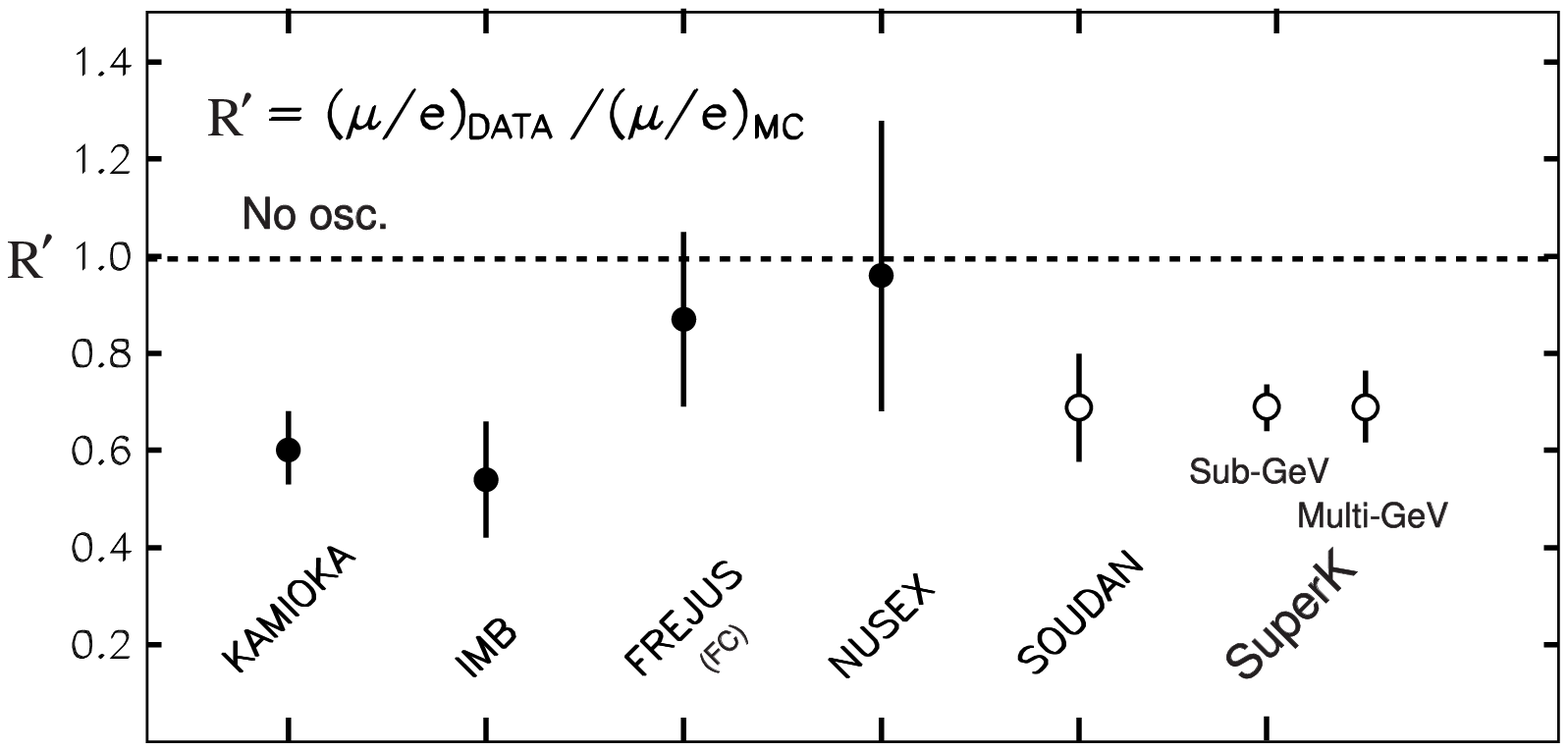,width=7.5cm}
\vspace*{13pt}
\fcaption{Compilation of the ${\mathcal{R}}'$ ratio (see text)
  measured by several atmospheric neutrino experiments (from ref.\cite{mann}).}\label{fig:atmo_ratios}
\end{center}
\end{figure}

Historically, the first results on what is nowadays known as the
{\it atmospheric neutrino anomaly} came from
experiments originally designed to search for proton decay, for which
the atmospheric neutrino 
flux constituted a background source. 
A compilation of results on ${\mathcal{R}}'$ from different
experiments is shown
in Fig.\ref{fig:atmo_ratios} (from ref.\cite{mann}): all measurements
are systematically below one and all 
but two give a value for ${\mathcal{R}}'$ close to $0.7$.

The first experiment to report, in 1986, a
discrepancy between the observed and the predicted number of
atmospheric neutrinos, although that was not immediately recognised as 
a possible effect of neutrino oscillations, was IMB,\cite{imb} a water-Cherenkov
detector located in the Morton mine, Cleveland, Ohio, USA. 
Two years later Kamiokande confirmed that the measured deficit of muon-like events 
was the order of $30\%$.\cite{kamiokande_atmo}
However, the same effect was not observed by two other proton decay
experiments, NUSEX,\cite{nusex} and Fr\'ejus,\cite{frejus}, both using 
fine-grained iron calorimeters.
This discrepancy led for some time to the belief that poorly understood
systematic effects, mainly related to an incomplete description of
neutrino interactions in iron and water, 
were inducing an intrinsic difference between the
two experimental techniques. Although later calculations 
had already shown that the neglected physical processes 
could account for at most a $10-15\%$ change
in the expected event rates,\cite{engel_vogel}
the issue 
was only definitively resolved when the Soudan2 experiment,\cite{soudan2_97}
another fine-grained iron calorimeter, located in the Soudan mine ($2100$~m.w.e.), Ely, 
Minnesota, USA, confirmed Kamiokande and IMB results.
It is now common belief that the original discrepancy was due to
fluctuations in NUSEX and Fr\'ejus data.
In 1994 Kamiokande showed a distortion of the zenith angle 
distribution in the multi-$\gev$ contained events (see below for a
definition),\cite{kamiokande_94} which seemed to confirm the
neutrino oscillation hypothesis. Kamiokande and IMB as well as 
Baksan,\cite{baksan} an experiment in the former Soviet Union, and
MACRO,\cite{macro_95} a proton decay experiment at the Gran Sasso
Laboratory, also tried to extract information from the sample of
particles travelling in the vertical direction, 
but the results were not consistent with each other and none of
them seemed conclusive.
 
The big change in the scientific community's perception of the
atmospheric neutrino anomaly certainly came from the results obtained
by the 
Super-Kamiokande experiment,\cite{superk_atmo1,superk_geomagnetic,sobel_nu2000} which in summer 1998
announced evidence for neutrino oscillations in the atmospheric
neutrino sample. However, 
before we discuss atmospheric neutrinos in Super-Kamiokande, we shall
first present some of the ideas and the status of the
atmospheric neutrino flux calculations.

\subsection{Atmospheric Neutrino Flux Calculations}
The neutrino flux on Earth (approximately
$1$~event/$\text{cm}^2$/sr/sec) can in principle 
be calculated either starting from the spectrum
of the primary protons and alpha particles which initiate the hadronic
cascade in the atmosphere, or from the measured muon flux at high
altitudes. In the first case it is necessary to parametrize the
hadronic shower development, thus requiring a relatively good
understanding of the pion production process at the relevant energies.
On the other hand, the muon acceptance has the disadvantage of being 
quite sensitive to variations of
the geomagnetic field. Moreover the exposure time for muon flux 
measurements is quite short, the relevant experiments being generally
performed on balloons,\cite{mass,caprice_muon,heat} 
while primary fluxes measurements, made at high altitudes or above the 
atmosphere, are usually made over relatively
longer intervals.\cite{mass,ams,bess,caprice_proton,imax} 
In any case a measurement of the muon flux gives a useful constraint on the 
the atmospheric neutrino flux calculations.

\begin{figure}[h]
\vspace*{13pt}
\begin{center}
\epsfig{file=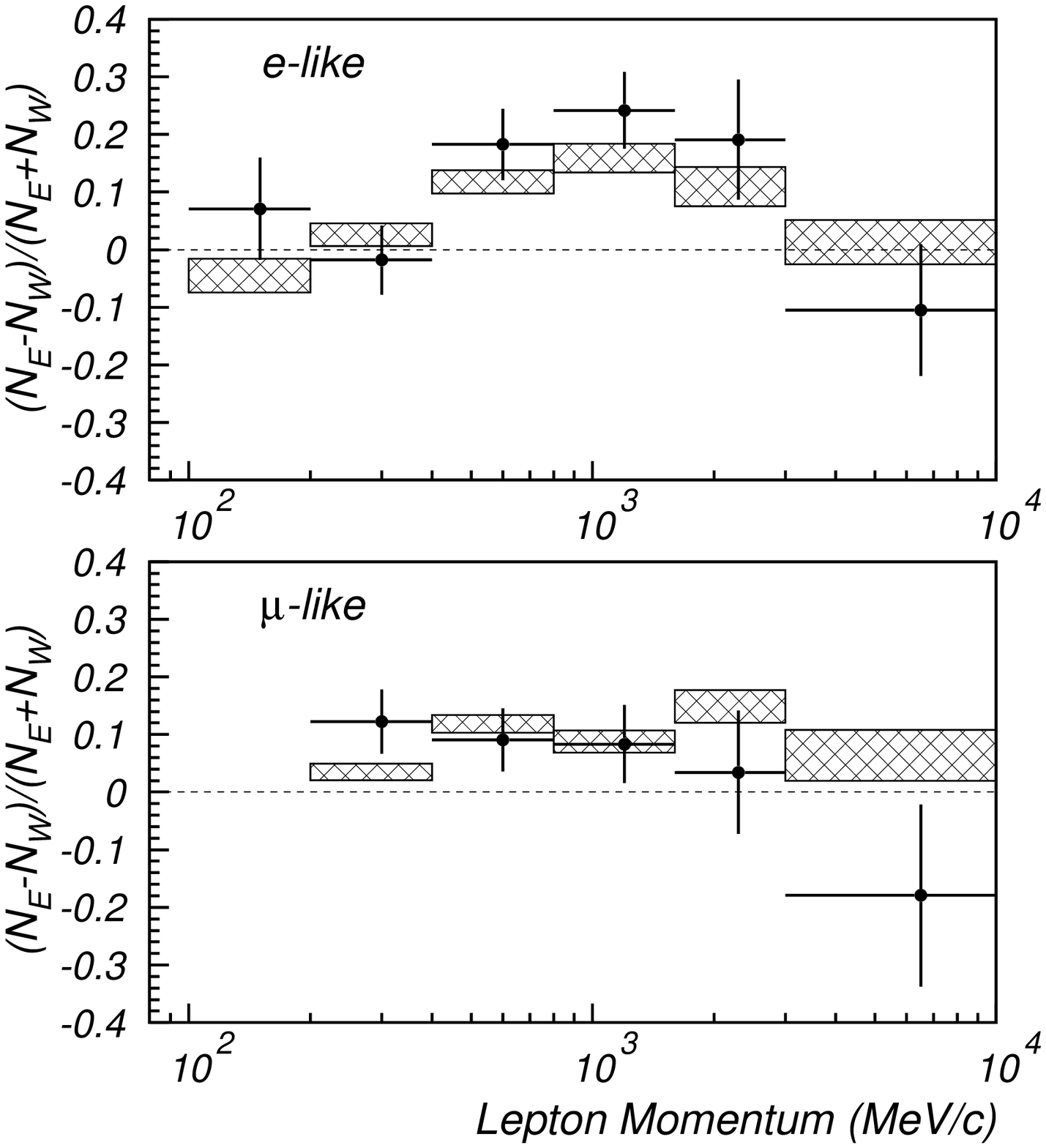,width=6cm}
\vspace*{13pt}
\fcaption{East-west asymmetry measured by Super-Kamiokande\cite{superk_geomagnetic} as a
  function of the lepton momentum (from ref.\cite{superk_geomagnetic}), 
  compared to expectations.\cite{honda} $N_E$ and $N_W$ are the number 
  of events from the east and the west respectively.}\label{fig:superk_east_west}
\end{center}
\end{figure}

At low geomagnetic latitudes the trajectories 
of low energy primary cosmic rays can 
be significantly deviated under the action of the geomagnetic field
and, in extreme cases (energies of the order of few $\gev$), they
would not reach the atmosphere at all. 
At intermediate energies ($10-20~\gev$), the flux of primary cosmic particles
exhibits a strong east-west asymmetry, while at high energies the flux
is essentially unaffected by the terrestrial magnetic field.
In summary, the
net effect of the geomagnetic field 
is a non-isotropy of the neutrino flux even in the absence of
oscillations. This has been observed by
Super-Kamiokande,\cite{superk_geomagnetic} which has measured an
east-west asymmetry (see
Fig.\ref{fig:superk_east_west}, from ref.\cite{superk_geomagnetic}) consistent with expectations,
thus confirming that
the details of the geomagnetic effects on the cosmic
radiation are well understood.
In Super-Kamiokande (geomagnetical latitude $25.8^0$~N, close to the
geomagnetic equator)
the cutoff momentum for protons arriving horizontally from the east is 
$\sim 50~\gev$.

A modulation of the low energy component of the 
primary cosmic ray flux is expected as a
function of the solar activity (cycle of $\sim 11$~years), which
induces fluctuations in the geomagnetic field (see for example
ref.\cite{longair_book}): the greater the solar activity, the lower
the primary cosmic ray flux on the Earth. The correlation between the relative
sun-spot number and the cosmic ray flux can be studied using the data
from the Mount Washington neutron monitor.\cite{shea_smart}
The influence of the solar wind 
is only sizeable ($\sim 15-20\%$) for neutrino energies of $1~\gev$ or
below and only at high geomagnetic latitudes, where the geomagnetic
effect is minimal, while it becomes negligible at $\sim 2~\gev$.\cite{webber_lezniak}

\begin{figure}[h]
\vspace*{13pt}
\begin{center}
\epsfig{file=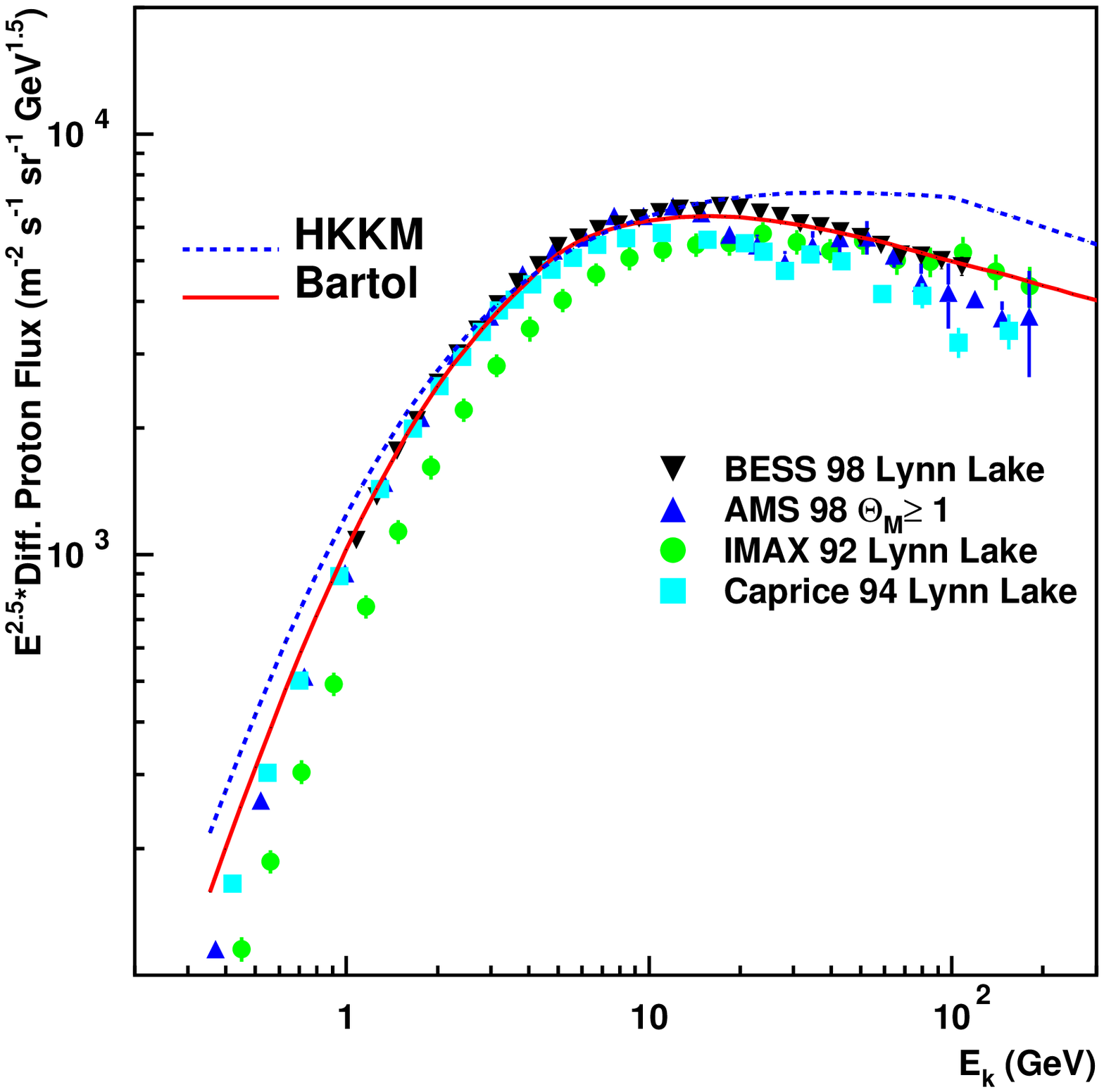,width=6cm} 
\vspace*{13pt}
\fcaption{Comparison between the HKKM\cite{honda} and the
  Bartol\cite{bartol} flux calculations with some of the most recent
  data on primary protons (from ref.\cite{battistoni_otranto}).}\label{fig:primary_flux_battistoni}
\end{center}
\end{figure}

The experimental results for atmospheric neutrinos are usually
presented in terms of the neutrino flux calculations performed by
Honda and collaborators (HKKM)\cite{honda} and by the Bartol
group.\cite{bartol} A comparison between the two models and a
compilation of some of the most recent data on primary protons is
shown in Fig.\ref{fig:primary_flux_battistoni} (from ref.\cite{battistoni_otranto}).

\begin{figure}[h]
\vspace*{13pt}
\begin{center}
\epsfig{file=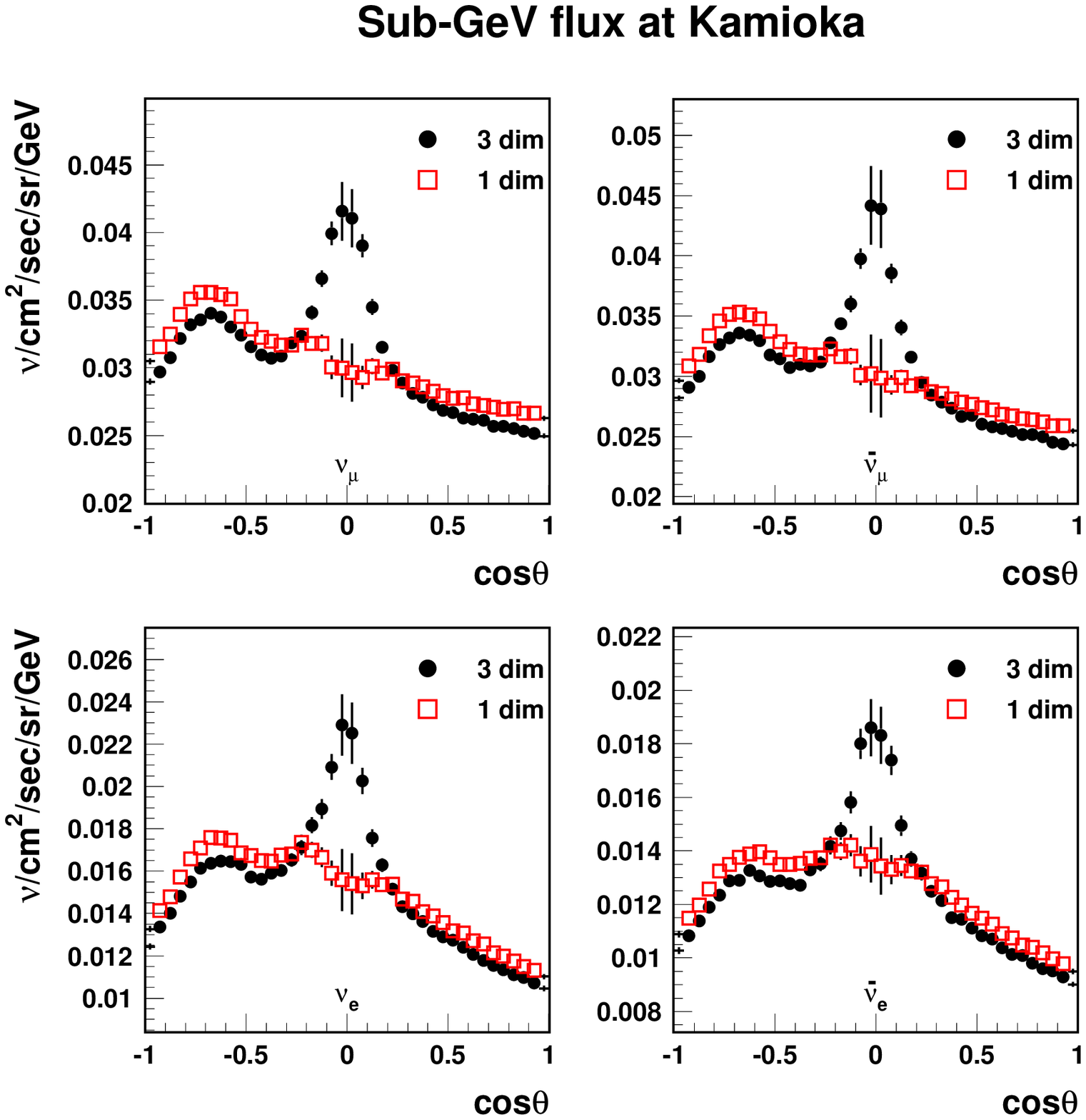,width=8cm} 
\vspace*{13pt}
\fcaption{Comparison between 1-D and 3-D predictions for the neutrino
  fluxes at the Kamioka site, as a function of the zenith angle (from ref.\cite{battistoni_3d}).}\label{fig:fluka3d_kamioka_flux}
\end{center}
\end{figure}

Both HKKM and the Bartol models are based on one-dimensional
computations, in the sense that the momenta of 
all secondary particles, hadrons,
charged leptons as well as neutrinos, are assumed to be collinear with 
that of primary cosmic ray (1-D calculation).
It has only very recently been realized that three-dimensional (3-D)
calculations can be important for a correct description of the
neutrino spatial distribution,
especially in the sub-$\gev$ energy
range.\cite{battistoni_3d,lipari_3d,hkkm_3d} The new calculations, 
originally motivated by the need of having a better description of the 
hadronic processes involved in the shower development,
are based on the 
the FLUKA simulation package.\cite{fluka} The effect of taking the
three-dimensional development of the shower 
at low energies ($\lesssim 1~\gev$) can be clearly
seen in Fig.\ref{fig:fluka3d_kamioka_flux}, where the results from
FLUKA-based 1-D and 3-D simulations are shown for the zenith angle
distribution at the Kamioka site
(figure from ref.\cite{battistoni_3d}).

The Bartol and FLUKA groups
have started a comparison of the two models, using the same input
primary spectrum.\cite{battistoni_otranto} 
The results of this study show that 
the two normalisations disagree by
approximately $20\%$ (Bartol is lower) and the difference between the two models appears 
to be energy-dependent. According to the author of
ref.\cite{battistoni_otranto}, a preliminary comparison with HKKM shows that, 
if they had used the same primary spectrum as an input, also their
predictions on the normalisation would have been lower than
FLUKA's.

In summary, even if the $e/\mu$ ratio is rather well known, the
uncertainties on the absolute neutrino fluxes are still quite
large. Among the dominant sources of systematic errors are the 
knowledge of the primary fluxes, still limited despite new
data has been recently made available, and the
uncertainties on the secondary particle production model. Other
effects, such as the atmosphere modelling and the detector altitude,
play only a marginal role. In order to improve the theoretical
calculations it is very important that further constraints on the
models are obtained from new experimental data.

Several experiments relevant for the atmospheric neutrino flux
calculations should be taking data over the next few years. 
HARP,\cite{harp} an experiment running at the CERN PS, aims to study
hadroproduction in a range of energy relevant for future projects such 
as neutrino factories and for the
determination of the atmospheric neutrino flux. More recently a
proposal for another experiment has been submitted to CERN:\cite{minos_na49} 
if approved, that experiment would make use of the 
NA49\cite{na49} apparatus to study hadroproduction in a range of 
energies ($120~\gev$) relevant for both atmospheric neutrinos and long-baseline
neutrino oscillation projects. 

Among the new
proposals to study cosmic ray muons 
for the determination of the neutrino fluxes, 
an interesting project, known as ADLER,\cite{adler} is now in
its design stage. The main idea is to fly a small detector on board an
aircraft to measure the atmospheric muon flux at different geomagnetic 
latitudes. 
This possibility, together with the relatively longer exposure times,
would be an obvious advantage with respect to balloon-borne
experiments. 

\subsection{Atmospheric Neutrinos in Super-Kamiokande}

\begin{figure}[h]
\vspace*{13pt}
\begin{center}
\psfig{file=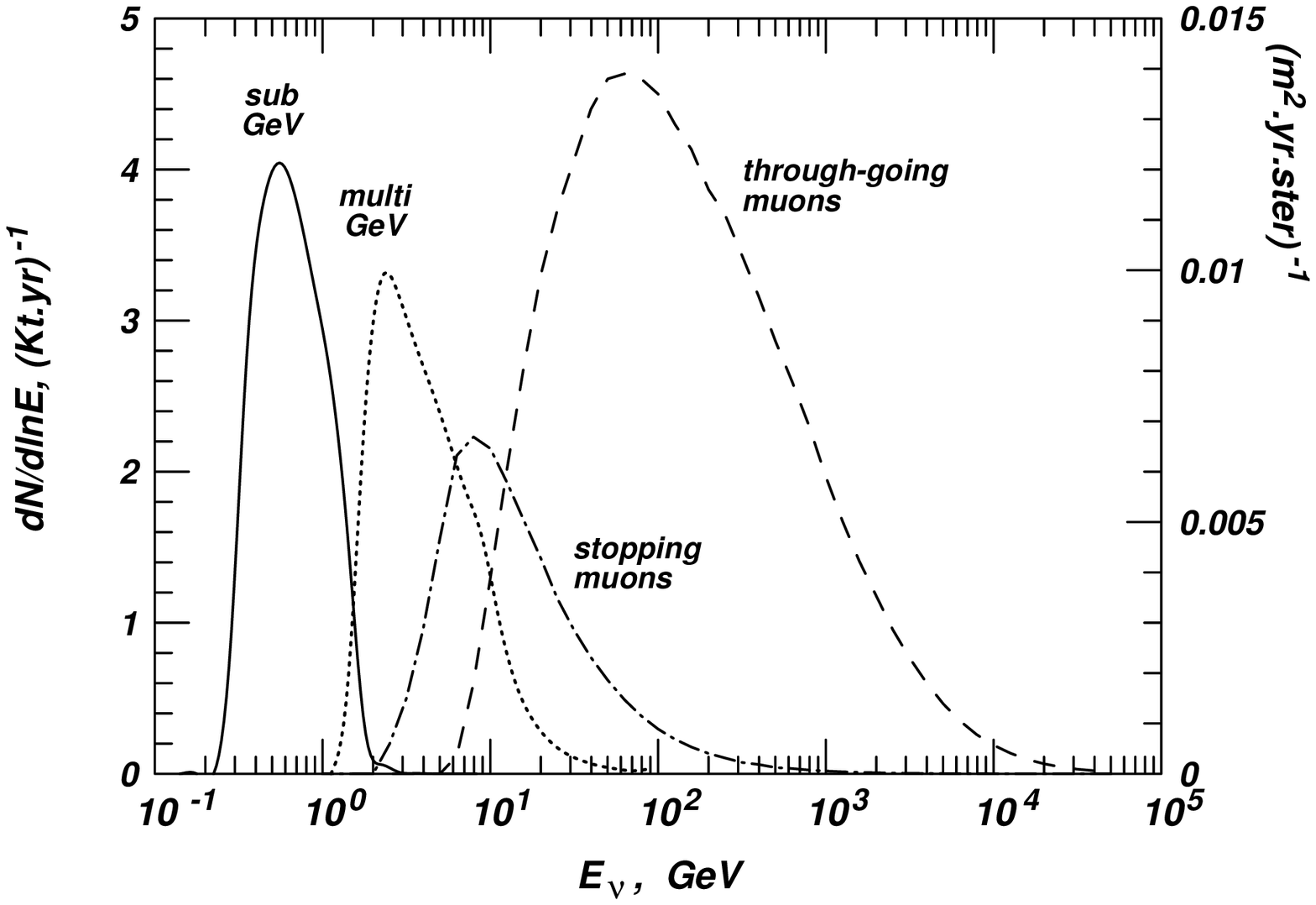,width=6cm}
\vspace*{13pt}
\fcaption{Event rates as a function of the neutrino energy for some of 
  the 
  different categories of neutrino interactions defined in the text (fig. from ref.\cite{engel_gaisser_stanev}).}\label{fig:atmonu_energies}
\end{center}
\end{figure}

We have already given a brief description of 
the Super-Kamiokande experiment and will not repeat it here, however it is 
worth summarising
some information relevant for the
following discussion. 
Atmospheric neutrinos are detected in Super-Kamiokande 
by measuring the Cherenkov rings
generated by the primary particles produced in the neutrino
CC interactions with the water nuclei. 
Thanks to the high PMT coverage, the experiment is characterised by an 
extremely good light yield ($\sim 8$ photo-electron per
$\mev$) and can detect events of energies as low as $\sim
5~\mev$.
The large detector mass and the possibility of clearly
defining a large inner volume allow to collect a high 
statistics sample of {\it fully contained} events (FC) of relatively high
energies (up to $\sim 5~\gev$), the FC events being defined as those
having both the
neutrino vertex and the resulting particle tracks entirely within the
fiducial volume. The contamination from downward-going 
cosmic muons is drastically
reduced by the containment requirement on the primary vertex coordinates.
Fully contained events can be further subdivided
into two subsets, the so-called {\it sub-$\gev$} and {\it multi-$\gev$} events,
with energies below and above $1.33~\gev$ respectively.
In Super-Kamiokande jargon FC events include only single-ring events,
while {\it multi-ring} ones (MRING) are treated as a separate category.
Another sub-sample, defined as the {\it partially contained} events (PC), is
represented by those charged-current interactions where the vertex
is still within the fiducial volume, but at least a primary particle, typically 
the muon, 
exits the detector without releasing all of its energy: 
for those events the energy resolution is therefore worse than
for FC interactions.
Finally, {\it upward-going muons} (UPMU), produced by neutrinos coming from
below and interacting in the rock, can also be used to independently
check the neutrino oscillation result.
In the literature they are sometimes further subdivided into {\it stopping
  muons}
and {\it through-going muons}, according to whether or not they stop in the
detector. The different samples defined above explore different 
ranges of the 
neutrino energy: this is shown in Fig.\ref{fig:atmonu_energies}
(from ref.\cite{engel_gaisser_stanev}), where the event rates for
sub-$\gev$ and multi-$\gev$ FC events, as well as for through-going and
stopping muons is shown. For PC events the neutrino energy is in the
multi-$\gev$ range.

Particle identification in Super-Kamiokande is performed using 
likelihood functions to parametrize the sharpness of
the Cherenkov rings, which are more diffused for electrons than for 
muons. The algorithms, which 
have been tested on cosmic muons, decay
electron samples 
and using test-beam data,\cite{superk_pid} are able to discriminate
the two flavours with very  high purity
(of the order of $98\%$ for single track events).

The most recent value for ${\mathcal{R}}'$ reported by the
Super-Kamiokande collaboration, based on $1289$~days of data, is 
$0.638^{+0.017}_{-0.017}\pm 0.050$ for the sub-$\gev$ sample and 
$0.675^{+0.034}_{-0.032}\pm 0.080$ for the multi-$\gev$ sample
(both FC and PC).\cite{superk_atmo_moriond2001} Thus the measured value of
${\mathcal{R}}'$ is different from unity by about $7~\sigma$.

\begin{figure}[h]
\vspace*{13pt}
\begin{center}
\epsfig{file=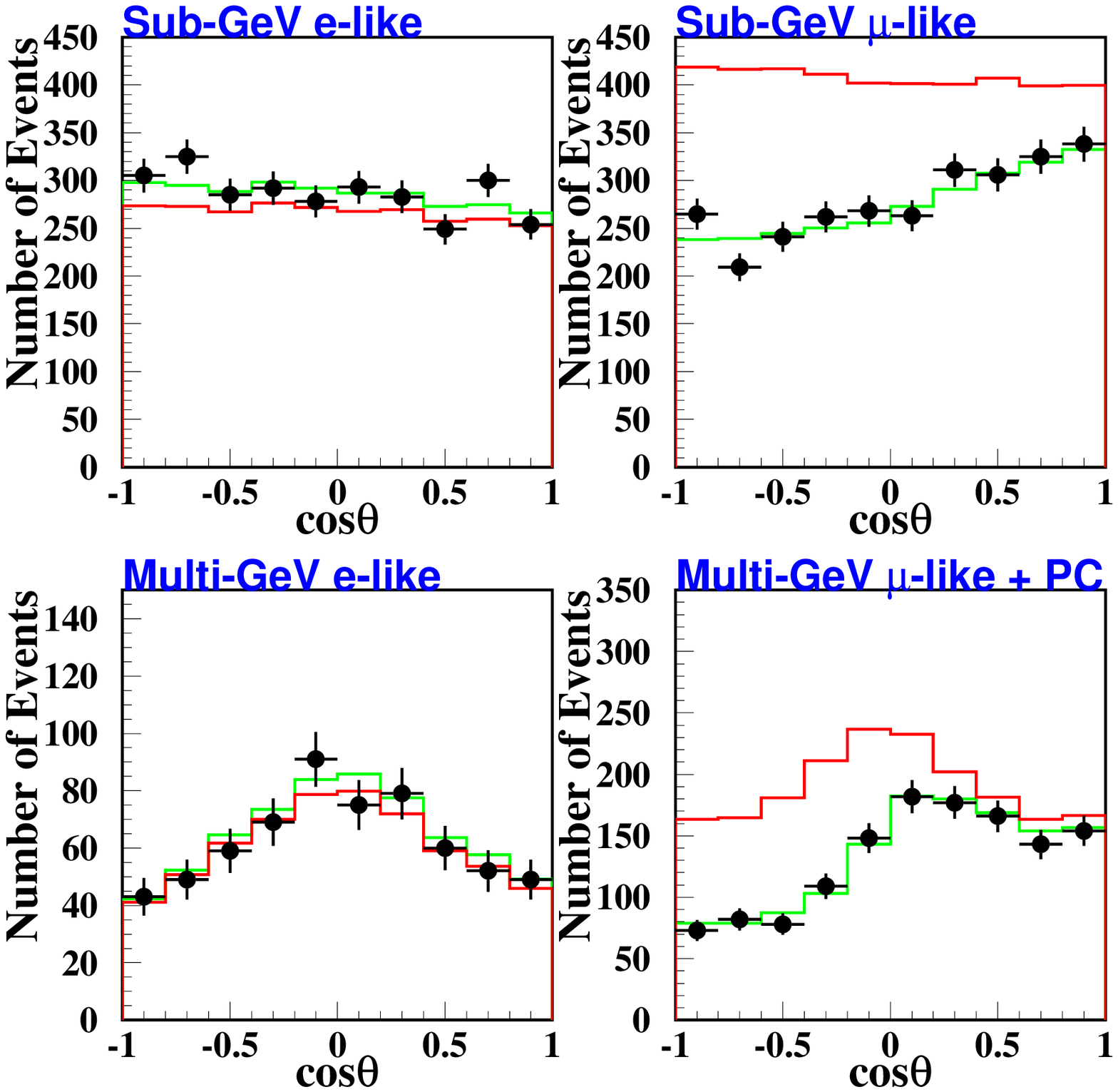,width=7cm}
\vspace*{13pt}
\fcaption{Distribution of the cosine of the zenith angle obtained by
  Super-Kamiokade from $1289$~live days data, for electron-like and
  muon-like contained events (from ref.\cite{superk_nashville}). The solid line is the distribution expected in
  absence of oscillations, while the hatched histogram is for
  $\nu_\mu\rightarrow\nu_\tau$ 
oscillations in the two-flavour mixing scheme, with maximal mixing
  and $\sin^22\theta=2.5\times 10^{-3}~\evolt^2$.}\label{fig:superk_atmo_zenith}
\end{center}
\end{figure}

The distributions of the cosine of the zenith angle for 
the sub-$\gev$ and the multi-$\gev$ samples are shown
in Fig.\ref{fig:superk_atmo_zenith} ($1289$~live days, from ref.\cite{superk_nashville}). In both subsets the electron-like events are in good
agreement with predictions in absence of oscillations,
while the muon spectrum, which strikingly disagrees with no-oscillation
expectations, is well described by an oscillation induced
modulation. It has to be noticed that, 
while the sub-$\gev$ muon sample is depleted over the
entire spectrum, the deficit being larger for
smaller values of $\cos\Theta$, the spectrum of 
neutrinos coming from above ($\cos\Theta > 0$) is very little changed
in the case of multi-$\gev$ muons. 
This is consistent with $\nu_\mu$ oscillations 
occurring with a value of $\Dm2$ which is better matched by low-energy 
neutrinos (sub-$\gev$), 
while for high energy neutrinos the effect of oscillations
becomes sizeable only for larger baselines (neutrinos coming from below).
However it has to be stressed that the angular resolution degrades at  
lower energies, where the lepton momentum direction becomes a worse 
approximation for the incoming neutrino direction.

\begin{figure}[h]
\vspace*{13pt}
\begin{center}
\begin{tabular}{cc}
\epsfig{file=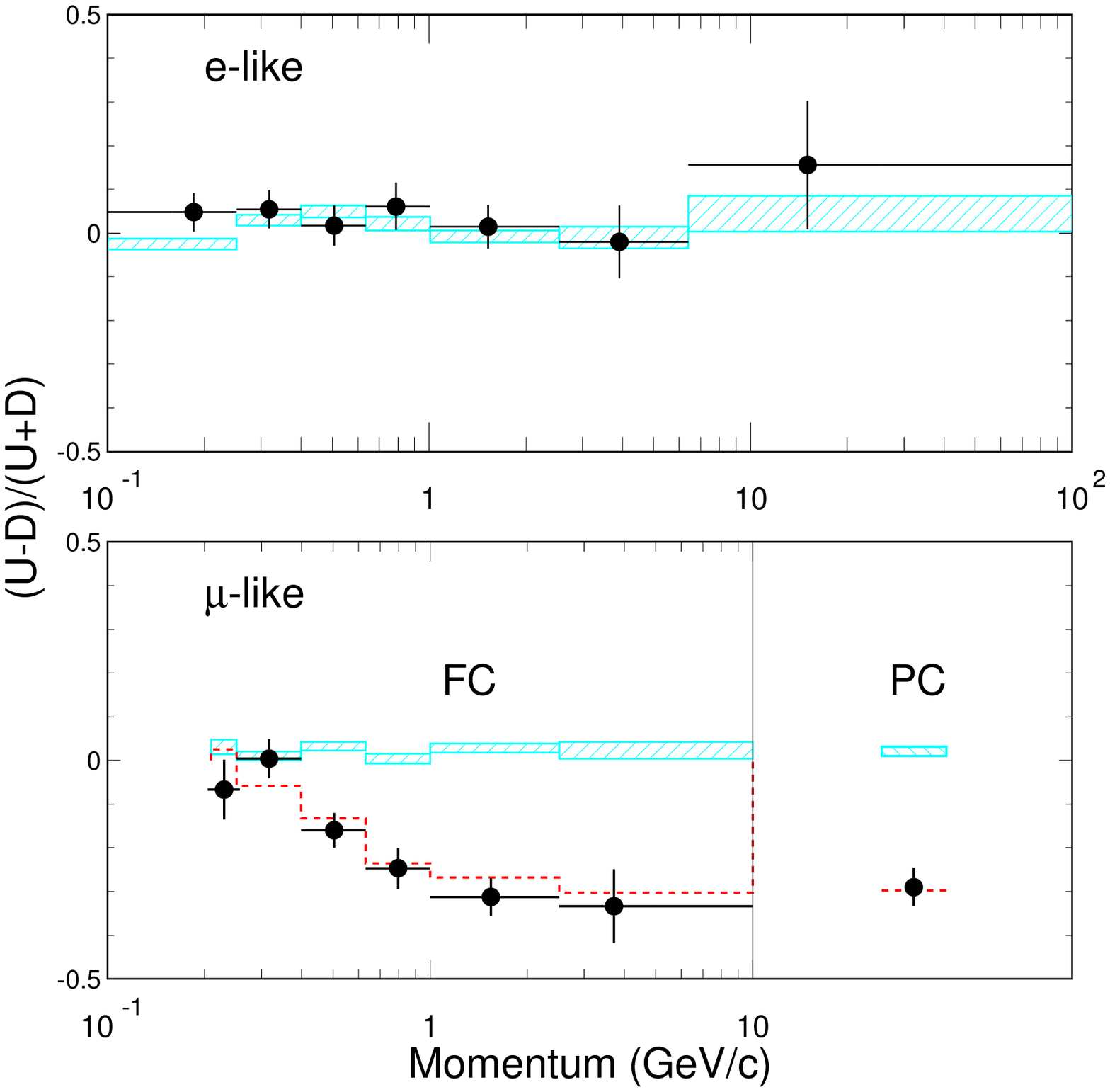,width=5.5cm} & \epsfig{file=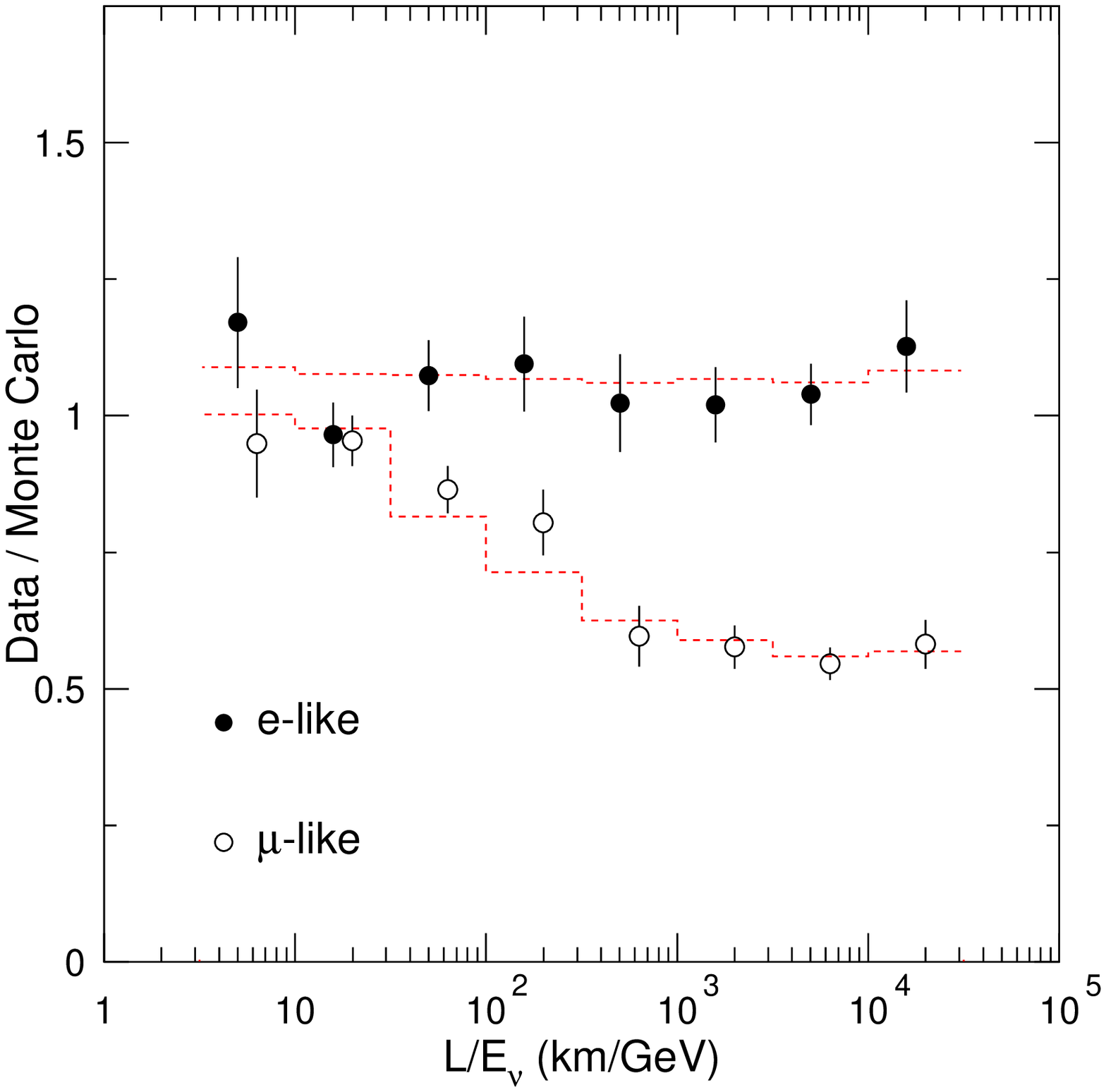,width=5.5cm}\\
\end{tabular}
\vspace*{13pt}
\fcaption{Super-Kamiokande ($1144$~days) up-down asymmetry for
  electron-like and muon-like events, as a function of the particle
  momentum(left) and ratio between observed and expected number of electron-like and
muon-like events as a function of 
$L/E$ (right). The dashed lines correspond to $\nm\rightarrow\nt$
oscillations with maximal mixing and $\Dm2=3.2\times
10^{-3}~\evolt^2$, while the hatched boxes indicate no-oscillation
expectations (figures from ref.\cite{sobel_nu2000}). }\label{fig:superk_updown_lovere}
\end{center}
\end{figure}

A useful dimensionless quantity which can be derived from the zenith
angle distribution is the so-called up-down asymmetry $A$, defined as the
difference between the number of upward-going (U, $\cos\Theta<-0.2$) and
the number of downward-going 
(D, $\cos\Theta>0.2$) neutrinos, divided by 
the sum of the two numbers:
\begin{equation}\label{eq:up_down_atmo}
A = \dfrac{U-D}{U+D}\,.
\end{equation}
\noindent
We have seen that, neglecting the geomagnetic effect corrections, which can
be taken into account in the simulation, there is no reason to have
any asymmetry in the neutrino flux. Any deviation of $A$ from zero 
is a signature for new physics and
can be interpreted in terms of neutrino oscillations.
The dependence of this quantity on the measured particle momentum is
shown in Fig.\ref{fig:superk_updown_lovere} (left) for electron-like and
muon-like events (both FC and PC). The uncertainty assumed on the absolute
normalisation is $25\%$, with $20\%$ contribution from the neutrino
flux calculation and $15\%$ from the neutrino interaction cross
section. Also here the experimental data are
consistent with neutrino oscillations in the muon sample, while 
electron neutrinos appear to be unaffected. In agreement with the
oscillation hypothesis, we see that  for
muon-like events the asymmetry
is essentially zero at low energies, where the oscillation length is small
and therefore both upward-going and
downward-going muons are equally depleted, while for increasing values 
of the muon momentum, when mostly neutrinos from below are affected by the 
oscillation, 
the asymmetry becomes more and more negative.
 
Another way of looking at the same effect is to consider the
ratio between observed and expected number of electron-like and
muon-like events as a function of the ratio
$L/E$: this is shown in Fig.\ref{fig:superk_updown_lovere} (right) for
$1144$~live day of Super-Kamiokande data. Also here electron-like
events show no
evidence for oscillations, while muon-like events show a 
dependence on $L/E$ which, although the oscillatory pattern is washed
out by angular and energy resolution, is consistent
with the oscillations hypothesis.

\begin{figure}[h]
\vspace*{13pt}
\begin{center}
\begin{tabular}{cc}
\epsfig{file=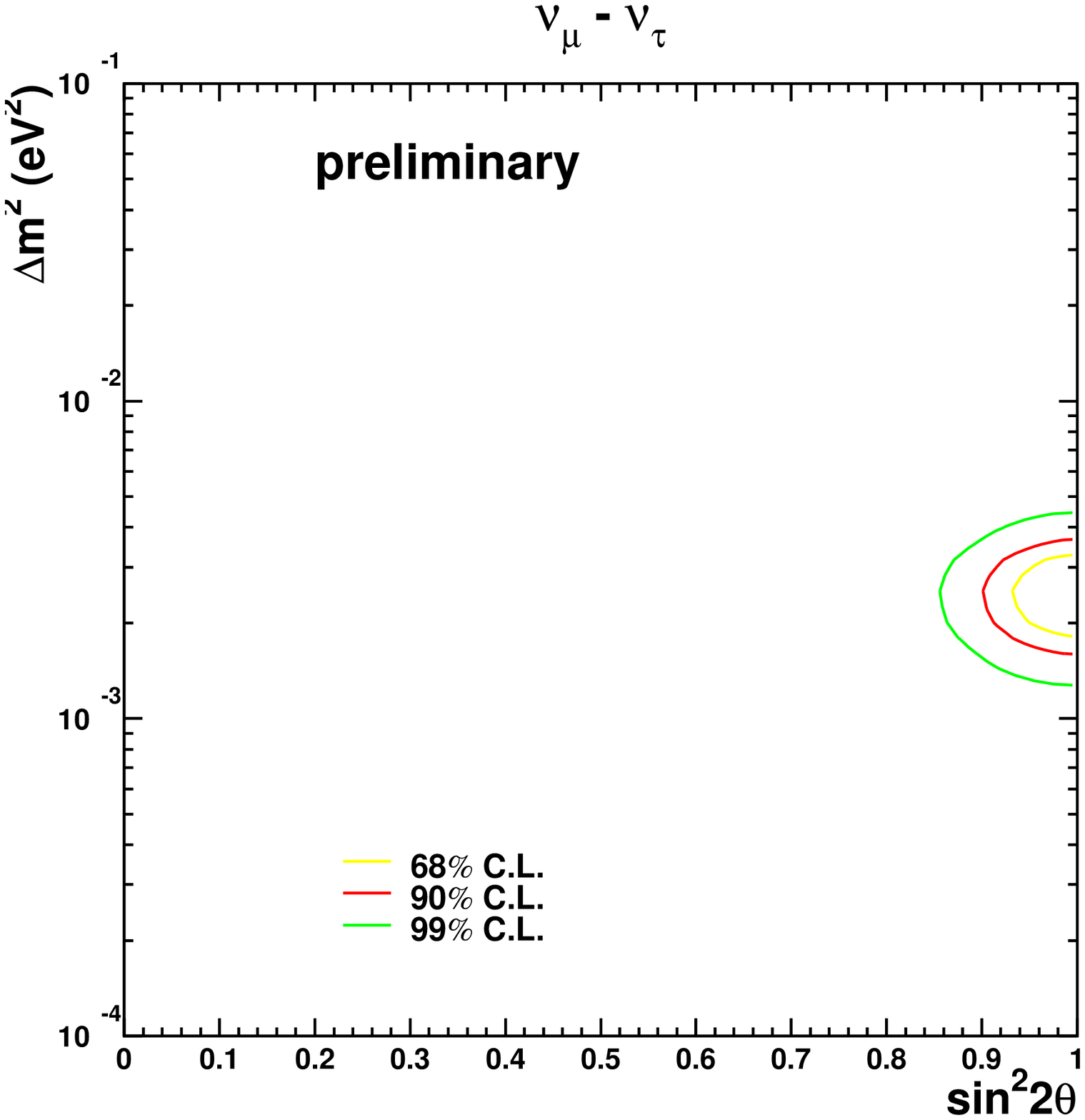,width=5cm} & \epsfig{file= 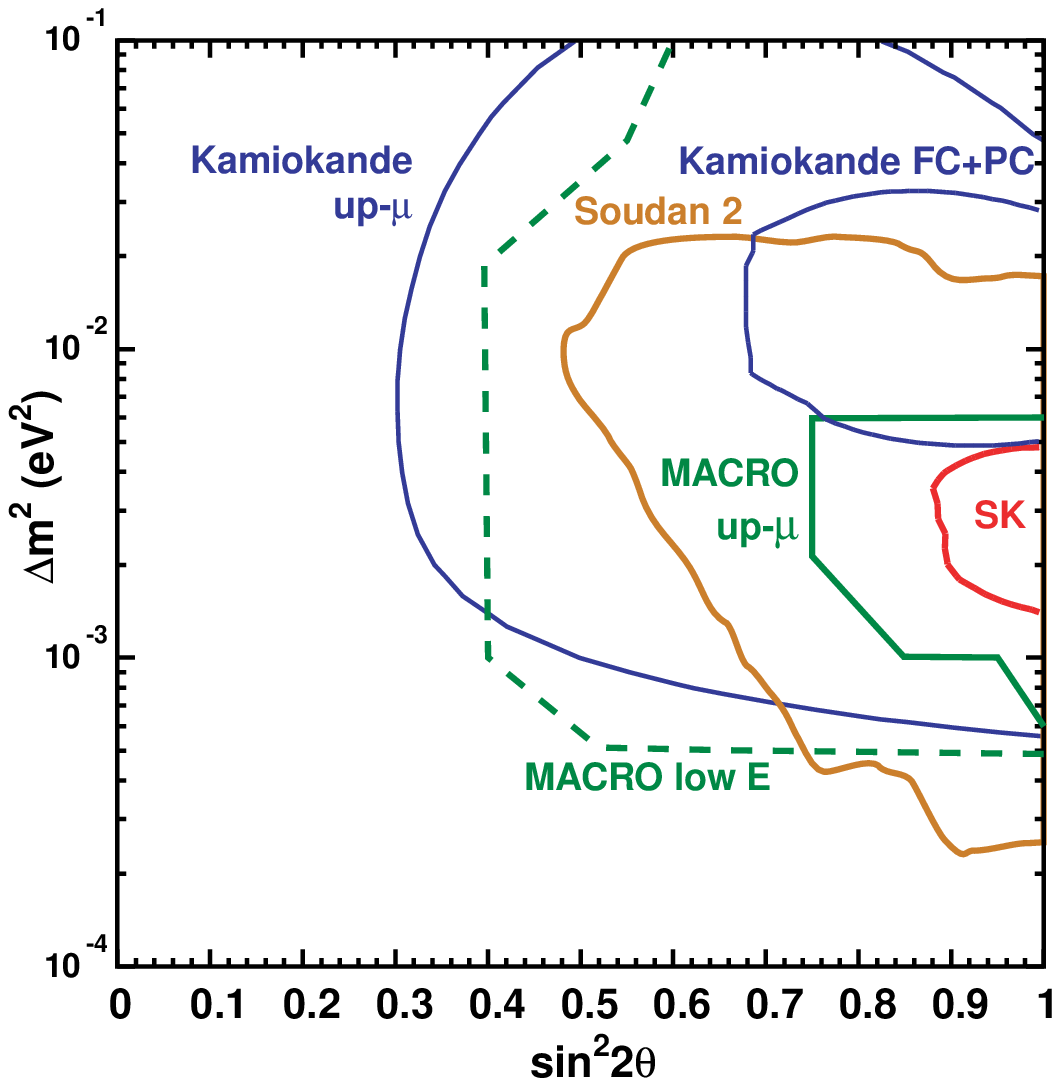,width=4.5cm}\\
\end{tabular}
\vspace*{13pt}
\fcaption{(Left) Allowed region at different C.L. as obtained by
  Super-Kamiokande ($1289$ days) in the two-flavour
  $\nu_\mu\rightarrow\nu_\tau$ hypothesis (from
  ref.\cite{superk_nashville}). (Right) An earlier contour from
  Super-Kamiokande\cite{scholberg} is compared to
  Kamiokande,\cite{kamiokande_atmo}, Soudan2\cite{soudan2_nu2000} and
  MACRO\cite{macro_2000} results (fig. from ref.\cite{scholberg}).}\label{fig:superk_atmo_solutions}
\end{center}
\end{figure}

Interpreting the muon-like event deficit as the result of $\nu_\mu\rightarrow\nu_\tau$ 
oscillations in the two-flavour mixing scheme, Super-Kamiokande extract 
an allowed domain for the oscillation parameters. 
Events are binned in a multi-dimensional space defined by  particle
type, energy and zenith angle, plus a set of parameters to account for 
systematic uncertainties. The result of the fit based on $1289$~days of data, 
using FC, PC, UPMU and MRING events,
is given in Fig.\ref{fig:superk_atmo_solutions}, left.
The best fit 
corresponds to maximal mixing and $\Dm2 =
2.5\times 10^{-3}~\evolt^2$.\cite{superk_nashville} For comparison, an earlier plot\footnote{the arguments outlined
  in the following do not change if the latest Super-Kamiokande results are considered.}\hspace{3pt}
of Super-Kamiokande allowed region ($90\%$~C.L.) is shown together with those obtained by
Kamiokande,\cite{kamiokande_atmo}, Soudan2,\cite{soudan2_nu2000} and MACRO.\cite{macro_2000}
In particular it is interesting to consider how the preferred solution for 
$\Dm2$ indicated by 
Super-Kamiokande is lower than that previously
presented by Kamiokande, with only a very small overlap between the
$90\%$~C.L. contours obtained by the two experiments. The reason for this effect is not fully
understood, although one  
possible explanation could be the different fitting procedure, which
for Kamiokande is based on the flavour ratio ${\mathcal{R}}'$
only, while in Super-Kamiokade a shape analysis of the energy and the
zenith angle spectra is performed.

The Soudan2\cite{soudan2_nu2000} and the MACRO\cite{macro_2000}
experiments, although with less sensitivity,
have confirmed an atmospheric neutrino anomaly consistent with that
observed by
Super-Kamiokande. 

The 
Super-Kamiokande interpretation of the atmospheric neutrino anomaly in 
terms of muon disappearance rather than electron appearance is corroborated by the results 
obtained by CHOOZ\cite{chooz_99} and Palo Verde,\cite{palo_verde} two
neutrino oscillation experiments at nuclear reactors, which have
studied $\bar{\nu}_e$ disappearance in the parameter region relevant
for atmospheric neutrinos. In particular CHOOZ limits (at $90\%$~C.L., 
$\Dm2 < 7\times 10^{-4}$ for maximum mixing and $\sin^22\theta < 0.10$ 
for large $\Dm2$) seem to exclude 
$\nu_\mu\leftrightarrow\nu_e$ mixing in the region of interest, at
least as a dominant mode.\cite{gonzalez-garcia}
However the
ambiguity between oscillations to tau or sterile neutrinos is not
resolved by the analysis of Super-Kamiokande data described previously.
In order to discriminate between the $\nu_\mu\rightarrow\nu_\tau$ and the
$\nu_\mu\rightarrow\nu_s$ hypotheses, it is necessary
to consider other physics effects able
to distinguish one mode from the other. First of all,
while the $\nu_\tau$ has standard neutral current 
interactions with matter, $\nu_s$ does not
couple to the $Z^0$ boson, with a consequent depletion of NC events
and an up-down asymmetry in the NC sample.
Moreover, in case of mixing with a sterile neutrino, matter effects -
see Eqs.(\ref{eq:mixing_angle_matter}), (\ref{eq:matter_prob}) and
(\ref{eq:matter_lambda}) - would induce a sizeable change in 
the oscillation probability for 
neutrino energies such that $E/|\Dm2 |\gtrsim 10^3 \gev / \evolt^2
$.\cite{lipari_lusignoli} Based on those considerations, 
Super-Kamiokande has published an analysis
of $1100$~live days data, showing that the
$\nu_\mu\rightarrow\nu_\tau$ hypothesis fits low energy CC interactions, but
does not fit the NC sample or high energy CC data.\cite{superk_sterile} Although they
cannot exclude more complicated schemes in which the mixing is
between $\nu_\mu$ and admixture of $\nu_\tau$ with a small component
of $\nu_s$, pure $\nu_\mu\rightarrow\nu_s$ is disfavoured by their
data at $99\%$~C.L..
Another way of testing the $\nu_\mu\rightarrow\nu_s$ hypothesis, which 
is again motivated by the $\nu_s$ not having standard electro-weak
interactions, is to
look for possible suppressions of $\pi^0$ production.
This technique is currently limited by the
systematic error on the $\pi^0$ production cross-section, but the situation
should improve in the near future, since the K2K experiment,\cite{k2k} 
which we shall discuss shortly, is expected to reduce the uncertainty
from the present value of $\sim 20\%$ down to $\sim 5\%$.

Super-Kamiokande has recently presented preliminary results on
$\nu_\tau$ appearance.\cite{superk_nashville} The experiment does 
not have the resolution to
unambiguously identify an event as being a $\nu_\tau$ CC interaction, 
however they have developed a likelihood-based analysis to extract a sample of
tau-like events. The observed number of selected events is $419$, for an
expected number of $31~\nu_\tau$'s plus $387$~background events; moreover the 
measured zenith angle distribution is better fitted by 
$\nu_\tau$'s plus background than by background only.

In summary, we have seen that Super-Kamiokande results, which have
been confirmed with less sensitivity by other atmospheric neutrino
experiments, represent a strong hint for neutrino oscillations with
near-to-maximal mixing ($\sin^22\theta > 0.88$) and
$\Dm2$ in the range $2-6\times 10^{-3}~\evolt^2$.
New experiments, like
ICARUS\cite{icarus_proposal} and MONOLITH,\cite{monolith}
will study or plan to study atmospheric neutrinos in the future.
However the
uncertainties on the atmospheric 
neutrino flux predictions are such that a firmer 
experimental proof of neutrino oscillations is required, together with
a better determination of the oscillation parameters if the
effect is confirmed. For this reason it is very important to check the 
atmospheric neutrino anomaly in controlled experimental conditions, by using
artificially produced neutrinos (i.e. accelerator beams).

Typical neutrino energies at accelerators are in the range
$\sim 1-30~\gev$, therefore, in order to match the $\Dm2$ domain
suggested by atmospheric neutrino results, it is necessary to
increase the detector-to-source separation to distances of the order
of hundreds of kilometres. Experiments of this sort are commonly
referred to as {\it long-baseline} experiments and we shall discuss
them in the next section.

\subsection{Long-baseline Accelerator Experiments}

The first accelerator-based long-baseline neutrino oscillation
experiment ever built 
is K2K (KEK-to-Kamioka),\cite{k2k} which makes use of an almost pure
$\nu_\mu$ beam ($98.2\%$) directed from KEK, in Japan, to the
Super-Kamiokande detector, $250$~km away. 
K2K, which has been taking data since June 1999, 
is a disappearance experiment, 
aiming to measure a deficit of muon neutrinos in
the far detector at Kamioka compared to the initial beam
intensity measured in the
near detector at KEK.
The neutrino beam, produced 
in the interactions of $12~\gev$ primary protons from the KEK-PS on an
aluminium target, has an average energy of 
$\sim 1.3~\gev$ (hence the maximum sensitivity to oscillations is for 
$\Dm2 \sim 5\times 10^{-3}~\evolt^2$). 
Pion and muon detectors, placed along the beamline,
are used to measure the secondary particles after the horn
focusing system and to monitor the beam centering and its
intensity.
The design proton intensity is $6\times 10^{12}$~protons per pulse
($1.1~\mu s$ spill/$2.2$~s period). The
results recently presented by the K2K collaboration are based on $2.29\times
10^{19}$~pot (protons-on-target), collected between June 1999 and June 
2000.
The alignment between the neutrino target and the far
detector is performed
using a GPS survey, giving a precision 
better than $0.01$~mrad, while the precision of the civil
construction is
better than $0.1$~mrad. GPS timing information is also 
used to correlate events in Super-Kamiokande with the KEK-PS spill.

In order to have a reliable measurement of the beam intensity at
production and to reduce the systematic uncertainties related to the
experimental technique adopted, the near detector has been built using 
the same principles and technology as for Super-Kamiokande.
Located $300$~m from the
production point, it is a $1$~kton water-Cherenkov detector 
with photo-multiplier
readout, the PMTs being arranged so to provide the same coverage as
for Super-Kamiokande ($40\%$). The water-Cherenkov detector 
is complemented by a scintillating
fibre tracker (with a $6$~ton water target), a lead-glass
electromagnetic calorimeter and a muon range detector. The combined
information from the different sub-detectors allows measurement of the
$\nu_\mu$ beam profile, its energy distribution and intensity, 
as well as the
determination the $\nu_e$ contamination ($\sim 1.3\%$) and the study of
neutrino interactions in the $1~\gev$ region. 

The beam has been
measured to be centred within systematic uncertainties ($0.7$~mrad)  
and the centre has been shown to be stable within $\pm 1$~mrad. The
muon energy and angular distributions are also monitored on a daily
basis and so is the 
beam intensity: they all 
appear to be stable over the period of data taking considered.
Given this stability, it is possible to predict the number of expected events at the far site in 
absence of oscillations by extrapolating the normalisation
measured at the near site to the far detector
location. The beam simulation used for the extrapolation is validated
by a comparison between the predicted secondary pion momentum and angular
distributions and those measured at the pion
monitor placed along the beamline.
The selection criteria 
used to identify muon-like and electron-like interactions at the 
near detector are essentially the same as those used in
Super-Kamiokande 
for the FC atmospheric neutrino analysis, with the
addition of a cut on the total number of photoelectrons to discard
spills with multiple events.

\begin{table}[h]
\tcaption{Number of expected and observed events in the far detector
  of the K2K experiment.}
\centerline{\footnotesize\smalllineskip
\begin{tabular}{c c c}\label{tab:k2k}\\
\hline
Sample & Expected Events (No osc.) & Observed Events\\
\hline
1-ring $\mu$-like & $20.9$ & $14$ \\
1-ring $e$-like   & $2.0$  &  $1$ \\
MRING             & $14.9$ & $13$ \\
\hline
Total             & $37.8\pm 0.2 (stat.)^{+3.5}_{-3.8}(sys.)$ & 28\\
\hline\\
\end{tabular}}
\end{table}

\begin{figure}[h]
\vspace*{-2cm}
\begin{center}
\epsfig{file=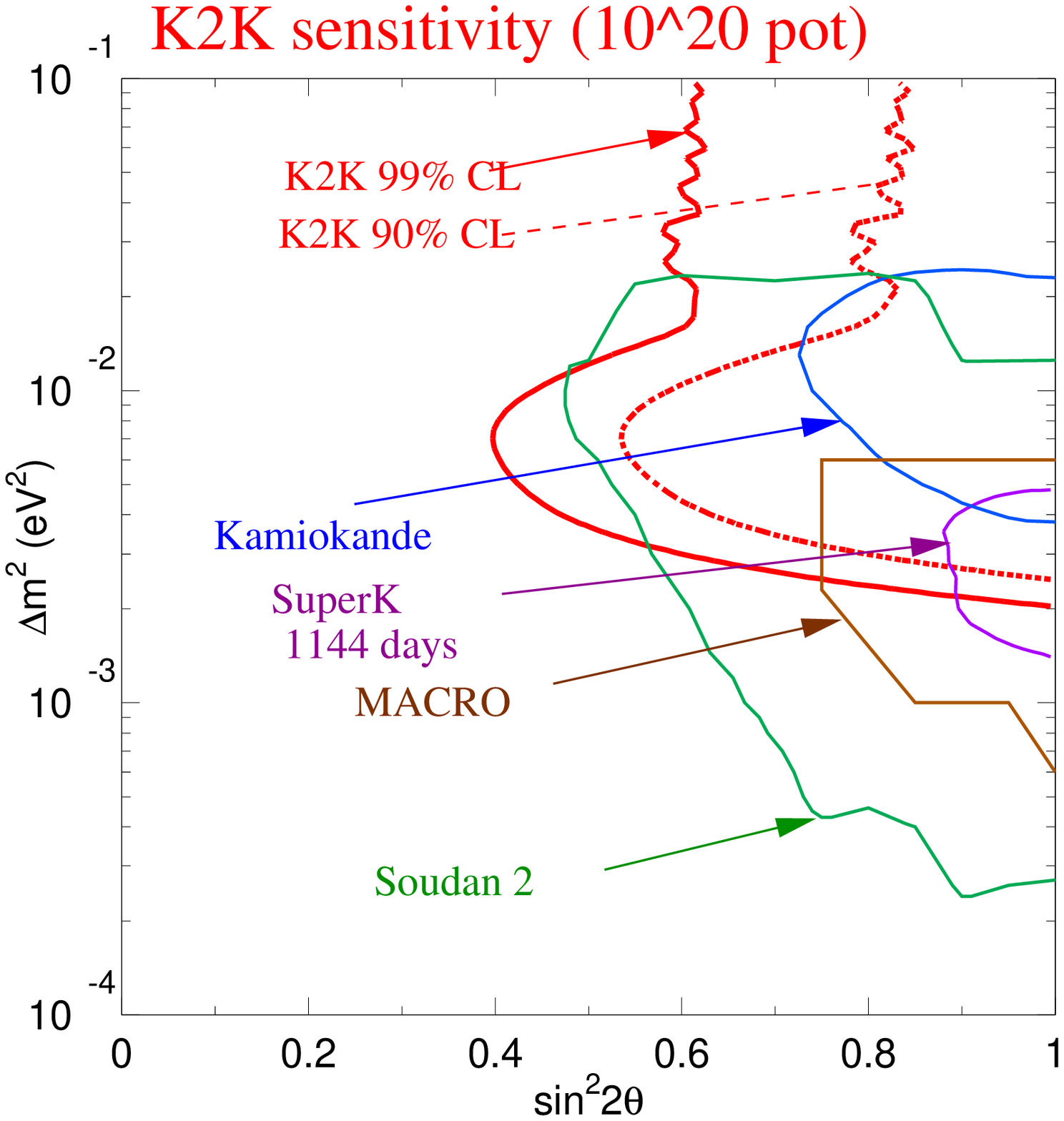,width=6cm}
\vspace*{-0.7cm}
\fcaption{K2K expected sensitivity for $10^{20}$~pot,
compared with the allowed regions in the oscillation 
parameter space obtained by
Super-Kamiokande, Soudan2 and MACRO (from ref.\cite{k2k_nashville}).}\label{fig:k2k_sens}
\end{center}
\end{figure}

The systematic uncertainty on the near
detector flux measurement is $5\%$, with a statistical error of
$<1\%$. For the far detector the systematic error
is $3\%$, while the uncertainty related to the near-to-far
extrapolation is estimated to be $^{+6}_{-7}\%$. Other uncertainties,
such as those on the beam spectrum and the neutrino interaction cross
sections, are negligible to first order, due to cancellations in
the near-to-far comparison. 
The expected number of events in absence
of oscillations is $37.8\pm 0.2 (stat.)^{+3.5}_{-3.8}$, while only
$28$ have been observed in real data. 
In Tab.\ref{tab:k2k} the predicted and observed number of events are
compared for the different categories (one-ring muon-like, one-ring
electron like and multi-ring): while essentially 
no conclusion can be drawn about
electron-like events, due to the small statistics, 
the muon neutrino sample appears to be significantly depleted, a fact
which might be interpreted as evidence for neutrino oscillations.
However, the K2K collaboration 
has not yet given any interpretation of their result 
and is waiting to have more statistics ($10^{20}$~pot by year 2005) and a better
understanding of systematic uncertainties before drawing any strong
conclusion from the observed anomaly. The expected final sensitivity 
for K2K is shown in Fig.\ref{fig:k2k_sens}, together with the allowed
solutions obtained by Kamiokande, Super-Kamiokande, MACRO and
Soudan2. From that plot it becomes clear that, unless the actual
$\Dm2$ is at the high end of Super-Kamiokande allowed region, K2K will 
not provide a definitive proof that the atmospheric neutrino anomaly
is due to neutrino oscillations.

Two other long-baseline projects are expected to
start operation over the next few years: one in the USA, the MINOS\cite{minos_proposal} 
experiment at Soudan, using the NuMI\cite{numi_tdr} beam from Fermilab, and the
other in Europe, using the CERN-to-Gran Sasso neutrino
beam\cite{ncgs_tdr} (NCGS) in conjunction 
with the two proposed experiments OPERA\cite{opera_proposal} and ICARUS/ICANOE.\cite{icarus_proposal,icanoe_proposal}

The MINOS experiment aims to investigate the atmospheric neutrino anomaly by
measuring the neutrino beam at
two different locations, one close to production (near detector, 
at Fermilab, $\sim 320$~m from the production point), where no effect is
expected from neutrino mixing, and the other far
away (far detector, in the Soudan mine, Ely, Minnesota, about $730$~km 
from Fermilab), where neutrino oscillations would modify the
beam flavour composition and induce distortions in the measured energy
spectra.

\begin{figure}[h]
\vspace{5cm}
\begin{center}
\includegraphics{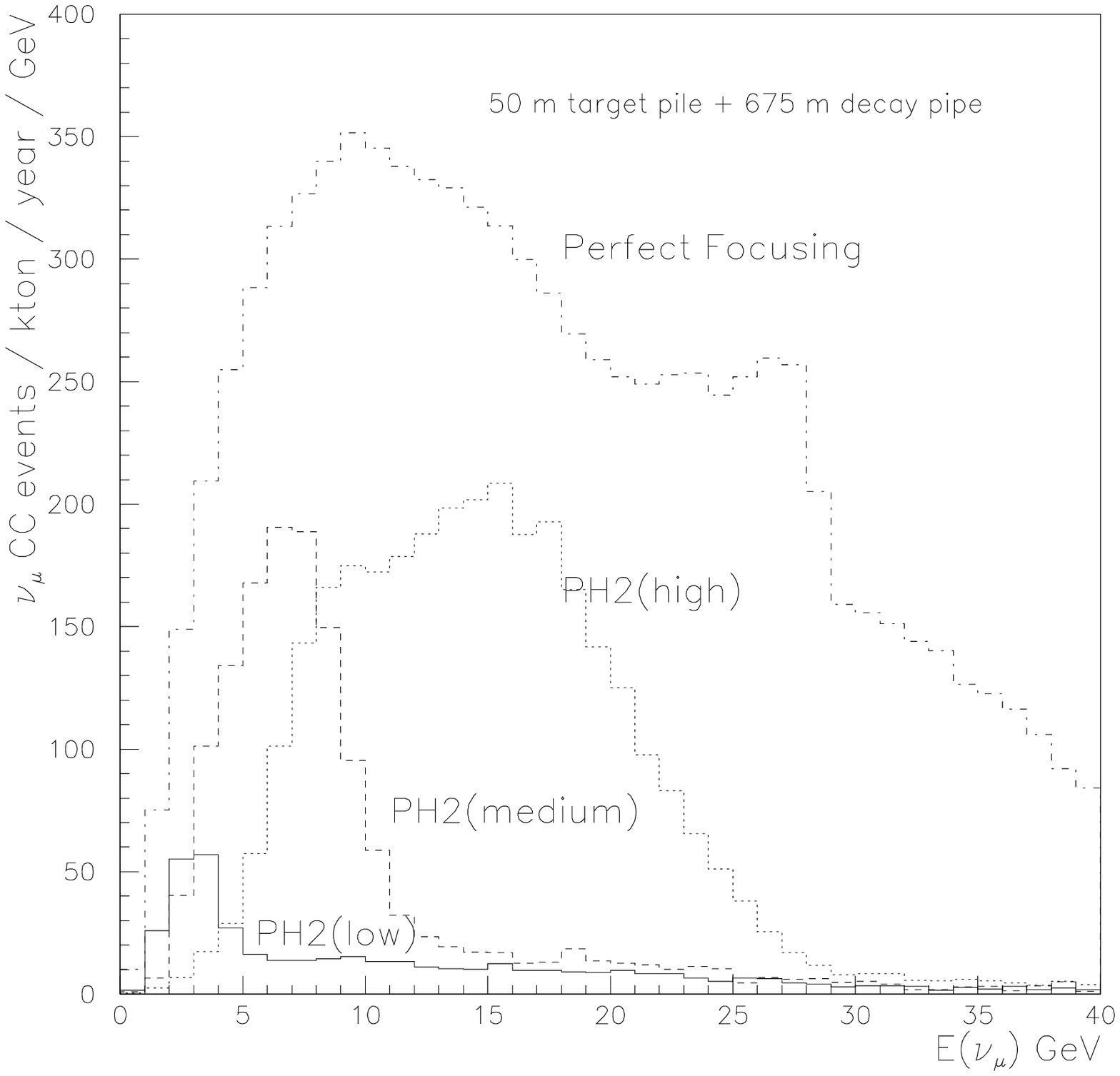}
\vspace*{13pt}
\fcaption{Predicted event rates at the MINOS far site for the three
  possible beam configurations, together with the spectrum expected
  for ``perfect focusing'', where all produced neutrinos reach the detector.}\label{fig:numi_spectra}
\end{center}
\end{figure}
The NuMI 
neutrino beam is obtained from the decay of secondary particles produced 
in the interactions of $120~\gev$ protons, delivered by 
the new Fermilab Main Injector, with a carbon target. The expected
proton flux  
is $4\times 10^{13}$ protons per pulse, with
repetition rate $1.9$~s and spill duration time $10~\mu$s. 
The secondary
mesons are focused by two parabolic horns and allowed to decay in a
$675$~m long decay tunnel. 
The relative distance between the two magnetic elements can 
be changed by moving the second horn with respect to the
other: by adjusting the horn positions and currents it is then possible to
select different regions of the phase space, thus providing the
flexibility to optimize the beam energy during the course of the experiment.
Three beam settings are possible 
(see Fig.\ref{fig:numi_spectra})
, corresponding to three different energy regimes: 
{\it low} ($<E_\nu>\sim 3~\gev$), {\it medium} ($<E_\nu>\sim 7~\gev$) and
{\it high} ($<E_\nu>\sim 15~\gev$) energy. In order to have optimal sensitivity 
in the low part of the atmospheric neutrinos allowed region, which
seems to be preferred by Super-Kamiokande data, MINOS
has decided to start taking data with the low-energy beam configuration.
Another focusing system is under consideration and might be included
in the final design of the NuMI beam, the so-called {\it hadron
  hose}. The basic idea is to have a 
wire carrying a $1$~kA current down the centre of the decay tunnel, to 
provide additional focusing for the charged secondary mesons, 
with the double advantage of
increasing the beam intensity and reducing the systematic
uncertainties on the beam extrapolation from the near to the far site.

In order to minimize the impact of systematic uncertainties, MINOS far
and near detectors ($5.4$~kton and $980$~ton mass respectively) 
have been designed to be as similar as possible.
They are both magnetised iron-scintillator sampling
calorimeters with photo-multiplier readout.  
The average toroidal magnetic field in the iron is
about $1.3$~T. The longitudinal and transverse granularity is the same 
for the two detectors: $2.54$~cm thick octagonal iron plates are
alternated with
scintillator planes, made of plastic slabs with $1\times 4~\text{cm}^2$ 
transverse section and lengths varying between $4$~m and $8$~m for the
far detector planes. The slabs orientation on successive planes
differs by $90^0$,
so to provide two coordinate measurements on the transverse plane.
A green wave-length shifting fibre, $1.2$~mm diameter, 
is glued into a grove at the centre
of the wide side of each slab: the light produced
in the scintillator is trapped into the fibre, shifted in frequency
and then 
routed to multi-pixel PMTs\footnote{16-pixel PMTs are
  used at the far detector, where the signal is optically multiplexed, 
  while at the near detector 64-pixel PMTs have been chosen and 
  multiplexing is applied only in the muon spectrometer section. }\hspace{3pt} through a clear fibre, optically
connected to the wave-length shifting one. The
readout is two-sided at the far detector and one-sided at the near
detector. While the far detector planes are all fully equipped with
scintillator strips, the near detector will be only partially
instrumented. In the forward part (the first $120$ plates, out of a total of
$180$), which logically defines the target and calorimeter sections,
only every fifth plane is fully instrumented, while $4/5$ of the plates are
instrumented in a limited region around the beam fiducial area.
In the remaining section, the muon spectrometer, every fifth plane is
fully instrumented.

A calibration detector, much smaller in size but with the same
longitudinal segmentation as the MINOS far and near detectors, is being 
tested at the CERN PS,\cite{minos_caldet} 
to characterize its response to muons, hadrons and
electrons and to study the performances of the different hardware
components used by MINOS. 

The main physics goals of MINOS, which is expected to
start taking data in 2004, are to
test the neutrino oscillations hypothesis, to verify that the dominant 
mode is $\nu_\mu\rightarrow\nu_\tau$, as suggested by Super-Kamiokande 
data, and to measure the oscillation parameters with a precision of
about $10\%$. Moreover the $\nu_\mu\rightarrow\nu_s$ and
$\nu_\mu\rightarrow\nu_e$ channels will also be investigated.

The criterion envisaged by MINOS to verify the oscillations hypothesis
is the so-called {\it T-test}, based on the parameter:\cite{minos_ttest}
\begin{equation}
T= 
\dfrac{{\mathcal{N}}_{1\mu}}{{\mathcal{N}}_{1\mu}+{\mathcal{N}}_{0\mu}}
\end{equation}
\noindent
where ${\mathcal{N}}_{1\mu}$ and ${\mathcal{N}}_{0\mu}$ are the number 
of muon-like
and muon-less events respectively.
If $\nu_\mu$'s
oscillate to another active neutrino species, say $\nu_\tau$'s,
the muon-like sample will be depleted ({\it disappearance} of
$\nu_\mu$) and the measured
${\mathcal{N}}_{1\mu}$ in the far detector will be less than that
expected in the absence of oscillations; on the other hand, 
$\nu_\tau$CC
events will look very much like neutral current interactions and,
for neutrino energies above the $\tau$ production threshold, will
populate the muon-less sample, so that the total number of events will remain
unchanged. Therefore, $\nu_\mu\rightarrow\nu_\tau$ oscillations would
manifest themselves as a low $T_{FAR}/T_{NEAR}$ ratio with respect to expectations
in absence of oscillations. If the $\nu_\mu$'s oscillated to sterile
neutrinos, the muon-like and the muon-less samples would be depleted
in the same manner, since sterile neutrinos do not couple with any of the
weak interactions mediating bosons. Therefore for pure 
$\nu_\mu\rightarrow\nu_s$ oscillations the ratio $T_{FAR}/T_{NEAR}$
would be the same as for the no-oscillations case. However, the NC
sample would also be affected by oscillations to sterile neutrinos,
which would induce a 
reduction in the number of neutral current interactions and a 
distortion of the NC energy spectrum: 
if none of those effects is  measurable, 
a limit can be set on the $\nu_\mu\rightarrow\nu_s$ 
oscillation amplitude.

\begin{figure}[h]
\begin{center}
\epsfig{file=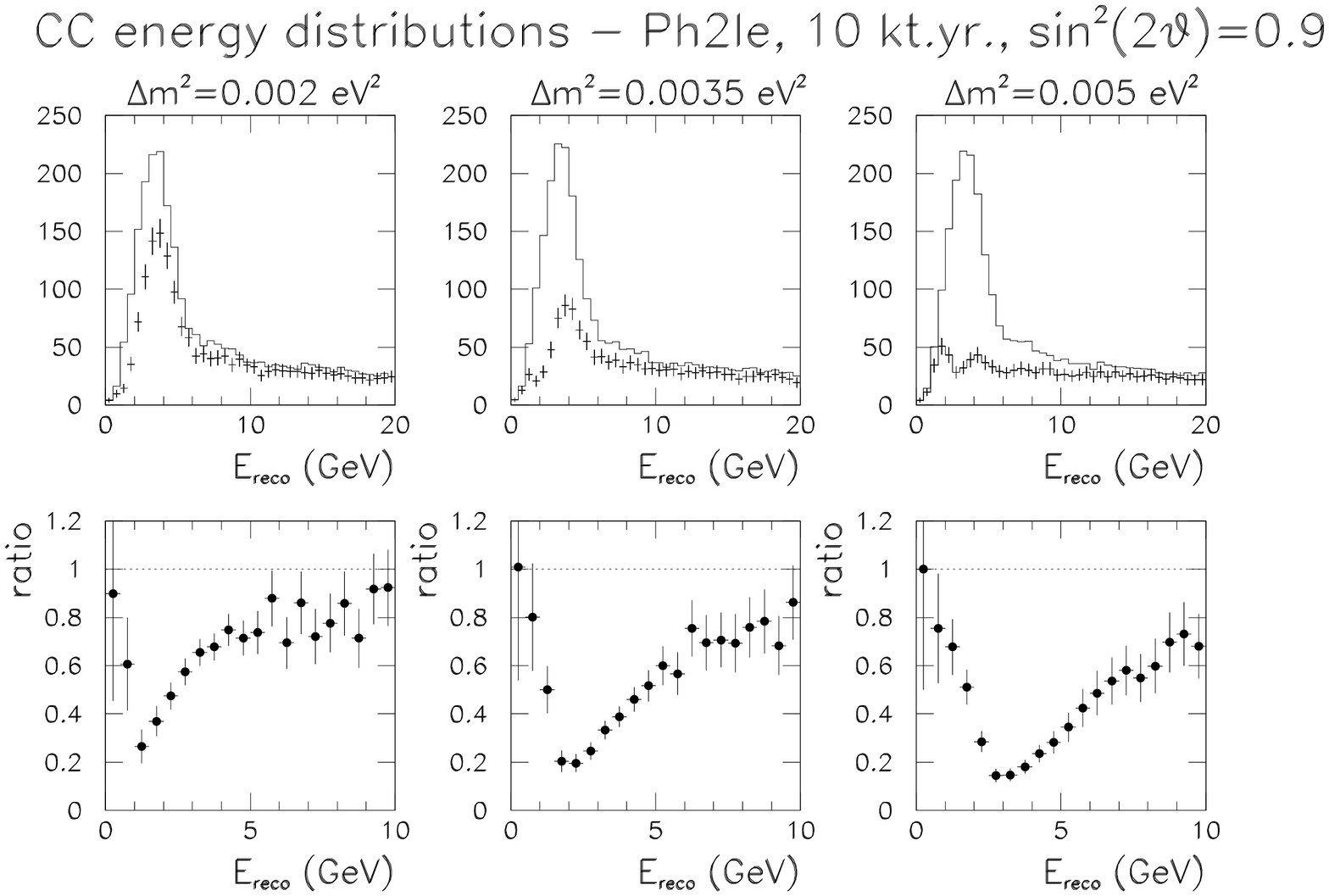,width=9.0cm}
\vspace*{13pt}
\fcaption{(Top) Energy distributions at the MINOS far detector, in absence of
  oscillations (solid line) and as would be measured for different
  values of $\Dm2$ and $\sin^22\theta = 0.9$. (Bottom) The ratio between the oscillated
  and the unoscillated spectra shows the characteristic dip shape, at an
  energy determined by the value of $\Dm2$ and whose depth depends on
  the oscillation amplitude.}\label{fig:minos_cc_spectra_le}
\end{center}
\end{figure}

If the oscillation hypothesis is confirmed by the T-test,
assuming the dominant oscillation mode is
$\nu_\mu\rightarrow\nu_\tau$, the oscillation parameters can be
extracted by comparing the $\nu_\mu$CC energy spectrum measured in the far
detector to expectations in absence oscillations as obtained from the
measured spectrum in the near detector. When considering the ratio
between the two distributions (see Fig.\ref{fig:minos_cc_spectra_le}\cite{minos_david_petyt}), 
a very characteristic dip appears, at an energy corresponding to 
the oscillation $\Dm2$ and whose depth is determined by the
oscillation amplitude $\sin^22\theta$.

\begin{figure}[h]
\vspace*{13pt}
\begin{center}
\begin{tabular}{cc}
\epsfig{file=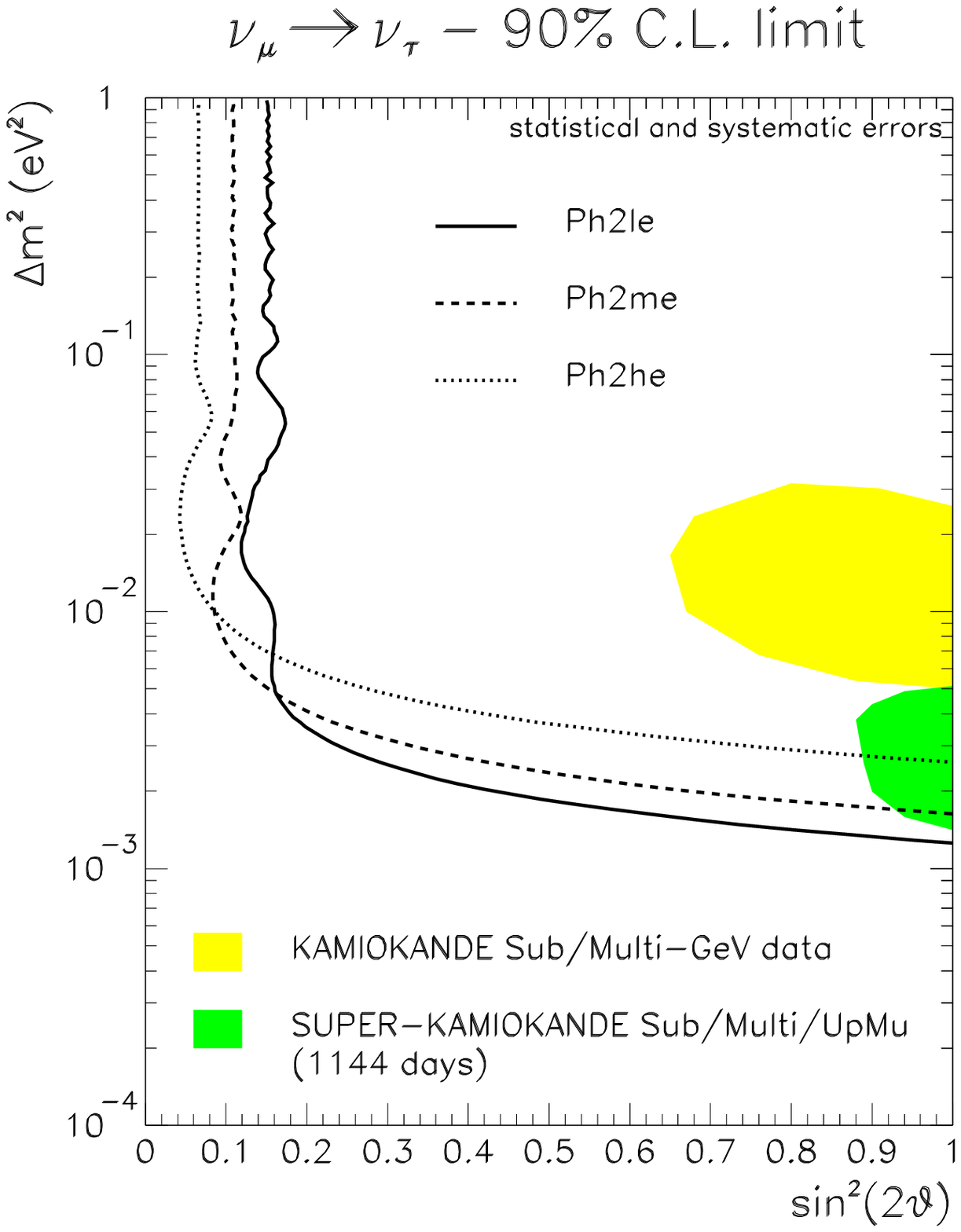,width=5.5cm}&
\epsfig{file=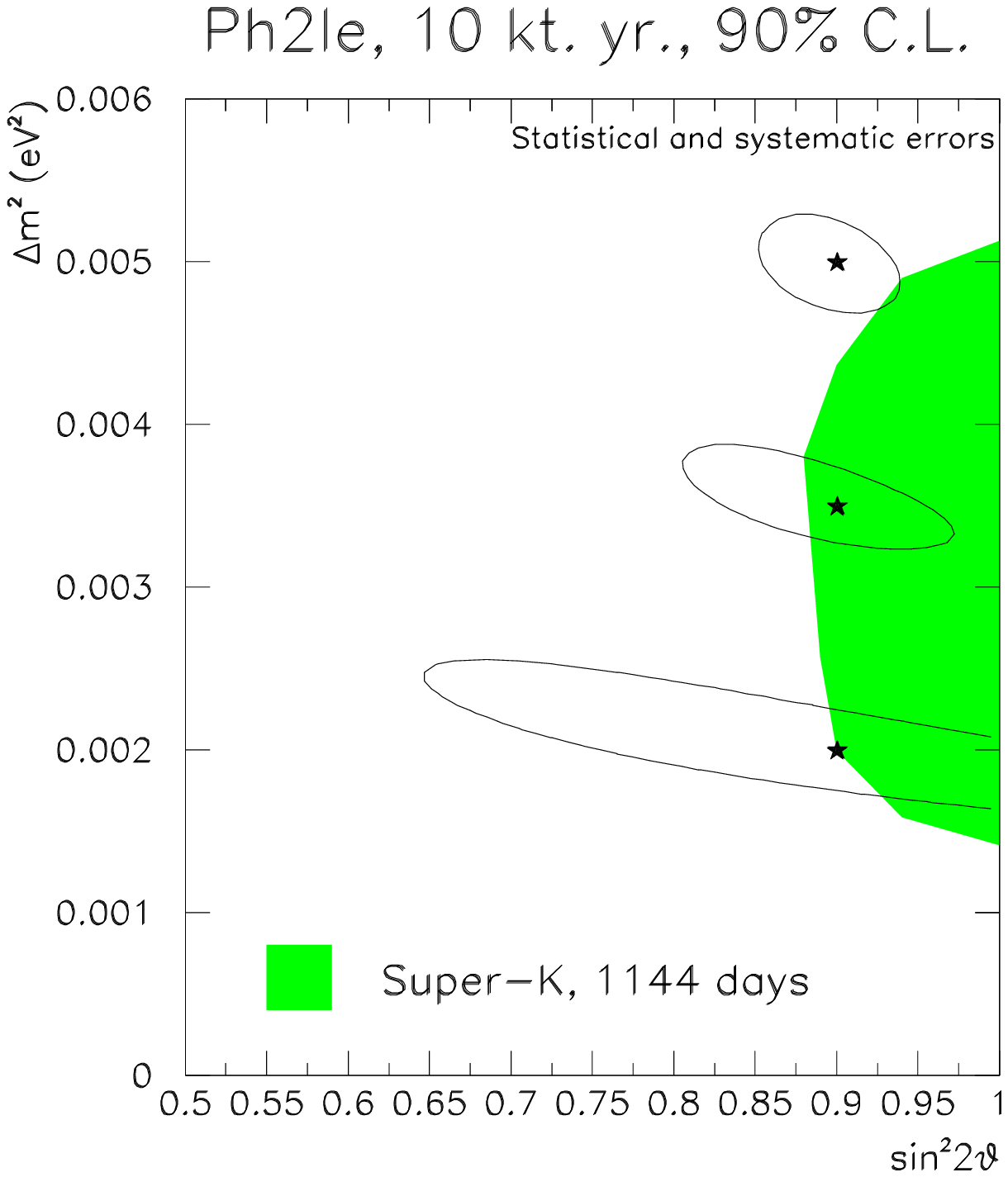,width=6cm}
\end{tabular}
\vspace*{13pt}
\fcaption{(Left) MINOS sensitivity at $90\%$~C.L. for
  $\nu_\mu\rightarrow\nu_\tau$ as obtained from the T-test; results are
  shown for $10\text{kton}\times\text{year}$ and all three possible beam configurations.(Right)
  $90\%$~C.L. 
  allowed regions from fits to reconstructed CC energy distribution, for
  $\sin^22\theta = 0.9$ and different values of $\Dm2$; results are
  shown for the low-energy beam only.}\label{fig:minos_nmnt}
\end{center}
\end{figure}

If no oscillation signal is found, the $90\%$~C.L. limits which MINOS
would extract from the T-test analysis, based on
$10~\text{kton}\times\text{year}$ of data, are shown in
Fig.\ref{fig:minos_nmnt} (left),\cite{minos_david_petyt} together with the allowed regions obtained by Kamiokande and
Super-Kamiokande. Using the low-energy beam, then, MINOS will 
cover the entire parameter space suggested by the 
atmospheric neutrino results. On the other hand, if an oscillation
signal is observed, a measurement of the oscillation parameters can be 
performed.
MINOS  
sensitivity to the oscillation parameters, as obtained
from the far-to-near comparison of the $\nu_\mu$CC energy spectrum
for $10~\text{kton}\times\text{year}$, is
shown in  Fig.\ref{fig:minos_nmnt} (right),\cite{minos_david_petyt} for $\sin^22\theta = 0.9$
and $\Dm2$ varying between $0.2\times 10^{-3}~\evolt^2$
and $0.5\times 10^{-3}~\evolt^2$: as expected, the higher the value of
$\Dm2$, the better the precision to which the parameters can be determined.

MINOS also plan to study atmospheric neutrinos and preliminary
results, based on a $18~\text{kton}\times\text{year}$ sample of simulated events, show that 
the experiment has the sensitivity to check the atmospheric neutrino
anomaly in the range of parameters suggested by Super-Kamiokande, 
with completely different systematics.\cite{minos_atmo}

Intense efforts have been recently made within the MINOS collaboration 
to optimize the experiment capabilities for detection of a
$\nu_\mu\rightarrow\nu_e$ signal. The main difference with respect to
the $\nu_\mu\rightarrow\nu_\tau$ search is that in case of
oscillations to $\nu_e$ an 
appearance analysis has to be performed, looking for an excess of
events in the far detector, 
above the expected background from prompt $\nu_e$CC
interactions and
NC events with a topology similar to that of $\nu_e$CC. 
The prompt $\nu_e$ contamination in the $\nu_\mu$ beam is  
quite small, of the order of $0.6-2\%$, depending on beam configuration.
The systematic uncertainties on background predictions in the far
detector should be drastically
reduced by the measurements of the same background events in the near
detector. 
Moreover results from hadron-production experiment should help to
reduce the systematic error on beam predictions.
Preliminary results\cite{minos_nmne} show that, if the total error 
can be kept within $10\%$, MINOS should be able
to improve the CHOOZ\cite{chooz_99} limit on
$|U_{e3}|^2$ by about a factor two.\footnote{in the three-generation model, for 
$|\Delta m_{23}^2|\gg|\Delta m_{12}^2|$, $m_3>m_2$ and 
$|U_{\mu 3}|^2 =|U_{\tau 3}|^2$.}

The CERN-to-Gran Sasso program is based on a completely different
strategy than that of NuMI-to-Soudan. Instead of
measuring the disappearance of muon neutrinos at the far location 
and extracting the
oscillation parameters from the characteristic pattern induced by
oscillations in
the neutrino energy distribution, they have decided to go for direct
detection ({\it appearance}) of tau neutrinos in an almost pure $\nu_\mu$ beam. The two approaches should be regarded as complementary
and equally valuable, especially in a field where redundancy is a
requirement rather than a choice.

\begin{figure}[h]
\begin{center}
\psfig{file=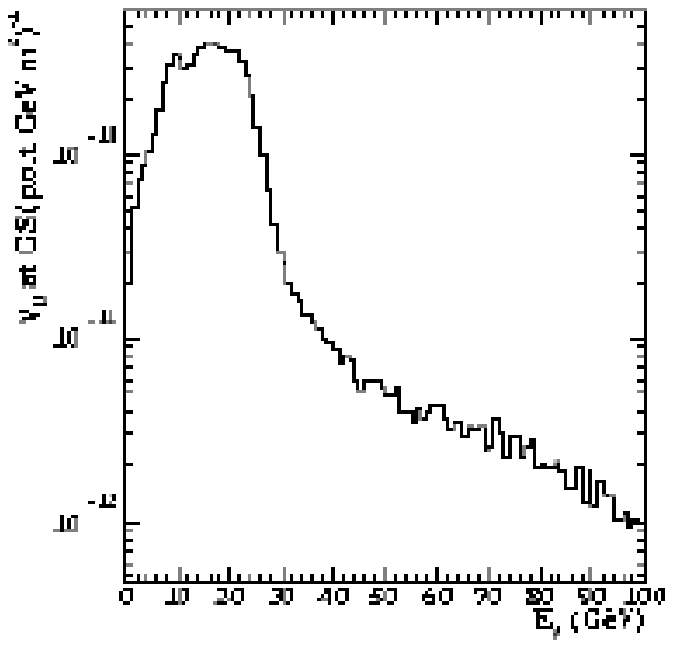,width=5cm}
\vspace*{13pt}
\fcaption{CNGS $\nu_\mu$ flux at the Gran Sasso site.}\label{fig:cngs_spectrum}
\end{center}
\end{figure}

CNGS is a wide-band neutrino beam of relatively high energy
($<E(\nu_\mu CC)>\sim 20~\gev$, see Fig.\ref{fig:cngs_spectrum}), which has been optimised to maximize 
the number of $\nu_\tau$CC events for appearance
experiments at Gran Sasso. It is expected to be commissioned by May 2005.
The beam design benefits from the
experience gained at CERN with the West Area Neutrino Beam
(WANF), in most recent times used by the short-baseline experiments
NOMAD\cite{nomad_nim} and CHORUS.\cite{chorus_nim} Primary protons of 
$400~\gev$ are extracted from the CERN-SPS and let interact with a
graphite target, where secondary particles, mostly pions and kaons, are produced. High energy
positively charged mesons are focused by means 
of two magnetic elements, the horn and the reflector, and then 
decay in a $900$~m long tunnel.  
The expected proton intensity is of $4.5\times 10^{19}$~pot/year.
The $\bar{\nu}_\mu$, $\nu_e$ and $\bar{\nu}_e$ contamination is very
low (in terms of CC interactions: $\sim 2\%$, $\sim 0.8\%$ and $0.05\%$
respectively), while the prompt $\nu_\tau$ contamination 
($O(10^{-6})$) is
essentially negligible for the purposes of $\nu_\tau$ appearance experiments.
Since the baseline is fixed (about $730$~km between the
production point at CERN and the Gran Sasso Laboratory), the low value 
of the $\Dm2$ suggested by atmospheric neutrino experiments would
require operation at the lowest energy possible; on the other hand,
neutrinos must be energetic enough to produce charged tau leptons in
charged current interactions. A compromise between these two competing 
requirements is achieved by adjusting the settings of the beam focusing system.
The expected number of $\nu_\tau$CC events at Gran Sasso per kton per
year are given in Tab.\ref{tab:rates_gran_sasso} 
(from ref.\cite{icanoe_proposal}).

\begin{table}[h]
\tcaption{Expected number of $\nu_\tau$ charged current interactions
  for an isoscalar target at Gran Sasso, for $\sin^22\theta = 1$ and
  different values of $\Dm2$, assuming $4.5\times 10^{19}$~pot/year
  (from ref.\cite{icanoe_proposal}). Detector efficiencies have not been taken 
  into account.}\label{tab:rates_gran_sasso}
\centerline{\footnotesize\smalllineskip
\begin{tabular}{c c c}\\
 {$\Dm2(\evolt^2)$}& \multicolumn{2}{c}{$N_{\nu_\tau CC}$~/kton/year}\\
\hhline{~--}
      {} & $E_\nu=1-30~\gev$ & $E_\nu = 1-100~\gev$\\
\hline\hline
$1\times 10^{-3}$ &  $2.34$ & $2.48$ \\
$3\times 10^{-3}$ & $20.7$ & $21.4$ \\
$5\times 10^{-3}$ & $55.9$ & $57.7$ \\
$1\times 10^{-2}$ & $195$  & $202$ \\
\hline
\end{tabular}}
\end{table}

OPERA\cite{opera_proposal} is an emulsion-based experiment, which aims 
to detect $\nu_\tau$CC events by identifying the $\tau$ produced 
through its decay kink. Emulsions are used as high precision trackers
rather than as the active target. The OPERA design, which largely
benefits from the experience acquired with the CHORUS\cite{chorus_nim}
and DONUT\cite{donut} experiments,
is an evolution of
the Emulsion Cloud Chamber (ECC) technique, where the high-precision
tracking capabilities of emulsions are
integrated with a passive material to provide a large target
mass. In the OPERA baseline design, an ECC cell
consists of a $1$~mm thick lead plate
followed by a thin film, made of two $50~\mu$m thick emulsion layers separated 
by a $200~\mu$m plastic base.  
The basic detector unit is obtained by stacking 
$56$ cells to form a compact {\it brick} ($10.2\times 12.7\times
\text{cm}^2$ transverse section, $10~X_0$ length, $8.3$~kg weight), 
which can in turn be
assembled into {\it walls}. An additional emulsion film is placed in
front of the first lead plate, to improve the track matching with the
upstream wall, while another film, the so-called Special Sheet,
separated from the most downstream cell by a $2$~mm plastic plate, is
used to have good $\tau$ detection efficiency also for events
occurring in the last lead plate.
After each wall, an electronic tracking detector is used to select the
brick where the neutrino interaction occurs. In the baseline design
the tracking detector in the target section  
follows very closely that of MINOS, with
scintillator strips coupled to wave-length shifting fibres and
multi-anode PMT readout, where the selected PMTs are the same as those used in
MINOS near detector. A wall of bricks and two planes of electronic
detectors constitute one OPERA {\it module}, whose structure is then
repeated $24$ times to form a {\it supermodule}, also including a
downstream muon spectrometer. Each muon spectrometer consists of a
dipolar magnet ($1.55$~T in the tracking region), made of two vertical 
walls of iron layers, interleaved with RPC detectors, and of drift
tubes, placed in front and behind the magnet as well as between the
two walls. Three supermodules form the full OPERA detector, for a total
target mass of about $1.8$~kton.

The data from the electronic detectors, which will be analysed {\it
  quasi-online}, are used to select bricks where neutrino
interactions occur and which have to be removed by an automated system. 
Additional bricks might need to be
removed for a complete reconstruction of the interesting events. Events
selected as $\tau$ candidates will then be sent to dedicated scanning
stations, where further studies can be performed to achieve the
required background suppression. The expected rate of $\nu_\mu$CC
events is of about $30$ events per day.

The $\tau$ decay channels considered by OPERA are listed in
Tab.\ref{tab:opera_decay_channels}, 
together with the corresponding
$\nu_\tau$ detection efficiencies and the expected number of background
 events\footnote{mostly from charm production, large angle muon scattering
   and hadron re-interactions.}\hspace{3pt} per year. New studies are
underway to improve the experiment sensitivity by adding also the
$\tau\rightarrow \rho\nu$ channel ($23.5\%$ B.R.) 
to the decay modes considered.
The expected number of $\tau$ events per year is given in
Tab.\ref{tab:opera_expected_ntau} as a function of possible values of $\Dm2$.

\begin{table}[h]
\tcaption{OPERA efficiency for the $\tau$ decay channels considered
  and the corresponding expected number of background events per year.
  The efficiencies, which also include the branching 
  ratio for each decay
  channel, have been obtained as weighted sums of 
  the efficiencies for DIS and non-scaling processes.
}\label{tab:opera_decay_channels}
\centerline{\footnotesize\smalllineskip
\begin{tabular}{c c c}\\
Decay mode & $\epsilon(\nu_\tau CC)$(\%) & $N_{BKGD}/$~year\\
\hline\hline
$\tau\rightarrow e$   & $3.7$ & $0.04$\\
$\tau\rightarrow\mu$  & $2.7$ & $0.03$\\
$\tau\rightarrow h$   & $2.3$ & $0.05$\\
\hline
\end{tabular}}
\vspace{1cm}
\tcaption{Expected number of $\tau$ events observed in OPERA per year as a
  function of possible values of $\Dm2$ (from ref.\cite{opera_nashville}).}\label{tab:opera_expected_ntau}
\centerline{\footnotesize\smalllineskip
\begin{tabular}{c c c}\\
$\Dm2 (\evolt^2)$ & $N_\tau$/year\\
\hline\hline
$1.5\times 10^{-3}$ & $0.82$\\
$2.5\times 10^{-3}$ & $2.82$\\
$3.2\times 10^{-3}$ & $3.66$\\
\hline
\end{tabular}}
\end{table}

\begin{figure}[h]
\begin{center}
\begin{tabular}{cc}
\epsfig{file=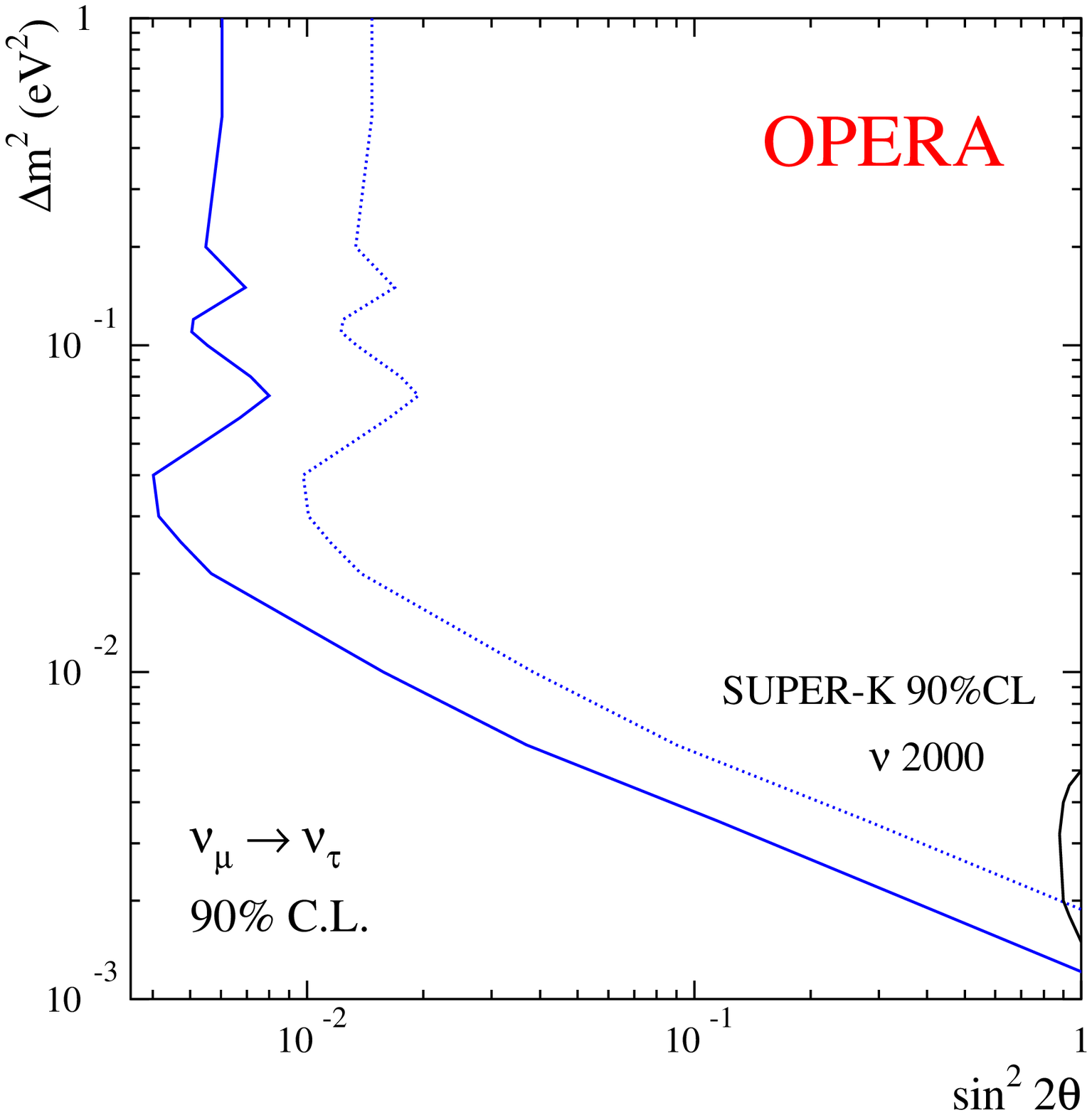,height=5.5cm}&
\epsfig{file=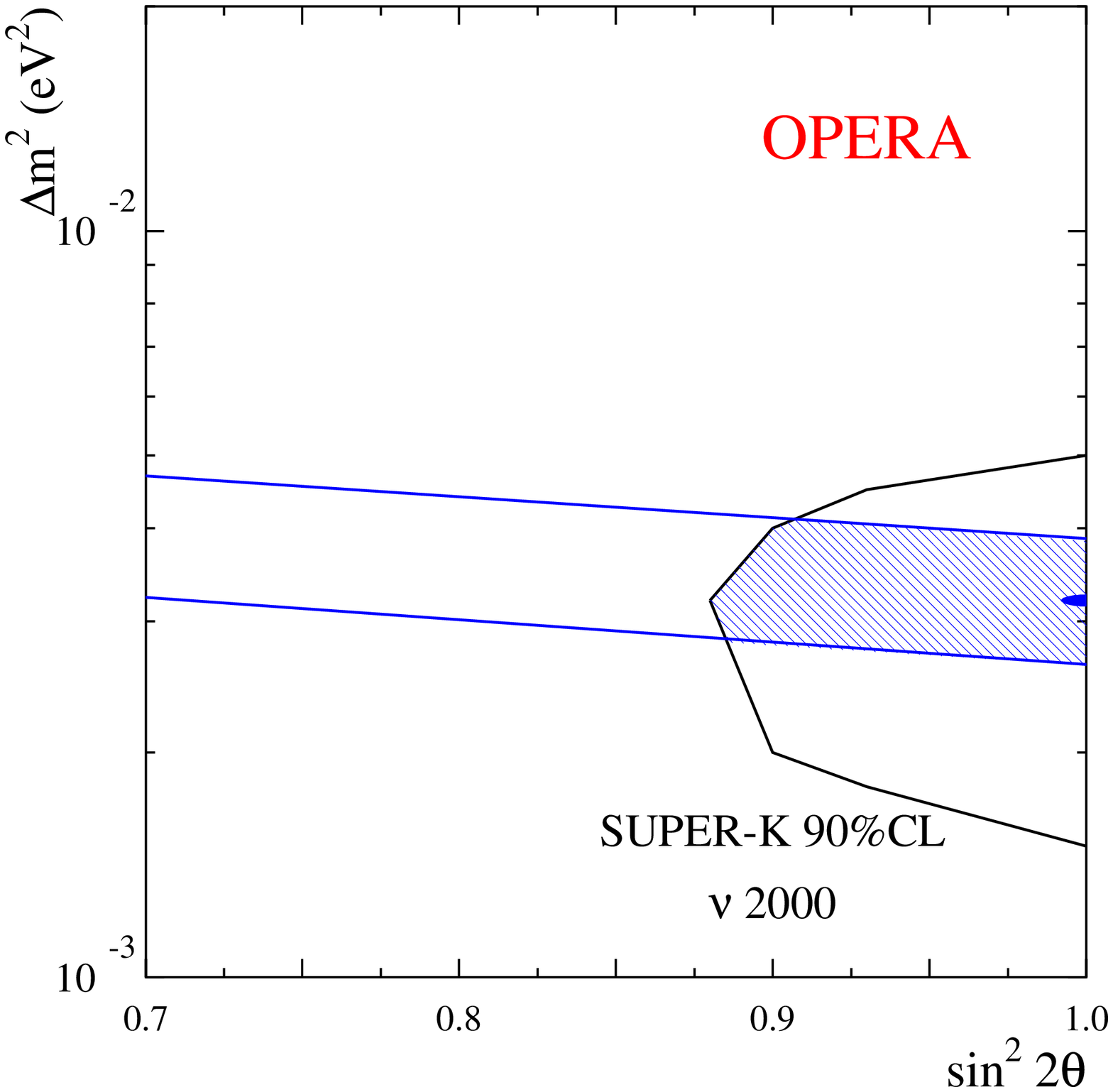,height=5.5cm}
\end{tabular}
\vspace*{13pt}
\fcaption{(Left) Average upper limit at $90\%$~C.L. which would be
obtained by OPERA if no oscillation signal is detected, after $2$ and
$5$ years of data-taking. (Right) $90\%$~C.L. allowed region for the oscillation
parameters as determined by OPERA after $5$ years of running.}\label{fig:opera_figures}
\end{center}
\end{figure}

If no signal is observed, the average upper limit at $90\%$~C.L. which would be
obtained by OPERA is shown in Fig.\ref{fig:opera_figures} (left) for
$2$ and $5$ years of exposure respectively. On the other hand, if
$\nu_\tau$ events are actually observed, a measurement of $\Dm2$ can 
be performed. Assuming maximal mixing and $\Dm2 = 3.2\times
10^{-3}~\evolt^2$, the $90\%$~C.L. allowed region for the oscillation
parameters as determined by OPERA after $5$ years of data taking is shown in
Fig.\ref{fig:opera_figures} (right).

\newpage
OPERA also plan to perform a $\nu_\mu\rightarrow\nu_e$ oscillation
analysis, by looking for $\nu_e$ appearance above the $\sim 1\%$ intrinsic
$\nu_e$ contamination, and to search for neutrino oscillations by
studying the ratio between NC and CC
interactions.\cite{opera_proposal} 
However the
study of the systematic uncertainties for those analyses is likely to
be somehow limited by the lack of a near detector.

The other experiment which is being planned at the Gran Sasso
laboratories to detect neutrinos from NCGS is based on the
ICARUS\cite{icarus_proposal,icarus_web_page} technology.
ICARUS is a liquid-Argon TPC which allows event reconstruction in
three dimensions, combining high spatial resolution
with excellent particle identification capabilities and precise
homogeneous calorimetry. ICARUS has been conceived as a multi-purpose
experiment with a very wide physics program, which goes from proton
decay searches to the study of solar and atmospheric neutrinos as well 
as of supernova neutrinos and, of course, of artificially produced
neutrinos from CERN.

The final phase of ICARUS foresees a 
detector with a sensitive mass of at least $5$~kton, but
a step-wise strategy has been adopted to demonstrate the technical
feasibility of such a challenging project. The first large-scale
prototype had a volume of $1,000\text{m}^3$ ($15$~ton) and was
built as a test-bench for the cryogenic system,
the internal detector mechanics and the liquid-Argon purification. The 
$15$~ton prototype was successfully tested between 1997 and 1999.
The next step towards the full-scale detector is the T600 module,\cite{icarus_t600} which is now being 
commissioned at Pavia and which will be transported to the Gran
Sasso Laboratory tunnel as soon as the technical tests are complete.
In 1999 a proposal was submitted for the ICANOE
experiment,\cite{icanoe_proposal} which combined the ICARUS
technology with a fine-grained iron calorimeter developed by the NOE
collaboration.\cite{noe} 

\begin{figure}[h]
\begin{center}
\epsfig{file=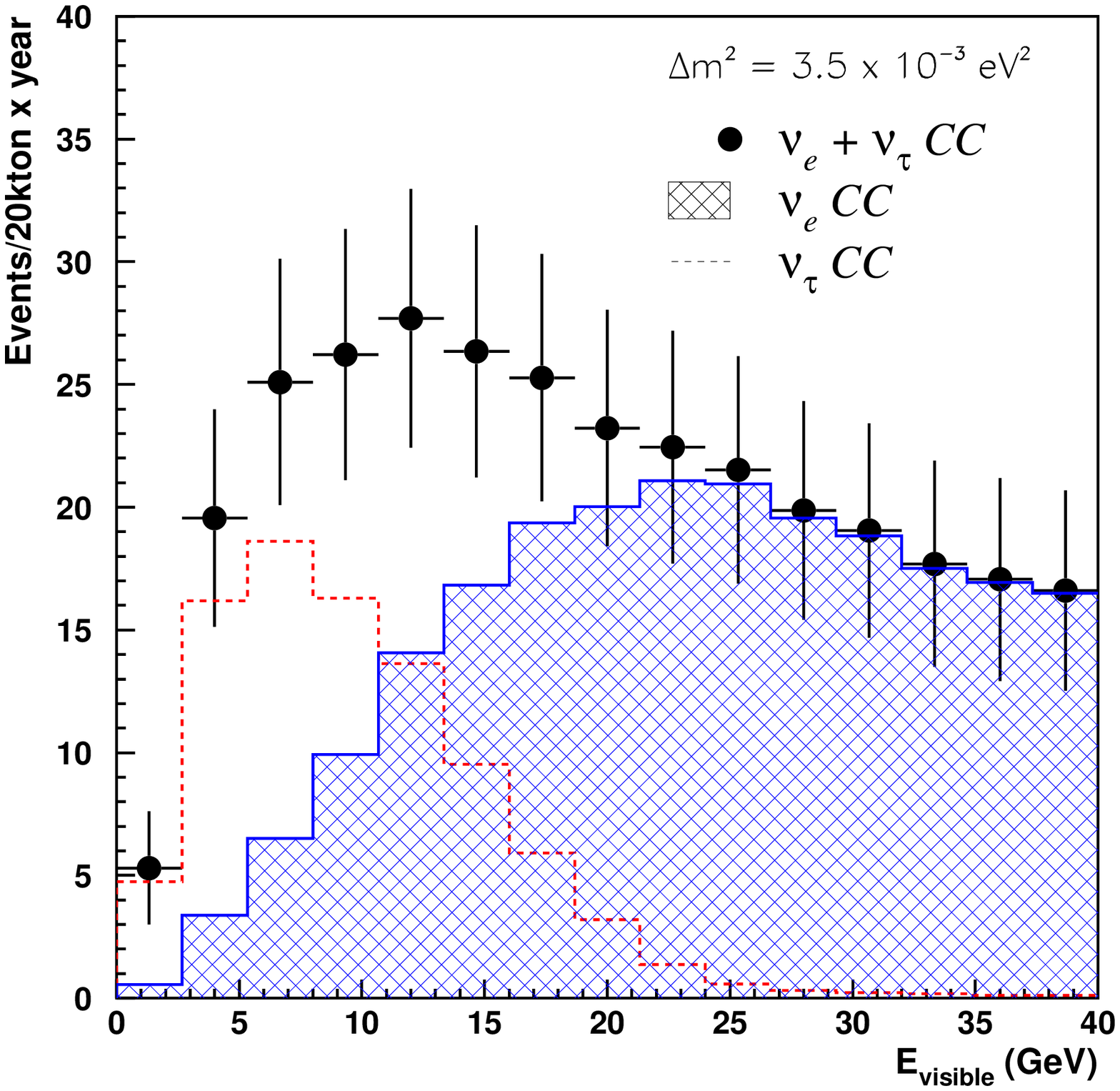,width=7cm,height=5.3cm}
\vspace*{13pt}
\fcaption{ICANOE expected visible energy distribution for events with
  a primary electron, in the case of 
  $\nu_\mu\rightarrow\nu_\tau$ oscillations with maximum mixing angle and
  $\Dm2 = 3.5\times 10^{-3}~\evolt^2$ (from ref.\cite{andre_nu2000}). Both the background from the prompt
  $\nu_e$ and $\bar{\nu}_e$ in the beam (hatched histogram) and the
  events from $\tau\rightarrow e$ decays (dashed line) have been
  considered. Errors are only statistical. }\label{fig:icanoe_taue_excess}
\end{center}
\end{figure}

Because of the high resolution on the event
kinematic variables, ICANOE could
search for $\nu_\tau$
appearance in the NCGS beam using kinematic selection criteria,
a strategy which has been successfully applied by the short-baseline
experiment NOMAD.\cite{nomad_tau} For ICANOE, as for NOMAD, the
$\tau\rightarrow e$ decay mode represents a golden channel.
After four years 
of running with intensity $4.5\times 10^{19}$~pot/year, 
the expected number of $\nu_\tau$CC events followed by a
$\tau\rightarrow e\nu\bar{\nu}$ decay would be about $110$ for $\Dm2
=3.5\times 10^{-3}~\evolt^2$, while the
total expected number of background events from prompt $\nu_e$ and
$\bar{\nu}_e$ is about $470$. This means that an excess of events
should be measured in the visible energy distribution 
even before applying any kinematic cut (see
Fig.\ref{fig:icanoe_taue_excess}).

The excellent electron identification in ICANOE would also allow
a search for $\nu_e$ appearance from $\nu_\mu\rightarrow\nu_e$
oscillations. Selection criteria
based on kinematic variables could be used to discriminate between events from 
$\nu_\mu\rightarrow\nu_e$ and events from $\tau\rightarrow e$ decays.
Assuming $\Delta m_{23}^2=3.5\times 10^{-3}~\evolt^2$ and $\theta_{23}=45^0$,
the statistical significance for a $\nu_\mu\rightarrow\nu_e$ signal
(after four years of running with intensity $4.5\times 10^{19}$~pot/year)
varies between $0.8~\sigma$ and $6.8~\sigma$ for $3^0<\theta_{13}<9^0$.\cite{andre_nu2000}

In summer 2000 the ICANOE collaboration ceased to exist as such,
but it is likely that as soon as the successful operation of the T600 
module is demonstrated, an updated version of the proposal will be
submitted to the relevant organisations.\cite{icarus_nashville}

Until now we have discussed the experimental hints for neutrino
oscillations coming from the study of naturally produced neutrinos
and we have described the experimental program which, in the near
future, should clarify many of the open issues concerning the solar and 
the atmospheric neutrino anomalies. There is a third hint for 
neutrino oscillations, which has the peculiarity of being the only one
coming from an accelerator experiment and the only one to represent 
evidence for neutrino oscillations in the appearance mode. This is the 
so-called {\it LSND effect}, which we shall discuss in the next section.

\section{The LSND Effect}

The LSND experiment\cite{lsnd_nim} at the Los
Alamos Meson Physics Facility (LAMPF), New Mexico, USA, 
was originally designed to
search for $\bar{\nu}_\mu\rightarrow\bar{\nu}_e$ oscillations using
$\bar{\nu}_\mu$ from $\mu^+$
decay at rest (DAR). In 1995
LSND reported an excess of electron-like events above 
background and gave an interpretation of that result in terms of neutrino oscillations.\cite{lsnd_dar_95}
The effect was later confirmed by an improved analysis with larger statistics,
both in DAR $\bar{\nu}_\mu$'s\cite{lsnd_dar_96} and
in the $\nu_\mu$ sample from $\pi^+$ decay in flight
(DIF).\cite{lsnd_dif_98} In the following we shall discuss the
results reported in a recent paper of the LSND
collaboration,\cite{lsnd_2001} where DAR and DIF events are treated in 
the same way, in a unified analysis of the two samples.

The neutrino beam used by LSND was obtained from the decay of the
secondary particles, mostly $\pi^+$ with a small fraction of $\pi^-$
($N_{\pi^-}/N_{\pi^+}\sim 1/8$), 
produced in the interaction of an intense ($\sim 1$~mA)   
$798~\mev$ proton beam with the primary target.

The decay chains
leading to the production of neutrinos and anti-neutrinos are:
\begin{align}
 \pi^+\rightarrow&\mu^+\nu_\mu\label{eq:lsns_pi+}\\
& \hookrightarrow e^+\nu_e\bar{\nu}_\mu \label{eq:lsns_mu+}
\end{align}
and the charge-symmetric sequence:
\begin{align}
 \pi^-\rightarrow&\mu^-\bar{\nu}_\mu\label{eq:lsns_pi-}\\
& \hookrightarrow e^-\bar{\nu}_e\nu_\mu\,. \label{eq:lsns_mu-}
\end{align}
\noindent
The fractions of $\pi^+$ DAR and DIF are $96.6\%$ and $3.4\%$ of the total
respectively. While only $\sim 0.05\%$ of the positive muons produced in
(\ref{eq:lsns_pi+}) decay in flight, most of them 
stop before decaying and then give
a standard Michel spectrum for $\nu_e$ and $\bar{\nu}_\mu$
($E_{max}(\bar{\nu}_\mu)=m_\mu/2 = 52.8~\mev$).
About $95\%$ of the  
$\pi^-$'s  are stopped in the target and immediately
captured, while the remaining $5\%$ decay in flight;
moreover, about $88\%$ of the muons from the
$\pi^-$ DIF are captured from atomic
orbit. In conclusion a suppression of the order of $\sim 1/8\times
0.05\times (1-0.88) = 7.5\times 10^{-4}$ is expected for 
the relative 
flux of prompt $\bar{\nu}_e$'s, which constitute a background for the 
$\bar{\nu}_\mu\rightarrow\bar{\nu}_e$ oscillation search. 
The $\nu_e$ flux from $\mu^+$ and 
$\pi^+$ DIF, representing a background for the
$\nu_\mu\rightarrow\nu_e$ oscillation search, is suppressed by the
long muon lifetime and by the small $\pi^+\rightarrow e^+\nu_e$ 
branching ratio ($1.23\times 10^{-4}$). 

In 1996 the production target, which was made of
water during
the running period 1993-1995, was upgraded: two 
upstream targets, used to provide beam to other
experiments, were removed and the water production target was replaced 
by tungsten. This resulted in a reduction of the DAR flux per 
pot and of the DIF flux per pot by $27\%$ and $34\%$ respectively.

The estimated systematic uncertainty on the beam predictions is $7\%$
for $\mu^+$ fluxes and $15\%$ for $\pi^\pm$ DIF and $\mu^-$ DAR fluxes.

The LSND detector was an approximately cylindrical tank ($8.3$~m long, 
$5.7$~m diameter), located $30$~m downstream from the neutrino source
(sensitivity to $\Dm2\sim 0.1-1~\evolt^2$) and filled with $167$~tons of 
mineral oil 
doped with scintillator ($0.031$~g/liter of butyl-PBD). Because of the low
scintillator concentration, both scintillation and Cherenkov light
could be detected by the $1220$ photo-multiplier installed on the
inside surface of the tank ($25\%$ coverage).
The Cherenkov cone and the time distribution of the light were used to 
tag electron events and to measure the event vertex and the $e^\pm$
direction. 
For any PMT activity above threshold, the pulse height, a digitised time
and the event position were recorded, so that
energy measurements and space and time correlation were
possible.
On all sides except the bottom the detector was surrounded
by an active shield (scintillator plus lead), providing a veto system
against charged particles ($10^{-5}$ 
inefficiency).
Triggers were not required to be
in-spill, but the beam state was recorded by the data acquisition and
a beam-on/beam-off comparison was used to perform 
a statistical subtraction of the residual cosmic ray background.

The identification of $\bar{\nu}_e$ events in LSND is based on the
detection of quasi-elastic scattering:
\begin{equation}\label{eq:lsnd_nuebar_signal}
\bar{\nu}_e + p \rightarrow e^+ + n
\end{equation}
\noindent
followed by the neutron capture process:
\begin{equation}\label{eq:delayed_gamma}
n+p \rightarrow d+\gamma (2.2~\mev)
\end{equation}
\noindent
Therefore the $\bar{\nu}_e$ signature in LSND is a delayed
coincidence between an electron signal and a spatially correlated 
$2.2~\mev$ photon.

Electron neutrinos are identified in LSND via the inclusive
charged-current reaction:
\begin{equation}
\nu_e C \rightarrow e^-X\,.
\end{equation}

Although the primary electron selection criteria have been optimize
for the DAR analysis, the same cuts are used to select also electrons
for the DIF analysis. The electron energy range defined for the
$\bar{\nu}_\mu\rightarrow\bar{\nu}_e$ search is $20<E_e<60~\mev$,
while events with $60<E_e<200~\mev$ are used for the
$\nu_\mu\rightarrow\nu_e$ analysis.
The cut at $20~\mev$ rejects 
the background events from the
$\beta$-decay of $^{12}B$ produced in the capture of cosmic $\mu^-$ on 
$^{12}C$, while events above $200~\mev$ are removed
to suppress the beam-related background from $\pi^+\rightarrow
e^+\nu_e$, which at those energies becomes
larger that any possible oscillation signal.
Cosmic ray events are further suppressed by using the veto information 
and by removing electron events with additional activity within
$8~\mu$s before the electron signal or $12~\mu$s after it.
Discrimination between correlated $2.2~\mev$ photons from neutron
capture and accidental $\gamma$'s from radioactivity is achieved by
means of a likelihood function, $R_\gamma$, which depends on the number of PMT
hits, the distance between the reconstructed $\gamma$ and $e^+$
positions and the time interval between the $\gamma$ and the $e^+$ 
events. The discrimination power of $R_\gamma$ has been checked by using 
$\nu_e C \rightarrow e^- N_{g.s.}$ exclusive interactions, where no recoil
neutron is present and which should therefore exhibit an $R_\gamma$
distribution consistent with that expected for purely accidental $\gamma$'s.

The two largest backgrounds to the $\bar{\nu}_\mu\rightarrow\bar{\nu}_e$ 
signal are $\bar{\nu}_e$CC 
interactions from  prompt
$\bar{\nu}_e$ in the beam, which are identical to the signal events,
or $\nu_\mu(\bar{\nu}_\mu)$CC
interactions with a correlated neutron signal and the muon 
misidentified as a positron. Both backgrounds have been estimated by using
the experiment Monte Carlo simulation and, whenever possible, by means 
of control samples from real data.

For the entire period 1993-1998 of data-taking,  
LSND observed a 
beam on-off excess of $117.9\pm 22.4$ events in the DAR channel. 
The corresponding expected number of background events from $\mu^-$ DAR
followed by $\bar{\nu}_ep\rightarrow e^+n$ scattering and from $\pi^-$ DIF
followed by $\bar{\nu}_\mu p\rightarrow\mu^+n$ scattering are $19.5\pm
3.9$ 
and $10.5\pm 4.6$ respectively, leading to a total excess of $87.9\pm
22.4\pm 6.0$ events above background expectations.\cite{lsnd_2001} The number of expected
$\bar{\nu}_e$ events from 
$\bar{\nu}_\mu\rightarrow\bar{\nu}_e$ oscillations with $100\%$
probability is $33000\pm 3300$, the error being a
combination of the systematic 
uncertainties on the neutrino flux ($7\%$) and on
the $e^+$ efficiency ($7\%$). The result then corresponds to an 
oscillation probability of $(0.264\pm 0.067(stat.)\pm 0.045(sys.))\%$,
\cite{lsnd_2001} to be compared with the previously published result of 
$(0.31\pm 0.12(stat.) \pm 0.05(sys.))\%$,
\cite{lsnd_dar_96} which was based on the 1993-1995 data sample.

The new DIF 
$\nu_\mu\rightarrow\nu_e$ 
oscillation analysis gives a beam on-off excess of 
$14.7\pm 12.2$ events. The estimated 
total background  
from the $\mu^+\rightarrow e^+\nu_e\bar{\nu}_\mu$, $\pi^+\rightarrow e^+$
and $\nu e\rightarrow \nu e$ processes is of $6.6\pm 1.7\ \nu_e$ events,
resulting in a total excess of $8.1\pm 12.2 \pm 1.7$ events above
background expectations. The
number of events expected for $\nu_\mu\rightarrow\nu_e$ oscillation
with $100\%$ probability is $7800$, thus giving an oscillation
probability of $(0.10\pm 0.16 \pm 0.04)\%$.\cite{lsnd_2001} This has
to be compared with the result obtained from the 1993-1995 data
analysis, which gave a probability of $(0.26\pm 0.10 \pm
0.05)\%$.\cite{lsnd_dif_98} 
Using the result from the DAR analysis and 
assuming that CP is conserved, the expected oscillation probability in 
the DIF sample is $\sim 0.26\%$ at high $\Dm2$ and $\sim 0.05\%$ at
low $\Dm2$.\cite{lsnd_2001}

A likelihood fit of the beam-on events which pass their oscillation cuts 
is used to extract the LSND allowed region in the
${\sin^22\theta,\Dm2}$ plane. The fit is performed over the entire
electron range $20-200~\mev$, so that both DAR and DIF events
are considered at the same time. The
probability density functions which define the likelihood function
${\mathcal{L}}$ 
used in the fit depend on the electron energy, the
electron longitudinal position along the tank, the cosine of the
reconstructed angle 
between the neutrino and the electron direction, the likelihood
ratio $R_\gamma$ for correlated $\gamma$ selection.
To account for the limited knowledge of the background, its 
contribution to ${\mathcal{L}}$ is allowed to vary with
gaussian profile around the expected central
value.\footnote{beam-related background are locked to vary together, while
  beam-unrelated backgrounds can fluctuate independently.}

\begin{figure}[h]
\vspace*{-1cm}
\begin{center}
\epsfig{file=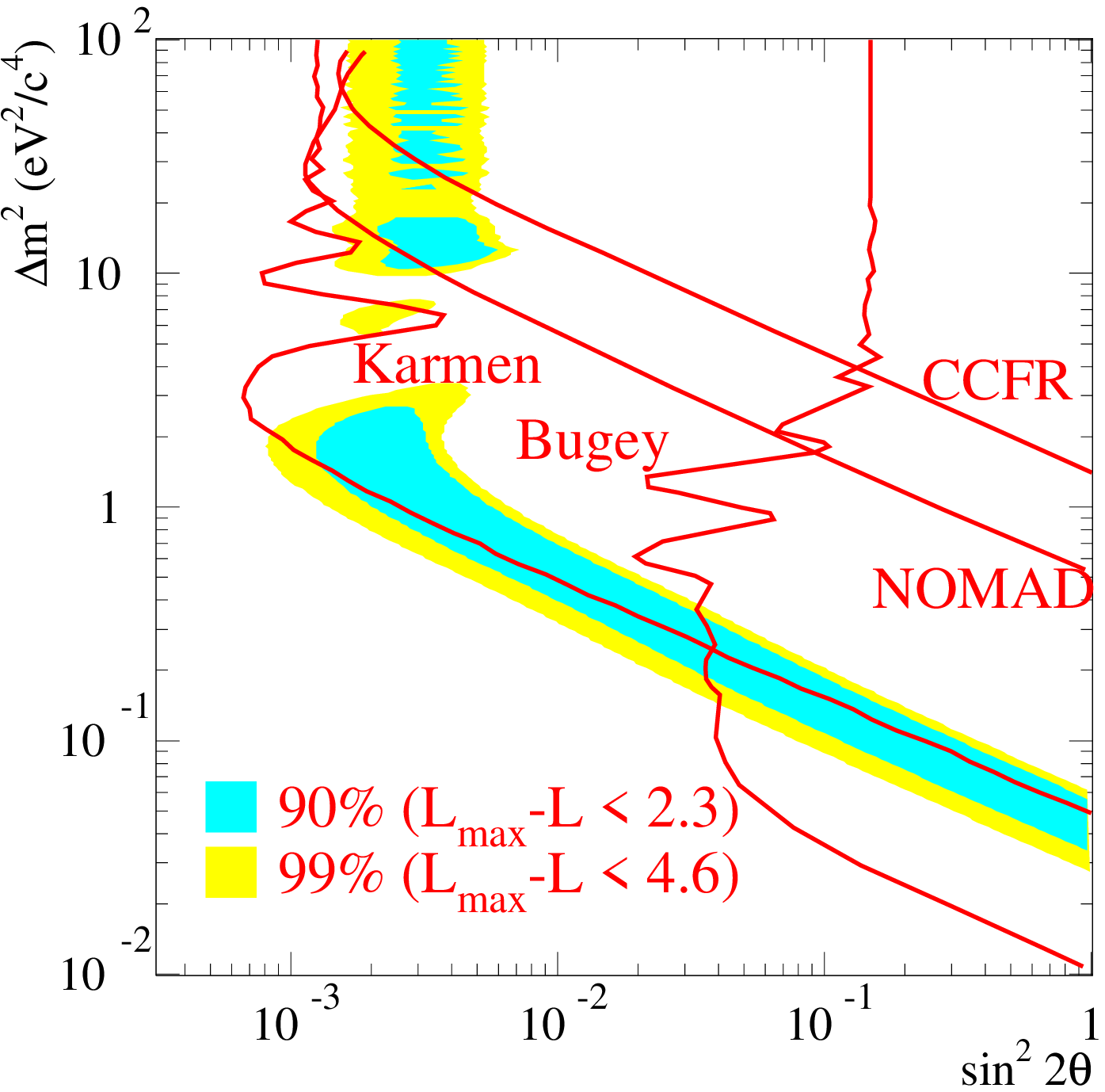,height=6cm}
\vspace*{8pt}
\fcaption{Allowed region obtained by LSND from a combined fit of the
  $\bar{\nu}_\mu\rightarrow\bar{\nu}_e$ and $\nu_\mu\rightarrow\nu_e$
  samples (from ref.\cite{lsnd_2001}; blue is $90\%$~C.L., yellow is $99\%$~C.L.).}\label{fig:lsnd_contour_2001}
\end{center}
\end{figure}

The result obtained is shown in Fig.\ref{fig:lsnd_contour_2001}, where 
the LSND allowed region is compared with limits obtained by other
experiments sensitive to $\nu_\mu\rightarrow\nu_e$
($\bar{\nu}_\mu\rightarrow\bar{\nu}_e$) oscillations in different
regions of the parameter space.\cite{karmen_nu2000,bugey,ccfr,nomad_nmne}
In particular it is interesting to briefly discuss the null result obtained by the
KARMEN2 collaboration, which has conducted a search for
$\bar{\nu}_\mu\rightarrow\bar{\nu}_e$ oscillations in approximately the 
same region of the parameter space as that investigated by LSND.

The KARMEN2 experiment was performed at the ISIS neutron spallation
facility at the Rutherford Appleton Laboratory, Chilton, Didcot, United
Kingdom.
As at LAMPF, neutrinos are obtained from the decay
at rest or in flight of secondary particles produced in the
interactions of $800~\mev$ primary protons with a massive beam stop
target. Also in this case the $\nu_\mu$'s from $\pi^+$ decays are
monoenergetic ($E_{\nu_\mu} = 29.8~\mev$), while both $\nu_e$'s and
$\bar{\nu}_\mu$'s have continuous energy spectra extending up to
$52.8~\mev$. One difference between the two beams is
that the one at ISIS has a well defined time structure: two very short
($\sim 100$~ns) proton pulses, separated by $225$~ns, are produced
with a repetition rate of $50$~Hz. Because of the different lifetimes
of their parent particles ($\tau_\pi = 26$~ns, $\tau_\mu = 2.2~\mu$s), 
the $\nu_\mu$-burst, which occurs essentially in coincidence with the 
proton pulses, can be clearly separated from the $\nu_e$- and
$\bar{\nu}_\mu$-induced events, which are instead characterised by a
much longer decay constant. Due to the accelerator duty cycle of
$10^{-5}$ (to be compared with the $6\times 10^{-2}$ duty factor for
LSND), the cosmic ray background is largely suppressed with
respect to the beam-related events. The proton 
beam intensity at ISIS is $0.2$~mA, that is to say a factor of five
lower than at LAMPF.

The KARMEN detector,\cite{karmen_detector} located $17.5$~m
downstream the neutrino source 
(about half the LSND baseline), consisted of a rectangular
tank filled with $56$~t liquid scintillator. The tank 
was segmented into $512$ optically independent modules,
wrapped with $\text{Gd}_2\text{O}_3$ coated paper to increase the
efficiency for the neutron capture process. The event position was
obtained from the time of the
PMT signal at the two ends of the hit module. Shielding against the
beam-related neutron background and the cosmic ray background was
provided by a $7$~kton steel blockhouse combined with a system
of two layers of veto counters. Moreover an additional veto counter,
added in 1996 for the KARMEN2 upgrade, was used to largely suppress the
background from energetic neutrons produced in $\nu_\mu$CC DIS
interactions.

The $\bar{\nu}_e$ detection in KARMEN2 was based on the delayed
coincidence between the positron signal from reaction
(\ref{eq:lsnd_nuebar_signal}) and either the $2.2~\mev$ photon 
of reaction (\ref{eq:delayed_gamma}) or the $8~\mev$ integrated photon
signal from Gd
de-excitation following neutron capture on gadolinium. No
particle identification was performed and the tagging of the positron event, 
which was expected within $2.2~\mu$s after the
beam-on-target signal, relied on the beam time structure
alone. Both time and spatial correlation between the
positron and the neutron capture events were required. 
An analysis of the data collected between February 
1997 and March 2000 shows that $11$ events have been
identified as $\bar{\nu}_e$ interactions, 
with an expected background\footnote{
Background events come from cosmic muon interactions, $\nu_e
^{12}C\rightarrow {e^-}^{12}N_{g.s.}$ sequences, prompt $\bar{\nu}_e$
contamination in the beam and neutrino-induced accidental
coincidences.}\hspace{3pt} of $12.3\pm 0.6$ events.\cite{karmen_nu2000} Therefore the KARMEN2 result is compatible with the
no-oscillation hypothesis and the corresponding exclusion plot at
$90\%$~C.L. is shown in Fig.\ref{fig:lsnd_contour_2001}.

Combining the LSND result with those obtained by all the other
experiments which, having studied the same transition, have obtained
null results, only a small portion of the LSND solutions is still
allowed to 
explain the experiment anomaly in terms of neutrino oscillation.
In particular, the skeptical reader may find the 
LSND and KARMEN2 results quite contradictory. 
However it has to be stressed that the two experiments have
sensitivities which peak at slightly different values of
$\Dm2$. Moreover, before any definitive conclusion can be drawn from
the various results, a combined analysis of all experimental data
should be performed. This has been tried for LSND and
KARMEN, but only on subset of the collected data.\cite{eitel_combined} 
In conclusion 
it is fair to say that no definitive 
answer has yet been given to the LSND puzzle and that more
experimental data is needed.

The MiniBOONE experiment,\cite{miniboone_proposal} which is now being
built at Fermilab, will address the LSND effect by searching for
$\nu_\mu\rightarrow\nu_e$ oscillations in the same region of 
parameter space. An almost pure $\nu_\mu$ beam ($\nu_e$ contamination
$<0.3\%$), with energy in the range $0.5-1~\gev$, will be obtained from the decay of secondary particles
produced in the interactions of $8~\gev$ protons 
with a beryllium target positioned in a magnetic horn. The protons
will be delivered by the Fermilab Booster, with an
expected intensity of $5\times 10^{20}$~pot per year. The decay
length for the positively charged secondary mesons, which are focused 
by the single horn, can be changed from $50$~m to $25$~m by inserting
an intermediate absorber inside the decay tunnel, thus allowing a
systematic study of the prompt $\nu_e$ contamination of the beam.
The MiniBOONE detector will be located about $500$~m from the neutrino 
production point, so that the $E/L$ ratio will be approximately
the same as LSND's. MiniBOONE will consists of a 
spherical tank ($12.2$~m
diameter) filled with $800$~tons of pure mineral oil. The Cherenkov
light produced in neutrino interactions in the inner part of the
detector ($445$~ton fiducial volume), as well as 
the modest amount of scintillation light from intrinsic
impurities in the oil, will be detected by $1280$ PMTs
($10\%$ coverage) 
installed on a support structure in the tank. 
This structure will also serve as an 
optical insulator for the 
inner part of the detector, 
while the outer $35$~cm thick oil layer, read out by $280$ additional
PMTs, can be used as a veto system.

The main background components will be due to the $\nu_e$
contamination in the beam and to the misidentification of muons and
$\pi^0$ as electrons.\footnote{ 
Given the higher neutrino energy
with respect to LSND, neutrons will not constitute a significant
background for MiniBOONE}\hspace{3pt}
The estimated systematic uncertainties on the
$\nu_e$ flux is about $5\%$ for the neutrino component originating from the
$\pi^+\rightarrow\mu^+\rightarrow\nu_e$ chain and $10\%$ for $\nu_e$'s
coming from $K_{e3}$ decays of secondary $K^+$ and $K^0_L$.
Particle identification will be performed using the
sharpness of the Cherenkov rings and the ratio between the amount of prompt
(Cherenkov) and late (scintillation) light, which is different for
electrons and muons. 
The misidentification uncertainty is expected to be less than $5\%$ for
muons and about $5\%$ for neutral pions: in the case of muons, the
the systematic error can be estimated using a sample of muon decays
from real data, while for $\pi^0$'s it will be obtained from a Monte
Carlo simulation, which can in turn be constrained using reconstructed
$\pi^0$'s from real data.

\begin{figure}[h]
\vspace*{7pt}
\begin{center}
\vspace{1.7cm}
\epsfig{file=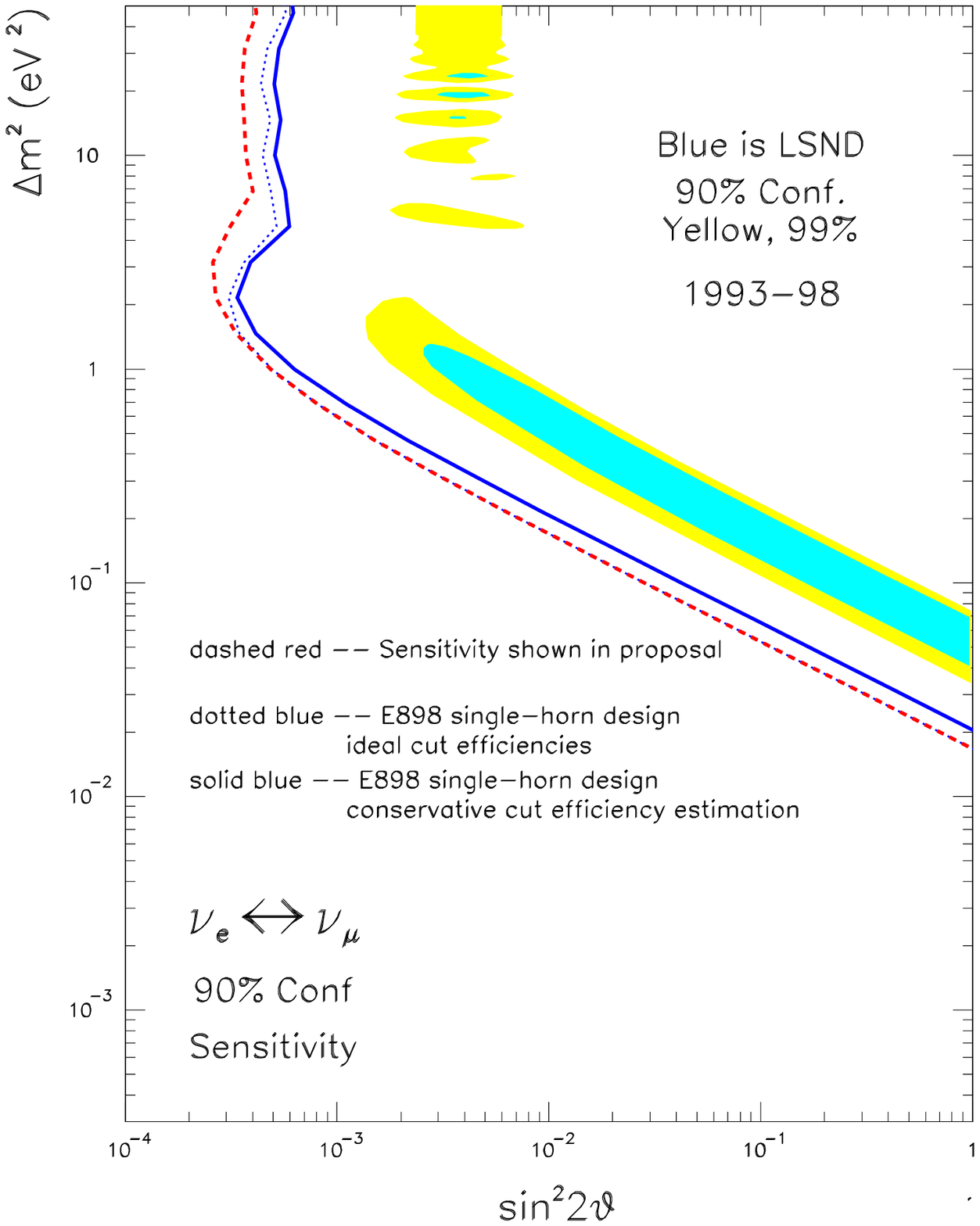,height=5cm}
\vspace{-1.cm}
\fcaption{Final $90\%$~C.L. expected sensitivity for
  MiniBOONE (figure from ref.\cite{miniboone_nu2000}). Different beam
  configurations and cut efficiencies have been considered.}\label{fig:miniboone}
\end{center}
\end{figure}

If oscillations occur with parameters compatible with LSND solutions,
after one year of running MiniBOONE should observe an excess of $\sim
1500~\nu_e$ events above background
expectations, with a
significance of $\sim 8-10\sigma$. On the other hand, 
in case of null result, after two years
of running
MiniBOONE should be able to completely exclude the entire LSND allowed 
region at $90\%$~C.L. (see Fig.\ref{fig:miniboone}).

\section{Neutrino Masses and Global Analysis of Neutrino Oscillation Data}

If we believe that all experimental hints on neutrino oscillations are 
correct and we accept that they are all manifestations of flavour 
mixing in the lepton sector, we are forced to introduce at least
another neutrino species in our framework. In fact,  
the $\Dm2$ regimes suggested by the 
the solar neutrino deficit, the atmospheric 
neutrino anomaly and the LSND effect are all well disconnected one from
the other:
\begin{equation}
\Dm2_{\odot}\ll
\Dm2_{atm} \ll
\Dm2_{LSND}\,.
\end{equation}
\noindent
On the other hand, only two
independent values of $\Dm2$ are possible for three different neutrino 
masses ($\Delta m^2_{31}=\Delta m^2_{32}+\Delta m^2_{21}$).
Therefore, recalling the LEP result on the number of active
neutrinos,\cite{pdg} in order to accommodate all experimental results,
an additional sterile neutrino would be required.

While both the solar and the atmospheric neutrino deficit
have been confirmed by more than one experiment, using different
experimental techniques, no other
experiment has confirmed LSND observations. For this reason, 
given also the enormous impact that accepting the LSND result would
have on our understanding of sub-nuclear phenomena, with the
introduction of a light sterile neutrino,
there is a 
tendency in the scientific community 
to adopt a ``conservative'' attitude and
to exclude LSND from most phenomenogical models and fits to neutrino
oscillation parameters, 
until more solid experimental data become 
available.
In the following we shall take the same 
approach and discuss some of the results obtained from three-neutrino 
analyses of the neutrino oscillation data, while 
only briefly commenting on
the possibility to have a fourth neutrino.

Neglecting possible CP violation phases, which are anyhow not
accessible to present experiments, the lepton mixing matrix
(\ref{eq:lepton_mixing}) depends only on five parameters, 
two $\Dm2$ and three
mixing angles, four of which can be identified with the parameters measured by
solar and atmospheric experiments:
\begin{align}
\Dm2_{\odot}   &\equiv \Dm2_{21}\notag\\
\Dm2_{atm}     &\equiv \Dm2_{32}\\
\theta_{\odot} &\equiv \theta_{12}\notag\\
\theta_{atm}   &\equiv \theta_{23}\notag
\end{align}
\noindent
while a limit on the fifth, $\theta_{13}$, is set by the CHOOZ reactor 
experiment.

The hierarchy $\Dm2_{\odot}\ll \Dm2_{atm}$ implies that
one of the three mass eigenstates, say $\nu_3$, is separated by a
larger gap 
$|\Dm2_{31}|\approx |\Dm2_{32}| \approx \Dm2_{atm}\sim 10^{-3}~\evolt^2$ from the other two, $\nu_1$ and
$\nu_2$, which are instead quasi-degenerate, being separated by
$|\Dm2_{21}|\approx \Dm2_{\odot}\lesssim 10^{-4}~\evolt^2$. 
However, since
oscillations determine only mass splittings and not the absolute value 
of the masses, it is still possible to shift the absolute mass scale
without affecting the oscillation phenomenology. 
Although different mass orderings are possible, here we consider 
the normal hierarchical scheme, corresponding to:
\begin{equation}\label{eq:normal_hierarchy}
m_1<m_2\ll m_3\,
\end{equation}
\noindent
and thus:
\begin{equation}
\Dm2_{21}\ll\Dm2_{32}\approx\Dm2_{31}\,.
\end{equation}
\noindent

It can be shown that,
given the strong hierarchical constraint 
$\Dm2_{\odot}\ll \Dm2_{atm}$, 
the solar neutrino data 
can be described in terms of only three parameters, $\Dm2_{21}$,
$\theta_{12}$ and $\theta_{13}$, while the atmospheric neutrino data
can be analysed in terms of $\Dm2_{32}$,
$\theta_{23}$ and $\theta_{13}$. 
For the CHOOZ data, the survival probability depends only on
$\theta_{13}$ and $\Dm2_{32}$.
Thus $\theta_{13}$ is 
the only parameter common
to the three data sets. In particular, 
in the limit $\theta_{13}=0$ atmospheric and solar neutrino
oscillation can be described by
two independent two-flavour oscillation analyses.

Several analyses of the available data have been performed 
in the three-neutrino scheme:\cite{gonzalez-garcia,fogli} 
in the following we shall refer to the results presented in
ref.\cite{gonzalez-garcia}. In particular it will be 
interesting to see how
the inclusion of different data in the fit affects the constraint
which can be extracted on the mixing angle $\theta_{13}$.

The three-dimensional analysis of ref.\cite{gonzalez-garcia} of the
solar neutrino data has shown that 
solutions with small values of the mixing angle $\theta_{13}$ are
preferred, the most favourable scenario being that of two-flavour
mixing ($\theta_{13}=0$). The 
upper limit which can be extracted on the mixing angle based on
solar neutrino data only is
$\tan^2\theta_{13}<2.4$ ($\theta_{13}<57^0$) at $90\%$~C.L.. 

If only atmospheric neutrinos are considered,
the analysis favours the $\nm\rightarrow\nt$
oscillation hypothesis. Also in this case
the best fit corresponds
to a small value of the mixing angle $\theta_{13}$ ($\theta_{13}=9^0$). 
However oscillations between
muon and electron neutrinos are still allowed as a sub-dominant mode,
with the following $90\%$~C.L.
allowed ranges for the relevant oscillation
parameters:
\begin{align}
1.6\times 10^{-3}~\evolt^2 < &\Dm2_{32} < 6\times 10^{-3}~\evolt^2 \\
0.43<&\tan^2\theta_{23}<4.2\\
&\tan^2\theta_{13}<0.34\end{align}

If the CHOOZ result is included in the atmospheric neutrino
analysis, because the full allowed region of $\Dm2_{32}$
lies within the parameter domain accessible to CHOOZ, a tighter
constraint can be obtained on the mixing angle $\theta_{13}$ and the
corresponding $90\%$~C.L. limit becomes even more stringent:
$\tan^2\theta_{13}<0.043$ ($\theta_{13}<12^0$).

Finally, a combined analysis of solar, atmospheric and reactor data
gives, at $90\%$~C.L., the following allowed ranges for $\Dm2_{32}$,
$\tan^2\theta_{23}$ and $\tan^2\theta_{13}$, where the limits on
$\tan^2\theta_{13}$ are given for the SMA solution of the solar
neutrino deficit as well as for the LMA or unconstrained
case:\footnote{in the unconstrained case no assumption is made about
  the solution to the solar neutrino deficit.}
\begin{align}
1.4\times 10^{-3}~\evolt^2 < &\Dm2_{32} < 6\times 10^{-3}~\evolt^2 \\
0.39<&\tan^2\theta_{23}<3.1\\
&\tan^2\theta_{13}<0.055\hspace{1cm}\text{(unconstrained or LMA)} \\
&\tan^2\theta_{13}<0.075\hspace{1cm}\text{(SMA)} 
\end{align}

\begin{figure}[h]
\vspace*{13pt}
\begin{center}
\epsfig{file=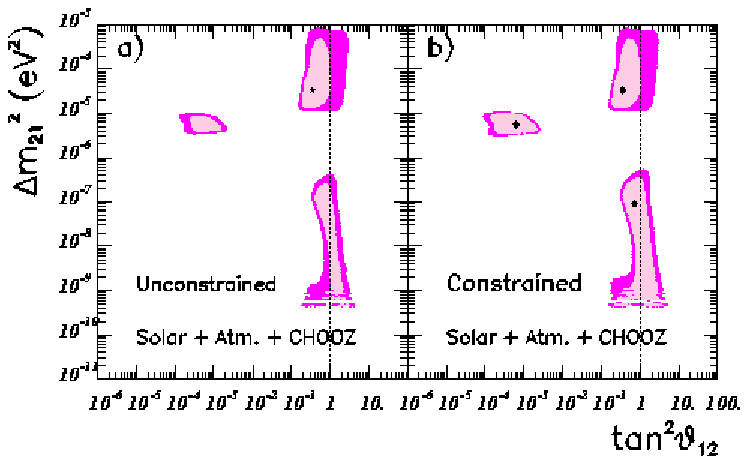,width=0.80\linewidth}
\vspace*{13pt}
\fcaption{Allowed regions in the plane
  $\{\tan^2\theta_{12},\Dm2_{21}\}$, as obtained from the global
  three-neutrino analysis of solar, atmospheric and reactor data of
  ref.\cite{gonzalez-garcia}, for both the unconstrained (left) and the 
  constrained fit (right). 
  The star denotes the global best fit solution 
  for the unconstrained fit and local best fit points for the
  constrained fit.}\label{fig:allowed_dm21_th12_gonzalez}
\end{center}
\end{figure}

The allowed regions for $\Dm2_{21}$ and $\theta_{12}$ obtained from
this global analysis are shown in
Fig.\ref{fig:allowed_dm21_th12_gonzalez}, both for the unconstrained
and constrained case. It is interesting to notice that, in the
unconstrained case, the best-fit solution lies in the LMA region.

In summary, the analysis in ref.\cite{gonzalez-garcia} 
shows that, although all solutions to the
solar neutrino deficit are still allowed, the LMA solution 
appears to be favoured. Moreover both solar and atmospheric
neutrino data seem to prefer small values of the mixing angle
$\theta_{13}$, the constraint becoming more stringent with the
inclusion of the CHOOZ result. Limits have been extracted also on the
other relevant oscillation parameters.

Ignoring CP violation, in the limit $U_{e3}\sim 0$ (that is
to say $s_{13}\sim 0$ and 
$c_{13}\sim 1$) the MNS matrix (\ref{eq:lepton_mixing}) 
can be written as:
\begin{equation}\label{eq:lepton_mixing_small_s13}
  U \simeq
  \begin{pmatrix}
    c_{12} & 
    s_{12} & 
    0  \\
    -s_{12}c_{23} &
    c_{12}c_{23}  &
    s_{23}\\
    s_{12}s_{23}  &
    -c_{12}s_{23} &
    c_{23}
  \end{pmatrix}\,,
\end{equation}

One appealing
realization of this matrix would be that corresponding to 
{\it  bi-maximal mixing} (see, for example, ref.\cite{fritzsch_xing}), 
which postulates that both
atmospheric and solar neutrino anomalies can be interpreted as
neutrino oscillations with maximal mixing (that it to say both
$\theta_{23}$ and $\theta_{12}$ 
equals $\pi/4$). In this case:
\begin{equation}\label{eq:bimaximal_mixing}
\begin{pmatrix}
\dfrac{1}{\sqrt{2}} & \dfrac{1}{\sqrt{2}} & 0 \\
-\dfrac{1}{2} & \dfrac{1}{2} & \dfrac{1}{\sqrt{2}}\\
\dfrac{1}{2} & -\dfrac{1}{2} & \dfrac{1}{\sqrt{2}}\\
\end{pmatrix}\,.
\end{equation}
\noindent
However it has to be noticed that this form of the mixing matrix seems
to be 
incompatible with the LMA solution of the solar neutrino deficit.\cite{giunti_bimax}

If one tries to define a coherent picture which accommodates also 
the LSND result, a
fourth neutrino has to be introduced. 
Essentially only two mass schemes can used 
to describe the mass spectrum, the so called $3+1$ and $2+2$ schemes. 
The $3+1$ scheme, which seems to be
disfavoured by experimental data,\cite{giunti_grimus_3+1,barger_pakvasa}
corresponds to having three relatively close neutrino masses, whose
spacing is determined by the two $\Dm2$ from the solar
and atmospheric neutrino
analyses, separated by about $1~\evolt^2$ from an isolated level, which
is responsible for the LSND anomaly. In the $2+2$ scheme, instead, two 
doublets of mass eigenstates, spaced by $\Dm2_{\odot}$ and
$\Dm2_{atm}$ respectively, are separated by the LSND gap. Several
analyses of the solar and atmospheric neutrino data in the $2+2$
four-neutrino framework have been recently presented, 
showing that, although Super-Kamiokande data disfavour both the pure
$\nu_\mu\rightarrow\nu_s$ and the pure $\nu_\mu\rightarrow\nu_e$
channels, it is not possible to exclude their occurrence 
with sizable oscillation amplitude.\cite{fogli_4nu,gonzalez-garcia_4nu}

\section{Conclusions}

In this paper we have reviewed the current status of the search for
neutrino oscillations. 
After an introduction of the formalism of
neutrino oscillation phenomenology, we 
have discussed the various experimental
results which, over the past three decades, 
more or less convincingly, have provided hints for 
the existence of flavour mixing in the neutrino sector:
the solar neutrino deficit, the atmospheric neutrino anomaly and the
LSND effect. If any of these indications will be
confirmed by future experiments, we would be forced to extend our
description of sub-nuclear phenomena beyond the Standard Model.

Having described the different experimental techniques
used to explore different domains of the oscillation parameter space,
and having discussed the sources of systematic error
intrinsic to each method, we hope we have given most of the
information necessary to discern the robustness of 
each method and result.
However, whatever the preference of the reader may be,
it is undeniable that more experimental data is urgently needed, before 
a clearer 
picture of neutrino oscillation phenomenology can be drawn.

The near and medium term experimental programme, which is supposed to
answer many of the as yet unresolved puzzles, has been described in the
previous sections. However it is possible that the experiments already
approved will not be able to provide a definitive 
solution to all our pending questions.
In particular, none of the experiments scheduled to
start running over the next few years will have the sensitivity to
address CP violation in the neutrino sector and, except if the actual
value of $\theta_{13}$ is very close to the present CHOOZ limit, 
none of them will be able to
measure that mixing angle. 
Very ambitious projects, such as {\it neutrino factories}\cite{nufact} or
super-intense neutrino beams ({\it
  super-beams}),\cite{superbeams} if ever approved, might hold the key
to the ultimate solution to the fascinating neutrino mystery.

To conclude, whatever the outcome of future experiments, 
there is no doubt this is an extremely exciting
time for neutrino physics.

\nonumsection{Acknowledgements}
\noindent
The author would like to thank J.~H.~Cobb and A.~Weber, for helpful and 
stimulating discussions during the preparation of this review and for
their careful reading of the manuscript, and L.~DiLella and C.~Yanagisawa, for
having kindly provided the OPERA sensitivity plots and the latest
Super-Kamiokande figures respectively.

\nonumsection{References}

\end{document}